\documentclass{aastex6}

\shorttitle{On the classification of GRBs and their occurrence rates}
\shortauthors{Ruffini et al.}

\begin{document}

\title{On the classification of GRBs and their occurrence rates}

\author{R. Ruffini\altaffilmark{1,2,3,4}, J.~A. Rueda\altaffilmark{1,2,4}, M. Muccino\altaffilmark{1,2}, Y. Aimuratov\altaffilmark{1,2}, L.~M. Becerra\altaffilmark{1,2}, C.~L. Bianco\altaffilmark{1,2}, M. Kovacevic\altaffilmark{1,2,3}, R. Moradi\altaffilmark{1,2}, F.~G. Oliveira\altaffilmark{1,2,3}, G.~B. Pisani\altaffilmark{1,2}, Y.~Wang\altaffilmark{1,2}}

\altaffiltext{1}{Dip. di Fisica and ICRA, Sapienza Universit\`a di Roma, Piazzale Aldo Moro 5, I--00185 Rome, Italy.}
\altaffiltext{2}{ICRANet, Piazza della Repubblica 10, I--65122 Pescara, Italy.}
\altaffiltext{3}{Universit\'e de Nice Sophia Antipolis, CEDEX 2, Grand Ch\^{a}teau Parc Valrose, Nice, France.}
\altaffiltext{4}{ICRANet-Rio, Centro Brasileiro de Pesquisas F\'isicas, Rua Dr. Xavier Sigaud 150, 22290--180 Rio de Janeiro, Brazil.}

\begin{abstract}
There is mounting evidence for the binary nature of the progenitors of gamma-ray bursts (GRBs). For a long GRB, the induced gravitational collapse (IGC) paradigm proposes as progenitor, or ``in-state", a tight binary system composed of a carbon-oxygen core (CO$_{\rm core}$) undergoing a supernova (SN) explosion which triggers hypercritical accretion onto a neutron star (NS) companion. For a short GRB, a NS-NS merger is traditionally adopted as the progenitor. We divide long and short GRBs into two sub-classes, depending on whether or not a black hole (BH) is formed in the merger or in the hypercritical accretion process exceeding the NS critical mass. For long bursts, when no BH is formed we have the sub-class of X-ray flashes (XRFs), with isotropic energy $E_{iso}\lesssim10^{52}$ erg and rest-frame spectral peak energy $E_{p,i}\lesssim200$~keV. When a BH is formed we have the sub-class of binary-driven hypernovae (BdHNe), with $E_{iso}\gtrsim10^{52}$ erg and $E_{p,i}\gtrsim200$~keV. In analogy, short bursts are similarly divided into two sub-classes. When no BH is formed, short gamma-ray flashes (S-GRFs) occur, with $E_{iso}\lesssim10^{52}$ erg and $E_{p,i}\lesssim2$~MeV. When a BH is formed, the authentic short GRBs (S-GRBs) occur, with $E_{iso}\gtrsim10^{52}$ erg and $E_{p,i}\gtrsim2$~MeV. We give examples and observational signatures of these four sub-classes and their rate of occurrence. From their respective rates it is possible that ``in-states'' of S-GRFs and S-GRBs originate from the ``out-states'' of XRFs. We indicate two additional progenitor systems: white dwarf-NS and BH-NS. These systems have hybrid features between long and short bursts. In the case of S-GRBs and BdHNe evidence is given of the coincidence of the onset of the high energy GeV emission with the birth of a Kerr BH.
\end{abstract}


\maketitle

\section{Introduction}\label{sec:1}

On February 1974, at the same AAAS meeting in San Francisco where the discovery of GRBs by the Vela satellites was publicly announced \citep{Strong1975}, the possible relation of GRBs with the ``moment of gravitational collapse'' leading to a BH formation was advanced \citep[see][]{GurskyRuffini1975}.
\citet{1975PhRvL..35..463D} considered, for definiteness, the vacuum polarization process occurring in an overcritical Kerr-Newman BH (KNBH). Evidence was given for: a) the formation of a vast amount of $e^+e^-$-baryon plasma; b) the energetics of GRBs of the order of $E_{\rm max}\approx10^{54} M_{\rm BH}/M_\odot$~erg, where $M_{\rm BH}$ is the BH mass, implying their cosmological origin; c) the ultra-high energy cosmic rays with energy up to $\sim10^{20}$~eV originating from such an extreme electrodynamical process. Soon after, the role of an $e^+e^-$ plasma for the origin of GRBs was also considered by \citet{CavalloRees}. It took almost thirty years to clarify some of the analogies and differences between these two processes of $e^+e^-$-pair creation leading, respectively, to the alternative concepts of ``fireball" and ``fireshell" \citep{2007PhRvL..99l5003A,2009PhRvD..79d3008A}.

Already in $1989$, well before the establishment of the GRB cosmological nature and energetics, \citet{Eichler1989} gave support to the cosmological interpretation of GRBs and indicated in merging NS binaries their possible origin.
They also pointed out the relevance of such NS--NS mergers for the occurrence of r-process, as well as for the emission of gravitational radiation, indicating the uncertainty in the determination of their rate of occurrence.

Following the launch of the Compton satellite and the observations by the BATSE detector \citep{Meegan1992}, a phenomenological classification based on the prompt $T_{90}$ duration was advanced: GRBs were classified into long GRBs for $T_{90}>2$~s, and short GRBs for $T_{90}<2$~s \citep{Klebesadel1992,Dezalay1992,Koveliotou1993,Tavani1998}.

Shortly after \citet{Narayan1992} indicated the possible cosmological origin of short GRBs originating in binary NS mergers. 
They also introduced the clear indication of the role of $\nu\bar{\nu}$ annihilation leading to the formation of an $e^+e^-$ plasma.
This paper was followed by a large number of theoretical works including the gravitational wave emission in Newtonian, post-Newtonian, and general relativistic treatments \citep[see, e.g.,][]{1999CQGra..16R...1R}, as well as the $\nu\bar{\nu}$ annihilation leading to an $e^+e^-$ plasma (see, e.g., \citealp{SalmonsonWilson2002} and \citealp{Rosswog2003} and references therein).

Soon after the paper by \citet{Narayan1992}, \citet{Woosley1993} also supported the cosmological origin of GRBs and introduced the concept of BH-Accretion-Disks, produced by the collapse of a very massive star. Such a system was indicated by its author as a \textit{collapsar} and was assumed to be the origin of ultrarelativistic jets expected to occur by the same author in long GRBs. For a recent review see \citet{MacFadyenWoosley,2001ApJ...550..410M,2006ARA&A..44..507W}.

After the determination of the cosmological nature of GRBs \citep{Costa1997,1997Natur.387..878M,vanParadijs1997} and the confirmation of their outstanding energy ($\approx10^{54}$ erg), we returned to our GRB scenario \citep{1975PhRvL..35..463D}.
In a period of four years, from 1997 to 2001, we developed a fully relativistic GRB theoretical model examining, as well, the dynamics of the $e^+e^-$ plasma originating the GRB emission \citep[the fireshell model, see, e.g.,][and Section~\ref{sec:fireshell}]{Preparata,RSWX2,RSWX,Ruffini2001c,Ruffini2001,Ruffini2001a}.
The fireshell model applies to both short and long GRBs.

The origin of short GRBs from NS--NS (or NS-BH) binaries as ``in-states'' has been confirmed by strong observational and theoretical evidences \citep[see, e.g.,][]{Goodman1986,Paczynski1986,Eichler1989,Narayan1991,Narayan1992,MeszarosRees1997_b,Rosswog2003,Lee2004,2014ARA&A..52...43B}.
In this article we address specifically some of the latest results within the fireshell model \citep{Muccino2012,2015ApJ...808..190R} on the possible presence or absence of a BH formation in NS--NS mergers, the consequent classification of short bursts into S-GRFs, when no BH is formed (see Section~\ref{sec:descr_S-GRFs}), and S-GRBs, when a BH is formed (see Section~\ref{sec:descr_SGRBs}), and the computation of their occurrence rate (see Section~\ref{sec:rates}).

The application of the fireshell model to the case of long GRBs followed a longer path for reaching a proper understanding of the overall phenomenon. 
The first application of our model to a long GRB was implemented on GRB 991216 \citep{Ruffini2001c,Ruffini2001,Ruffini2001a,Ruffini2002,Ruffini2004,2006AdSpR..38.1291R}. 
In these papers a clear difference between the thermal component observed at the transparency of the $e^+e^-$ plasma, the Proper GRB (P-GRB) emission \citep{RSWX2,RSWX}, and the non-thermal remaining part, later called prompt emission \citep{Ruffini2001}, was evidenced. 
This fully relativistic approach was not readily accepted by the GRB community, also in view of its objective technical complexity and novelties in the theoretical physics scenario. 
Some authors attempted to describe the GRB phenomenon by simplified Newtonian approaches,e.g., those based on the concept of \textit{magnetars} \citep{1992Natur.357..472U,1998A&A...333L..87D,1998PhRvL..81.4301D,1998ApJ...505L.113K,2001ApJ...552L..35Z}.
As the detailed observations of the X-ray afterglow by the \textit{Swift}-XRT \citep{2007A&A...469..379E} were obtained, as well as the high energy emission by the Fermi-LAT \citep{Atwood2009}, our model has correspondingly evolved pointing out the precise common power-law behavior of the rest-frame $0.3$--$10$ keV X-ray luminosity light curves \citep{Pisani2013}, as well as the nesting properties \citep{Ruffini2014}. 
As pointed out in the present article, the concept of long GRBs has evolved into XRFs and BdHNe, depending on the possible presence or absence of a BH in their formation process \citep[see also][]{2015ApJ...798...10R}.

It is appropriate to recall that the quest for having progenitors for the collapsar hypothesized by \citet{Woosley1993} led to an interesting direction of research dealing with a binary system composed of two very massive stars of $\approx50$~M$_\odot$ each. The large masses involved in these systems were introduced in order to form a BH at the end of their evolution. Similarly, the large amount of angular momentum of the system would guarantee the formation of an accretion disks needed in the collapsar model \citep{1999ApJ...526..152F}.
Up to six different scenarios were there envisaged leading to a \textit{collapsar}, as well as a few leading, alternatively, to a variety of binary compact systems. 
The need for choosing low-metallicity massive stars followed from the expectation of the formation large BHs in their evolution \citep{1999ApJ...526..152F}. The elimination of H from metal-rich massive stars would follow naturally, but the formation of a BH was not expected in their final stage of evolution \citep{Woosley1993}.

The spatial and temporal coincidence of a long GRB explosion with an optical SN, first observed in the association between GRB 980425 and SN 1998bw, created a profound conceptual turmoil in the GRB community. Woosley and collaborators postulated the birth of a SN out of a collapsar \citep[see, e.g.,][and references therein]{2006ARA&A..44..507W}.

In our approach, GRBs were supposed to originate from the BH formation, while SNe were expected to lead only to NSs \citep[see, e.g.,][and references therin]{2015ARep...59..591R}. 
We consequently introduced a new paradigm to explain the coincidence of these two qualitative and quantitative different astrophysical events in space and time: the birth of a SN and the occurrence of a GRB.
The induced gravitational collapse (IGC) paradigm was then introduced \citep[see, e.g.,][]{Ruffini2001c,2006tmgm.meet..369R,Ruffini2007b,2008mgm..conf..368R,2012A&A...548L...5I,2012ApJ...758L...7R,2014ApJ...793L..36F}.
This approach differs from alternative descriptions of the GRB-SN coincidences occurring, e.g., in the \textit{magnetars} and the \textit{collapsar} models, where the two events are coming from a single progenitor star, and takes full advantage of the recent observations of SNe Ibc in interacting binary systems \citep{2009ARA&A..47...63S}.

In a first formulation we considered a finely tuned process: the GRB triggering the explosion of a binary companion star very close to the onset of the SN \citep{Ruffini2001c}. 
This scenario soon led to the alternative IGC paradigm in which a CO$_{\rm core}$ undergoes a SN explosion in presence of a NS companion in a tight binary system. 
This is also by itself an unlikely event which in order to occur needs a fine tuning of the initial conditions of the binary system.
This scenario was shown to be consistent with population synthesis analysis \citep{1999ApJ...526..152F,2015arXiv150502809F}. 
The SN explosion induces a hypercritical accretion of its ejecta onto the companion NS, leading to the formation of a more massive NS (MNS), when the NS critical mass $M_{\rm crit}$ is not reached, or to the formation of a BH with the associated GRB emission in the opposite case \citep[see, e.g.,][]{2012ApJ...758L...7R,2014ApJ...793L..36F}.
The IGC scenario was first tested and verified in GRB 090618 \citep{Izzo2012,2012A&A...548L...5I}.
It soon became clear that the occurrence of a GRB is far from being a single event, but it is part of an authentic laboratory composed of a variety of astrophysical relativistic phenomena preceeding and following the prompt GRB emission.

We adopted as progenitor of our CO$_{\rm core}$--NS binary system the massive binaries independently considered in \citet{1999ApJ...526..152F} and \citet{1994Natur.371..227N,1995PhR...256..173N}. 
In our case, the late evolution of such massive binary systems do not lead to a collapsar, nor to an hypernova, but to a much richer and vast number of possibilities, made possible by our IGC paradigm.
Consistently with the considerations by \citet{2005ApJ...629..311S}, indicating that XRFs, X-ray rich bursts, and all long GRBs are part of a same population which we show to originate in the hypercritical accretion process of the SN ejecta onto a binary companion NS.
 
In agreement with the considerations by \citet{Soderberg2006Nature,Guetta2007,Liang07} for a sub-classification of long bursts into low-luminous and high-luminous GRBs, we have divided the long bursts into two different scenarios depending on the distance between the CO$_{\rm core}$ and the NS binary companion \citep{2015ApJ...812..100B}.
Correspondingly two different sub-classes of long bursts, both originating in the hypercritical accretion process of the IGC scenario, have been shown to exist \citep{2015ApJ...798...10R,2015ApJ...812..100B}: the XRFs, which clearly include low-luminous GRBs, such as GRB 060218 \citep{Campana2006}, when no BH is formed (see Section~\ref{sec_3_prot_rat}), and the BdHNe, such as GRB 130427A \citep{2015ApJ...798...10R}, when a BH is formed (see Section~\ref{sec:descr_BdHNe}).
Their occurrence rates have been computed (see Section~\ref{sec:rates}). 
Instead of proposing just a new classification, we also give the description of its underlying physical origin: the hypercritical accretion process of the SN ejecta onto a binary companion NS, with the full associated theoretical treatments at the basis of the IGC paradigm.

To the above four sub-classes of long and short bursts, we have recently added a new hybrid sub-class of ultrashort GRBs (U-GRBs) which, as recently pointed out in \citet{2015arXiv150502809F}, can originate during the further evolution of the BdHNe out-states. Indeed, nearly 100$\%$ of the NS-BH binaries which are the outcome of BdHNe remain bound. Their orbital velocities are high and even large kicks are unlikely to unbind these systems. They represent a new family of NS-BH binaries unaccounted for in current standard population synthesis analyses \citep[see, e.g.,][and Section~\ref{sec:rate5}]{2015arXiv150502809F}.

The above considerations based on the IGC paradigm and the NS-NS paradigm as progenitors encompass and classify into sub-classes most of the known astrophysical systems related to GRBs. 
We finally recall the existence of a class of long GRBs occurring in a low density circumburst medium (CBM) with density $\sim10^{-3}$ cm$^{-3}$, with hybrid short/long burst properties in their $\gamma$-ray light curves: 1) an initial spike-like harder emission and 2) a prolonged softer emission observed up to a hundred seconds. These bursts do not have an associated SN, even though in the case of a low value of the cosmological redshift its detection would not be precluded. The prototype of such systems is GRB 060614 \citep{DellaValle2006sn}. The progenitor for this class of long bursts has been identified in a binary system composed of a NS and a white dwarf (WD) \citep{Caito2009}. Their merger leads to a MNS with additional orbiting material, but not to an authentic GRB. We refer to these systems, historically addressed as disguised short GRB, as gamma-ray flashes (GRFs), see, e.g., GRB 060614, \citealp{Caito2009} and GRB 071227, \citealp{Caito2010}).

\begin{figure*}[!hbtp]
\centering
\includegraphics[width=0.9\hsize,clip]{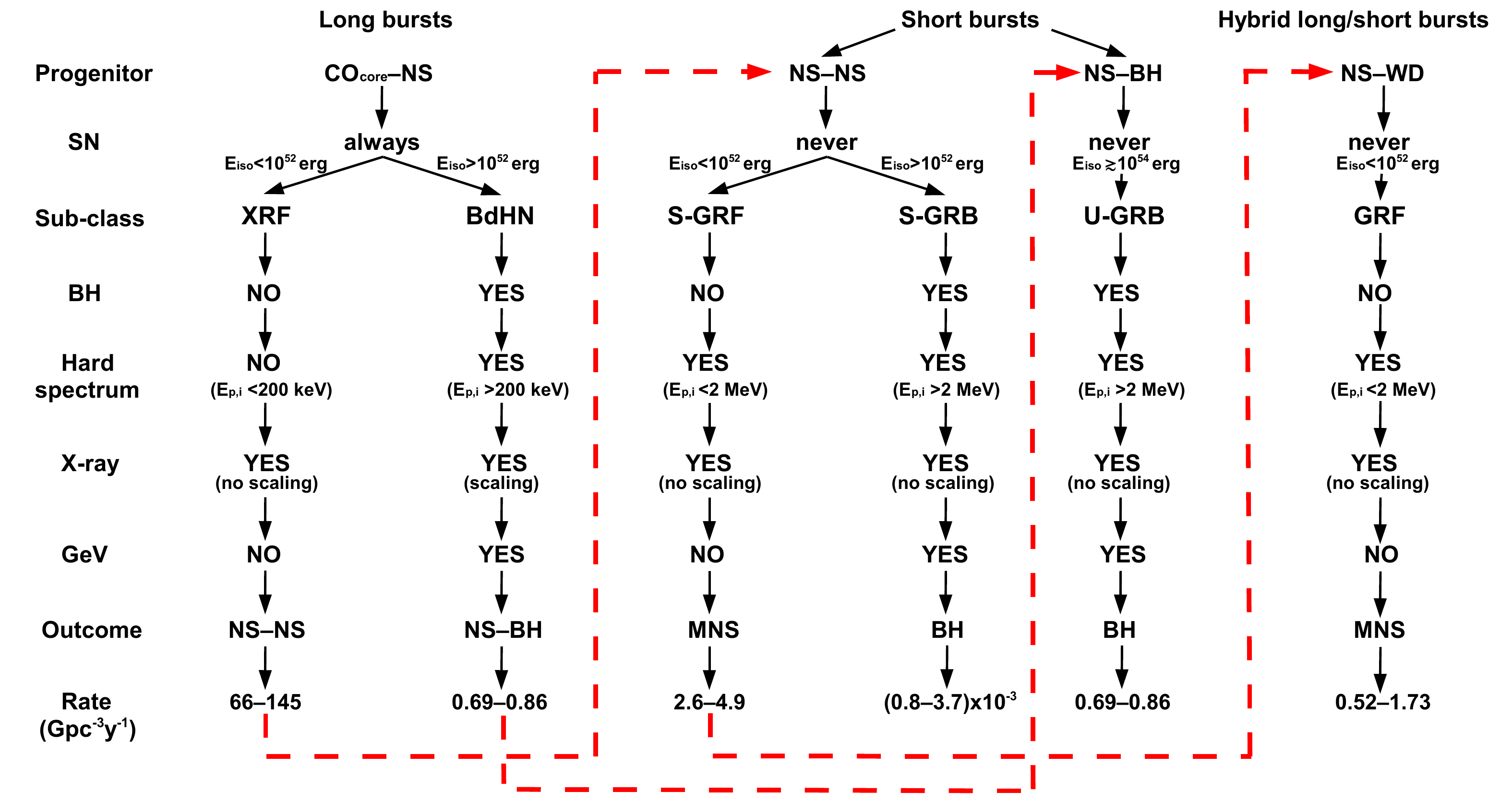}
\caption{Summary of the properties of long, short and hybrid long/short burst sub-classes discussed in the Introduction. The red dashed lines indicate the evolutionary tracks linking the out-states and the in-states of some of the sub-classes considered in this work.}
\label{fig:summary}
\end{figure*}

In the following we adopt the term \textit{burst} only for those systems leading to the BH formation, namely to S-GRBs, U-GRBs and BdHNe. We refer to the term \textit{flash}, instead, only for those systems not leading to the BH formation, namely to S-GRFs, GRFs and XRFs (see Fig.~\ref{fig:summary}).

The main topic addressed in the present article is to estimate the rates of occurrence of the XRFs, BdHNe, S-GRFs, S-GRBs, U-GRBs, and GRFs and to give a generel description of these GRB sub-classes. 
In Section~\ref{sec:fireshell} we present a short summary on the fireshell model. 
In Section~\ref{sec:1052erg} we discuss the $10^{52}$~erg lower limit in binary systems leading to BH formation.
After describing the observational properties of the above sub-classes, their interpretation within the IGC paradigm, the NS-NS merger scenario and the fireshell scenario, we present some prototypes (see Sections~\ref{sec_3_prot_rat}, \ref{sec:descr_BdHNe}, \ref{sec:descr_S-GRFs}, \ref{sec:descr_SGRBs}, \ref{sec:rate5}, and \ref{sec:DS}, respectively).
We then proceed in Section~\ref{sec:rates} to estimate their observed occurrence rates and to compare and contrast our results with those outlined in the literature \citep[see, e.g.,][]{Soderberg2006Nature,Guetta2007,Liang07,Vir09,Virgili2011,2010tsra.confE.204R,2010MNRAS.406.1944W,2015MNRAS.448.3026W,Kovacevic2014,2015ApJ...812...33S}. 
We then draw some general conclusions in Section~\ref{sec:conclusions}.

A standard flat ${\Lambda}$CDM cosmological model with $\Omega_M=0.27$, $\Omega_\Lambda=0.73$, and $H_0=71$ km s$^{-1}$ Mpc$^{-1}$ is adopted throughout the paper. A summary of acronyms used throughout the paper is shown in Table~\ref{acronyms}.
\begin{table}
\centering
\begin{tabular}{lc}
\hline\hline
Extended wording & Acronym \\
\hline
Binary-driven hypernova & BdHN \\
Black hole                    & BH \\
Carbon-oxygen core      & CO$_{\rm core}$ \\ 
Circumburst medium     & CBM \\
EQuiTemporal Surfaces  & EQTS \\
Gamma-ray burst         & GRB \\
Gamma-ray flash          & GRF \\
Induced gravitational collapse & IGC \\
Kerr-Newman black hole & KNBH \\
Massive neutron star     & MNS \\
Neutron star                & NS \\
New neutron star          & $\nu$NS \\
Proper gamma-ray burst & P-GRB \\
Short gamma-ray burst  & S-GRB \\
Short gamma-ray flash  & S-GRF \\
Supernova                  & SN \\
Ultrashort gamma-ray burst & U-GRB \\ 
White dwarf                & WD \\
X-ray flash                  & XRF \\
\hline
\end{tabular}
\caption{Alphabetic ordered list of the acronyms used in this work.}
\label{acronyms}
\end{table}

\section{Summary of the fireshell model}\label{sec:fireshell}

The fireshell model for GRBs \citep[see, e.g.][]{Ruffini2001c,Ruffini2001,Ruffini2001a} has been introduced to explain the GRB phenomenon as originating in the gravitational collapse leading to the formation of a BH 
\citep{1975PhRvL..35..463D}. The GRB emission results by taking into proper account relativistic magneto-hydrodynamical effects, quantum-electrodynamical process, and relativistic space-time transformations.

The role of the relativistic magneto-hydrodynamical effects arising in the gravitational collapse of a globally neutral magnetized plasma has been first considered in \citet{1975PhRvD..12.2959R}, where the occurrence of a local charge separation, during a globally neutral accretion process, led to the development of overcritical electric fields at the onset of a KNBH formation.\footnote{Overcritical electric fields are defined as larger than the critical value $E_c=m_e^2c^3/(\hbar e)$, where $m_e$ is the electron mass, $c$ the speed of light in the vacuum, $\hbar$ the reduced Planck constant, and $e$ the electron charge.}
These overcritical fields and, consequently, the vacuum polarization process leading to the creation of an $e^+e^-$ plasma, have been considered in \citet{1975PhRvL..35..463D}, for the sake of definiteness in a KNBH, as the energy source of GRBs:\footnote{The role of an $e^+e^-$ plasma for the origin of GRBs was also considered independently by \citet{CavalloRees}.} the pair creation process is fully reversible and as a result a highly efficient energy extraction mechanism occurs, which may deliver as much as $E_{\rm max}\approx10^{54} M_{\rm BH}/M_\odot$~erg.

Later on, the concept of \textit{dyadotorus} for a KNBH \citep{Cherubini,RRKerr} has been introduced to describe the region where pair creation occurs, leading to the formation of a BH.
The dynamics of an optically thick \textit{fireshell} of $e^+e^-$ plasma of total energy $E_{e^+e^-}^{\mathrm{tot}}$, i.e., its expansion and self-acceleration due to its own internal pressure, has been described in \citet{RSWX2}. 
The effect of baryonic contamination, i.e., the remnant of the collapsed object, on the dynamics of the fireshell has been then considered in \citet{RSWX}, where it has been shown that even after the engulfment of a baryonic mass $M_B$, quantified by the baryon load $B=M_Bc^2/E_{e^+e^-}^{\mathrm{tot}}$, the fireshell remains still optically thick and continues its self-acceleration up to ultrarelativistic velocities \citep{2007PhRvL..99l5003A,2009PhRvD..79d3008A}. 
When the fireshell reaches transparency condition, a flash of thermal radiation termed P-GRB is emitted \citep{RSWX2,RSWX}.
The dynamics of the fireshell up to the transparency condition is fully described by $E_{e^+e^-}^{\mathrm{tot}}$ and $B$: solutions with $B\leq10^{-2}$ are characterized by regular relativistic expansion; for $B>10^{-2}$ turbulence and instabilities occur \citep{RSWX}.

The P-GRB emission is followed by the prompt emission \citep{Ruffini2001}. 
The prompt emission originates in the collisions of the accelerated baryons left over after transparency, moving at Lorentz factor $\Gamma\approx100$--$1000$, with interstellar clouds of CBM \citep{Ruffini2002,Ruffini2004,Ruffini2005}.
These interactions give rise to a modified blackbody spectrum in the co-moving frame \citep{Patricelli}. 
The resulting observed spectral shape, once the constant arrival time effect is taken into account in the EQuiTemporal Surfaces \citep[EQTS, see][]{Bianco2005b,Bianco2005a}, is in general non-thermal, as the result of the convolution of a large number of modified thermal spectra with different Lorentz factors and temperatures.
To reproduce the prompt emission light curve and spectra three additional parameters, all related to the properties of the CBM, are required: the CBM density profile $n_{\mathrm{CBM}}$, the filling factor $\mathcal{R}$ that accounts for the size of the effective emitting area, and a low-energy power-law index $\alpha$ of the modified black body spectrum \citep{Patricelli}. These parameters are obtained by running a trial-and-error simulation of the observed prompt emission light curves and spectra.

To describe the dynamics of such an $e^+e^-$-baryon plasma from the vicinity of a BH all the way up to ultrarelativistic velocities at the infinity, both in the P-GRB and the prompt emission, the appropriate relative spacetime transformation paradigm has been discussed in \citet{Ruffini2001a}. It relates the observed GRB signal to its past light cone, defining the events on the worldline of the source that is essential for the interpretation of the data. Particular attention has been there given to the explicit equations relating the comoving time, the laboratory time, the arrival time, and the arrival time at the detector corrected by the cosmological effects, consistently with the equation of motion of the system \citep[see also][]{Bianco2004,Bianco2005b,Bianco2005a,2006ApJ...644L.105B}, compared and contrasted with the corresponding treatments in the literature \citep[see, e.g.,][]{Sari1997,Sari1998,Waxman1997,Panaitescu1998,PanaitescuMeszaros,ReesMeszaros1998,Granot1999,Chiang1999}.

As recalled above, the evolution of a baryon-loaded pair plasma, is generally described in terms of $E_{e^+e^-}^{\mathrm{tot}}$ and $B$ and it is independent of the way the pair plasma is created. Given this generality, in addition to the specific case of the dyadotorus mentioned above, these concepts can be applied as well in the case of a pair plasma created via $\nu \bar{\nu}\leftrightarrow e^+e^-$ mechanism in a NS merger as described in \citet{Narayan1992}, \citet{SalmonsonWilson2002}, and \citet{Rosswog2003}, or in the hyper-accretion disks around BHs as described in \citet{Woosley1993} and \citet{2011MNRAS.410.2302Z}, assuming that the created pair plasma is optically thick. The relative role of neutrino and weak interactions vs. the electromagnetic interactions in building the dyadotorus is currently topic of intense research.

In conclusion, the deeper understanding of the GRB phenomenon, occurring under very different initial conditions, has highlighted the possibility of using the general description of the dyadosphere (dyadotorus) to any source of an optically thick baryon-loaded $e^+e^-$ plasma and, consequently, to apply the above fireshell treatment in total generality.

The generality of the fireshell approach clearly differs from alternative treatments purporting late activity from a central engine (see, e.g., the \textit{collapsar} model in \citealt{Woosley1993}, \citealt{1999ApJ...518..356P}, \citealt{2006ARA&A..44..507W} and references therein, and the Newtonian \textit{magnetar} model in \citealt{2001ApJ...552L..35Z}, \citealt{2006Sci...311.1127D}, \citealt{2011MNRAS.413.2031M}, \citealt{2012MNRAS.419.1537B}, \citealt{2013ApJ...771L..26G}, \citealt{2014ApJ...785...74L}, and references therein), and proposes a different explanation for the afterglow observations in long GRBs \citep[see][and Aimuratov et al., in preparation]{Pisani2013}.

\section{On the $10^{52}$~erg lower limit in binary systems leading to BH formation}\label{sec:1052erg}

During the hypercritical accretion process onto the NS the total energy available to be released, e.g. in form of neutrinos and photons, is given by the gain of gravitational potential energy of the matter being accreted by the NS \citep{Zeldovich1972,RRWilson1973,2012ApJ...758L...7R,2014ApJ...793L..36F}. 
The total energy released in the star in a time-interval $dt$ during the accretion of an amount of mass $dM_b$ with  angular  momentum $l \dot{M}_b$ is given by \citep[see, e.g.,][]{2000AstL...26..772S,2015ApJ...812..100B}:
\begin{equation}
L_{\rm acc}=\left(\dot{M}_b - \dot{M}_{\rm NS}\right)c^2=\dot{M}_b c^2 \left[1-\left(\frac{\partial M_{\rm NS}}{\partial J_{\rm NS}}\right)_{M_b}\,l -\left(\frac{\partial M_{\rm NS}}{\partial M_b}\right)_{J_{\rm NS}}\right]\ ,
\label{eq:DiskLuminosity}
\end{equation}
where $J$ is the NS angular momentum. The last two terms of the above equation take into due account the change of binding energy of the NS while accreting both matter and angular momentum. 
We assume, as a norm, a typical NS mass of $\approx1.4$~M$_\odot$, a value observed in galactic NS binaries \citep{2011A&A...527A..83Z,2014arXiv1407.3404A} and characteristic of the XRFs \citep{2016arXiv160602523B}. 
We also assume a NS critical mass $M_{crit}$ in the range from $2.2$~M$_\odot$ up to $3.4~M_\odot$ depending on the equations of state and angular momentum \citep[see][for details]{2016arXiv160602523B,2015ApJ...812..100B,2015PhRvD..92b3007C}. $L_{\rm acc}$ is clearly a function both of the NS mass and of $M_{\rm crit}$.

Since $L_{\rm acc}\propto\dot{M}_b$, it evolves with time similarly to $\dot{M}_B$. 
We have shown that the accretion luminosity can be as high as $L_{\rm acc}\sim 0.1 \dot{M_b} c^2\sim 10^{47}$--$10^{51}$~erg~s$^{-1}$ for accretion rates $\dot{M_b}\sim 10^{-6}$--$10^{-2}~M_\odot$~s$^{-1}$ \citep[see][for details]{2016arXiv160602523B,2015ApJ...812..100B}. 
The duration of the accretion process is given approximately by the flow time of the slowest layers of the SN ejecta to the NS companion. If the velocity of these layers is $v_{\rm inner}$, then $\Delta t_{\rm acc}\sim a/v_{\rm inner}$, where $a$ is the binary separation. 
For $a\sim 10^{11}$~cm and $v_{\rm inner}\sim 10^8$~cm~s$^{-1}$ we obtain $\Delta t_{\rm acc}\sim 10^3$~s, while for shorter binary separation, e.g.~$a\sim 10^{10}$~cm ($P\sim 5$~min), $\Delta t_{\rm acc}\sim 10^2$~s. 
These estimates are validated by our numerical simulations \citep[see, e.g.,][]{2016arXiv160602523B,2015ApJ...812..100B,2015arXiv150502809F,2014ApJ...793L..36F}. 
From the above results we obtain that for systems with the above short orbital periods the NS collapses to a BH, namely BdHNe \citep{2016arXiv160602523B}, and a total energy larger than the separatrix energy of $\approx10^{52}$~erg is released during the hypercritical accretion process. 
For systems with larger separations, in which the hypercritical accretion is not sufficient to induce the collapse of the NS into a BH, namely the XRFs \citep{2016arXiv160602523B}, the value of $\approx10^{52}$~erg represents a theoretical estimate of the upper limit to the energy emitted by norm in the hypercritical accretion process.
These considerations are derived from theoretical expectations based on the above mentioned masses for the accreting NSs and $M_{crit}$.
Indeed, they are in satisfactory agreement with the observations of $20$ XRFs and $233$ BdHNe (considered up to the end of 2014) which we have used in our sample (see Table~\ref{tab:XRFs} and \ref{tab:BdHNe}, respectively).
The upper limit for the XRFs is $(7.3\pm0.7)\times10^{51}$~erg (see Sec.~\ref{sec:XRFs1}), while the lower limit for the BdHNe is $(9.2\pm1.3)\times10^{51}$~erg (see Sec.~\ref{sec:BdHNe1}).

The same arguments apply to the fusion process of two NSs in a binary NS merger \citep{2015ApJ...808..190R}. 
Therefore, from these general arguments, we can conclude that the energy emitted during the merger process leading to the formation of a BH should be larger than $\approx10^{52}$~erg.
Indeed, we find the upper limit for the S-GRFs of $(7.8\pm1.0)\times10^{51}$~erg (see Sec.~\ref{sec:SGRFs1}) and the the lower limit for the S-GRBs of  $(2.44\pm0.22)\times10^{52}$~erg (see Sec.~\ref{sec:SGRBs1}).

Such a separatrix energy is clearly a function of the initial NS mass undergoing accretion, by norm assumed to be $\approx1.4$~M$_\odot$. It is also a function of the yet unknown precise value of $M_{crit}$ for which only an absolute upper limit of $3.2$~M$_\odot$ has been established for the non-rotating case \citep{Rhoades1974}.
As already pointed out in \citet{2015ApJ...808..190R} for the case of binary NS mergers, the direct observation of the separatrix energy between S-GRFs and S-GRBs, and also (in this case) between BdHNe and XRFs, gives fundamental informations for the determination of the actual value of $M_{crit}$, for the minimum mass of the newly-formed BH, and for the mass of the accreting NS.
It is appropriate to notice that a value of the mass of the accreting NS binary, larger than $\approx1.4$~M$_\odot$, is \textit{a priori} possible and would give interesting observational properties: an exceptional accreting NS with mass close to $M_{crit}$ would lead to a BdHN with a value of the energy lower than the theoretical separatrix of $\approx10^{52}$~erg. Conversely, the accretion on a NS with mass smaller than $\approx1.4$~M$_\odot$ should lead to an XRF with energy larger than $\approx10^{52}$~erg. 
These rare possibilities will be precious in further probing the implications of the IGC paradigm, in estimating the NS masses, as well as in deriving more stringent limits on $M_{crit}$ directly from observations.

Our theory of the hypercritical accretion, applied in the GRB
analysis through the IGC paradigm in binary systems, introduces
substantial differences with respect to the traditional ones. To
appreciate these differences and gain familiarity in this novel
approach, we recommend the reading of all the
references quoted in this section.

\section{The X-ray flashes}\label{sec_3_prot_rat}

\subsection{General properties}\label{sec:XRFs1}

\begin{figure*}
\centering
\includegraphics[width=0.45\hsize,clip]{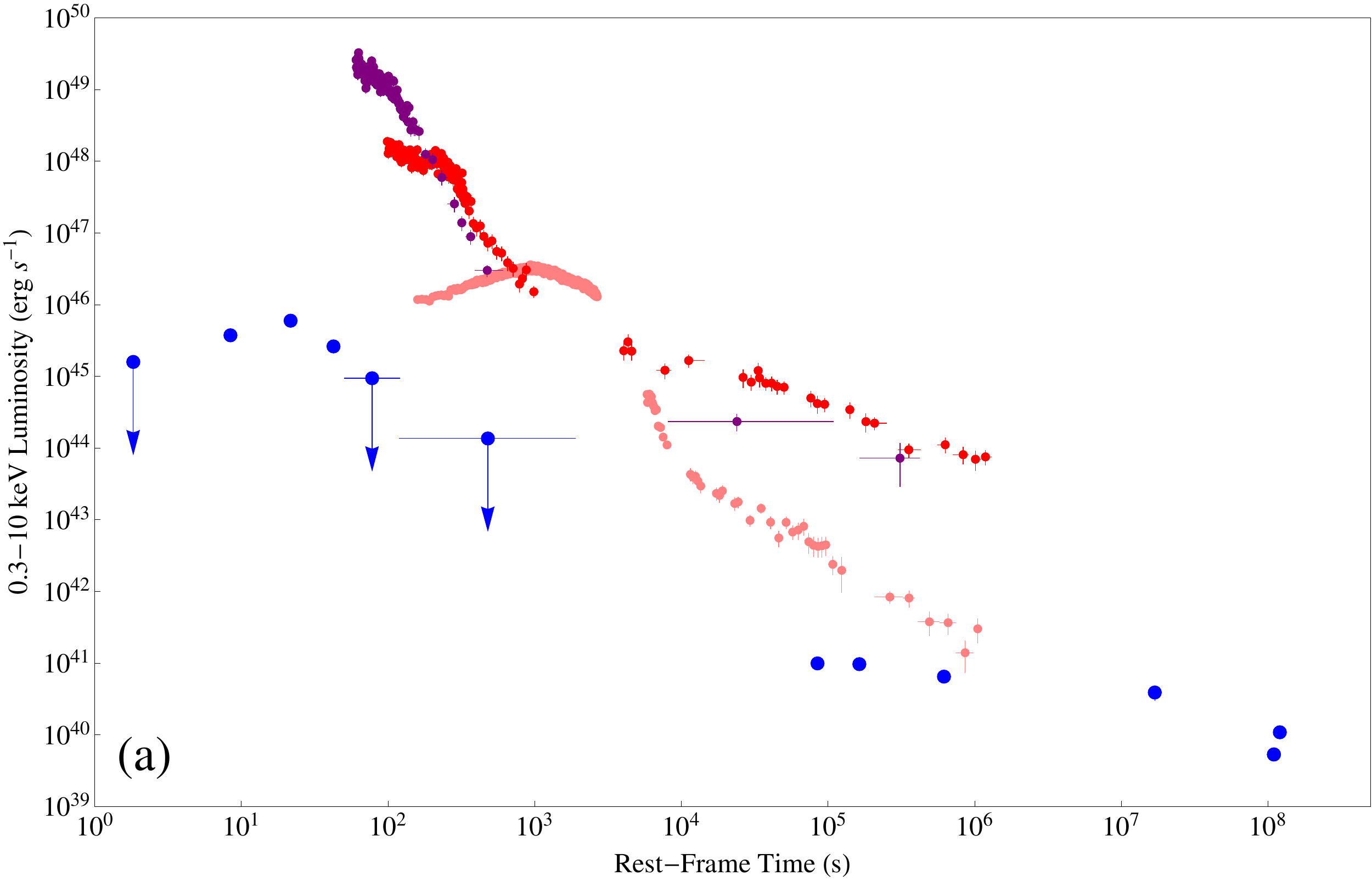}
(b)\includegraphics[width=0.45\hsize,clip]{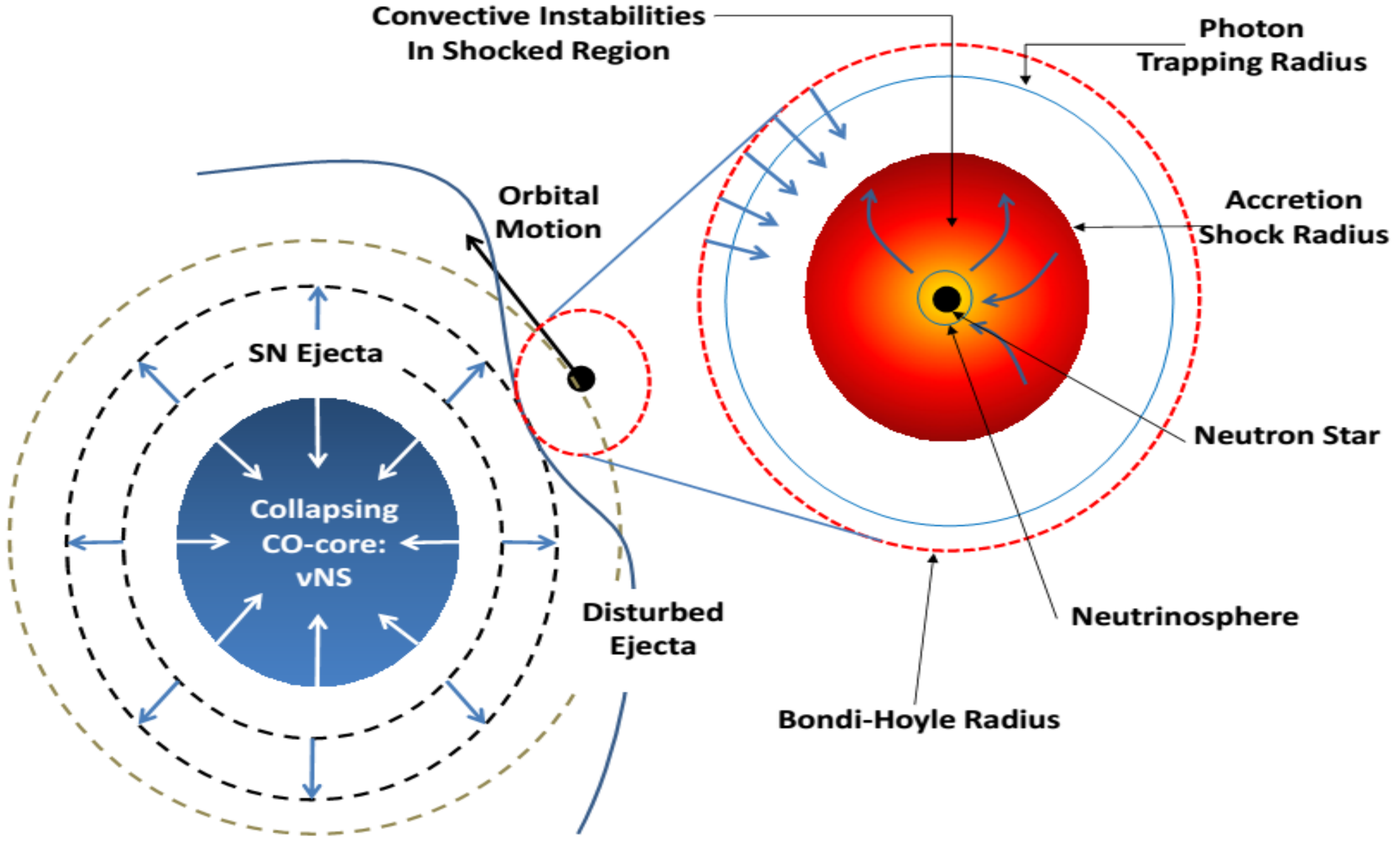}
\null\hfill\hfill\null\\
(c)\includegraphics[width=0.45\hsize,clip]{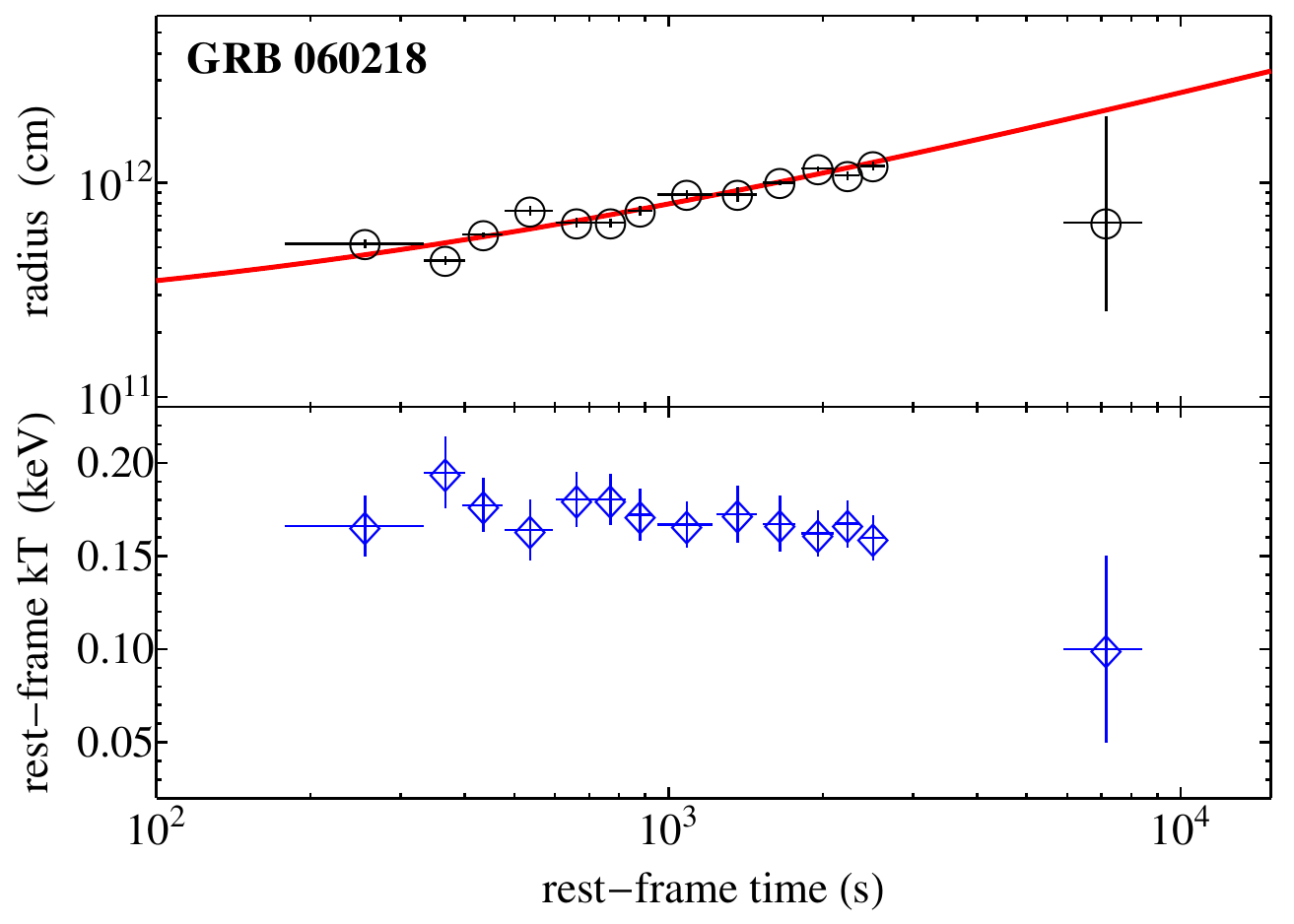}
(d)\includegraphics[width=0.45\hsize,clip]{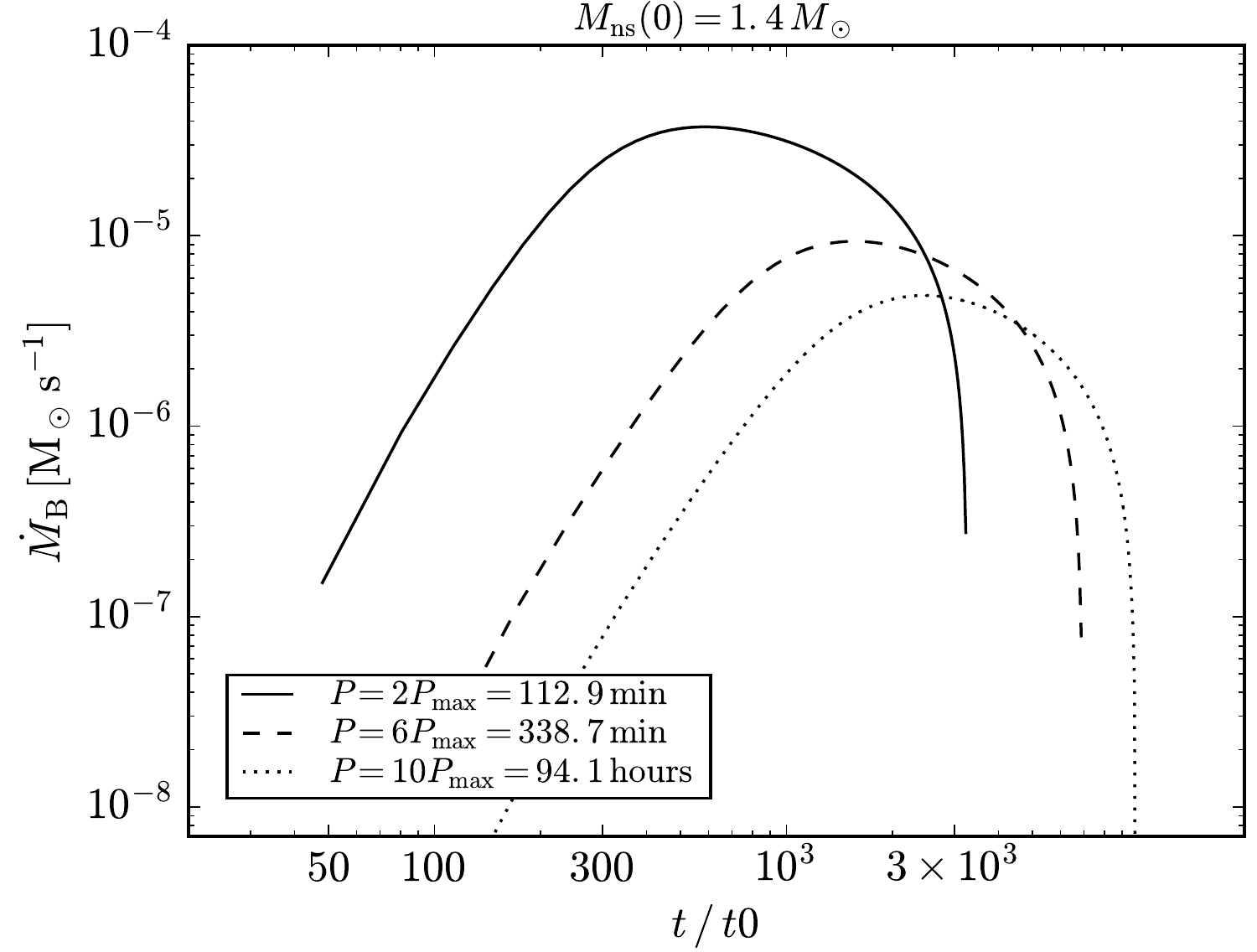}
\null\hfill\hfill\null
\caption{(a) Rest-frame $0.3$--$10$~keV luminosity light curves of four selected XRFs: GRB 980425 (blue), GRB 060218 (pink), GRB 070419A (purple), and GRB 101219B (red). (b) A sketch of the CO$_{\rm core}$-NS binary progenitor and the hypercritical accretion process in the IGC scenario (reproduced from \citealp{2014ApJ...793L..36F}). (c) Upper panel: the evolution of the radius of the thermal component detected in GRB 060218 (black circles) and its linear fit (solid red curve). Lower panel: the decay of the corresponding rest-frame temperature (blue diamonds). Reproduced from \citet{Campana2006}. (d) Mass accretion rate $\dot{M}_B$ of the SN ejecta onto a NS companion of initial mass $1.4~M_\odot$, as function of time. Three cases are plotted for various selected orbital periods $P>P_{\rm max}$ (see legend).}
\label{fig:fam1}
\end{figure*}

The observational features of long bursts with energy below $\approx10^{52}$~erg are listed below and summarized in Fig.~\ref{fig:fam1}. 
These bursts are interpreted within the theoretical framework of the IGC as a new class which we indicate as XRFs.

The upper limit on the energetic of the XRFs is $(7.3\pm0.7)\times10^{51}$~erg as measured in GRB 110106B.

The isotropic energies are in the range $(6.4\pm1.6)\times10^{47}\lesssim E_{\rm iso}\lesssim(7.3\pm0.7)\times10^{51}$~erg \citep[see Fig.~\ref{fig:ffff} and][]{2013IJMPD..2230028A,2015ApJ...798...10R}.

The spectral peak energies are in the range $4\lesssim E_{\rm p,i}\lesssim 200$~keV \citep[see Fig.~\ref{fig:ffff} and][]{2013IJMPD..2230028A,2015ApJ...798...10R} and increase monotonically with $E_{\rm iso}$.

The cosmological redshifts are in the range $0.0085\leq z\leq1.096$, with an average value of $\approx0.43$ (see Table~\ref{tab:XRFs}).

The prompt emission phase has a duration ranging between $\sim10^2$--$10^4$~s (see Fig.~\ref{fig:fam1}~(a)) with a spectrum generally characterized by a thermal component and power-law component. The radii of the thermal emitter are in the range of $10^{10}$--$10^{12}$~cm and the temperatures vary in the range of $0.1$--$2$ keV \citep[see, e.g.,][and Fig.~\ref{fig:fam1}~(c)]{Campana2006}, depending on the values of the binary period and separation of the progenitor systems.

The long lasting X-ray afterglow does not exhibit any specific common late power-law behavior (see Fig.~\ref{fig:fam1}~(a)).

For all XRFs at $z\lesssim1$, an optical SN with a luminosity similar to the one of SN 2010bh \citep{2012ApJ...753...67B}, occurs after $10$--$15$ days in the cosmological rest-frame.

No high energy emission has ever been observed.

In view of the observed values of $E_{p,i}$ which occur in the X-ray domain and also because of the low values of their $E_{iso}<10^{52}$ erg, we adopted the name XRFs for these soft and less energetic long bursts, a terminology already used in the literature on purely morphological grounds \citep[see, e.g.,][]{2003AIPC..662..229H,2004A&A...426..415A,Soderberg2006Nature}.

\subsection{Theoretical interpretation of XRFs within the IGC paradigm}\label{sec:XRFs1b}

In the IGC paradigm an XRF occurs when the CO$_{\rm core}$-NS binary separation $a$ is so large (typically $a>10^{11}$~cm, see e.g., \citealp{2015ApJ...812..100B}) that the accretion of the SN ejecta onto the NS is not sufficient for the NS to reach $M_{\rm crit}$.
Correspondingly, there is a critical or maximum value of the orbital period $P_{\rm max}$ (e.g. $P_{\rm max}\approx 28$~min for a NS with initial mass of $1.4~M_\odot$) for which the NS collapses to a BH, namely for $P>P_{\rm max}$ the accretion rate is not sufficient to induce the gravitational collapse of the companion NS into a BH (see Figs.~\ref{fig:fam1}~(d)).

The hypercritical accretion of the SN ejecta onto the NS binary companion occurs at rates $<10^{-2}~M_\odot$~s$^{-1}$ and can last from several hundreds of seconds all the way up to $\sim10^4$~s, until the whole SN ejecta flies beyond the NS binary orbit (see Fig.~\ref{fig:fam1}~(a)).
The photons are trapped in the accreting material and the accretion energy is lost through a large associated neutrino emission \citep[see, e.g.,][and references therein]{Zeldovich1972,RRWilson1973,2012ApJ...758L...7R,2014ApJ...793L..36F}.
The upper limit of $10^{52}$ erg of these sources is explainable by estimating the gravitational energy of the matter accreted onto the NS reaching a mass below $M_{\rm crit}$ at the end of the accretion process (see Sec.~\ref{sec:1052erg}).

The resulting emission, dubbed Episode 1, exhibits a spectrum composed of a thermal component, possibly originating from the outflow within the NS atmosphere driven out by Rayleigh-Taylor convection instabilities, and a power-law component. The shorter the binary period, the larger the accretion rate (see Figs.~\ref{fig:fam1}~(f)) and the values of $E_{\rm iso}$ and $E_{\rm p,i}$, and correspondingly the shorter the prompt emission duration (see Fig.~\ref{fig:fam1}~(a)). The excess of angular momentum of the system necessarily leads to a jetted emission, manifested in the power-law spectral component \citep{2015ApJ...812..100B}. Indeed in the IGC simulations the typical radii inferred from the evolving thermal component coincide with the observed ones of $10^{10}$--$10^{12}$~cm.

In the IGC paradigm the in-state is represented by an exploding CO$_{\rm core}$ and a companion NS. The out-state is multiple system composed of a MNS, resulting from the accretion of part of the SN ejecta onto the binary companion NS, a $\nu$NS, originating from the SN event, and the remaining part of the SN ejecta shocked by the hypercritical accretion emission of the XRF. This energy injection into the SN ejecta leads to the occurrence of a broad-lined SN Ic \citep[\textit{hypernova}, see, e.g.,][]{2003ApJ...598.1163M} with a kinetic energy larger than that of the traditional SNe Ic.
The presence of $^{56}$Ni in the SN ejecta leads to the observed SN emission after $\approx10$--$15$ days in the cosmological rest-frame which is observable for sources at $z\lesssim1$.

Clearly the absence of hard $\gamma$-ray and GeV emissions is implicit in the nature of the hypercritical accretion process not leading to a BH and the corresponding rate of neutrino emission (see also Appendix~\ref{Appendix}).

\subsection{Prototypes}\label{sec:XRFs2}

In Fig.~\ref{fig:fam1}(a) we reproduce the rest-frame $0.3$--$10$~keV luminosity light curves of four selected XRFs: GRB 980425 \citep{2000ApJ...536..778P,2004ApJ...608..872K,2004AdSpR..34.2711P}, associated with SN 1998bw \citep{Galama1998}, GRB 060218 associated with SN 2006aj \citep{Campana2006,Soderberg2006Nature}, GRB 070419A \citep{2007A&A...469..379E,2009MNRAS.397.1177E} with an optical SN bump \citep{Hill2007}, and GRB 101219B \citep{2007A&A...469..379E,2009MNRAS.397.1177E} associated with SN 2010ma \citep{Sparre2011}.
Their prompt emissions are represented by the above mentioned Episode 1.
In Fig.~\ref{fig:fam1}~(c) we plot the evolution of both temperature and radius inferred from the thermal component observed in the Episode 1 emission of GRB 060218. The increasing radius and almost constant temperature are obtained from the thermal component observed in GRB 060218 \citep{Campana2006}. Details will appear in forthcoming publications (Pisani et al., and Becerra et al., in preparation). A complete list of XRFs is shown in Table~\ref{tab:XRFs}.
\begin{table*}
\centering
\begin{tabular}{lcc|lcc}
\hline\hline
GRB & $z$ & $E_{\rm iso}/$($10^{50}$ erg) &  GRB & $z$ & $E_{\rm iso}/$($10^{50}$ erg)\\
\hline
970508	&	$0.835$		&	$65\pm13$					&	081007	&	$0.5295$	&	$17\pm2$\\
980425	&	$0.0085$	& $0.0064\pm0.0016$	&	100316D	&	$0.059$		& $0.59\pm0.05$\\
980613	&	$1.096$		&	$50\pm10$					&	100816A	&	$0.8049$	&	$71\pm9$\\
990712	&	$0.434$		&	$69\pm13$					&	101219B	&	$0.55$		&	$63\pm6$\\
020819B	&	$0.41$		&	$69\pm18$					&	110106B	&	$0.618$		&	$73\pm$\\
020903	&	$0.251$		& $0.24\pm0.06$     &	120121B	&	$0.017$   & $0.0139\pm0.0002$\\
031203  &	$0.105$ 	& $0.99\pm0.10$     & 120422A	&	$0.283$		&	$2.4\pm0.8$\\
050416A	&	$0.6528$	&	$11\pm2$	        &	120714B	&	$0.3984$	&	$8.0\pm2.0$\\
060218	&	$0.033$		& $0.54\pm0.05$     &	130702A	&	$0.145$		&	$6.5\pm1.0$\\
070419	&	$0.97$		&	$24\pm10$         &	130831A	&	$0.4791$	&	$46\pm2$\\
\hline
\end{tabular}
\caption{List of the XRFs considered in this work up to the end of 2014. For each source (first columns) the values of $z$ and $E_{\rm iso}\lesssim10^{52}$~erg (second and third columns, respectively) are listed.}
\label{tab:XRFs}
\end{table*}

\section{The binary-driven hypernovae}\label{sec:descr_BdHNe}

\subsection{General properties}\label{sec:BdHNe1}

\begin{figure*}
(a)\includegraphics[width=0.45\hsize,clip]{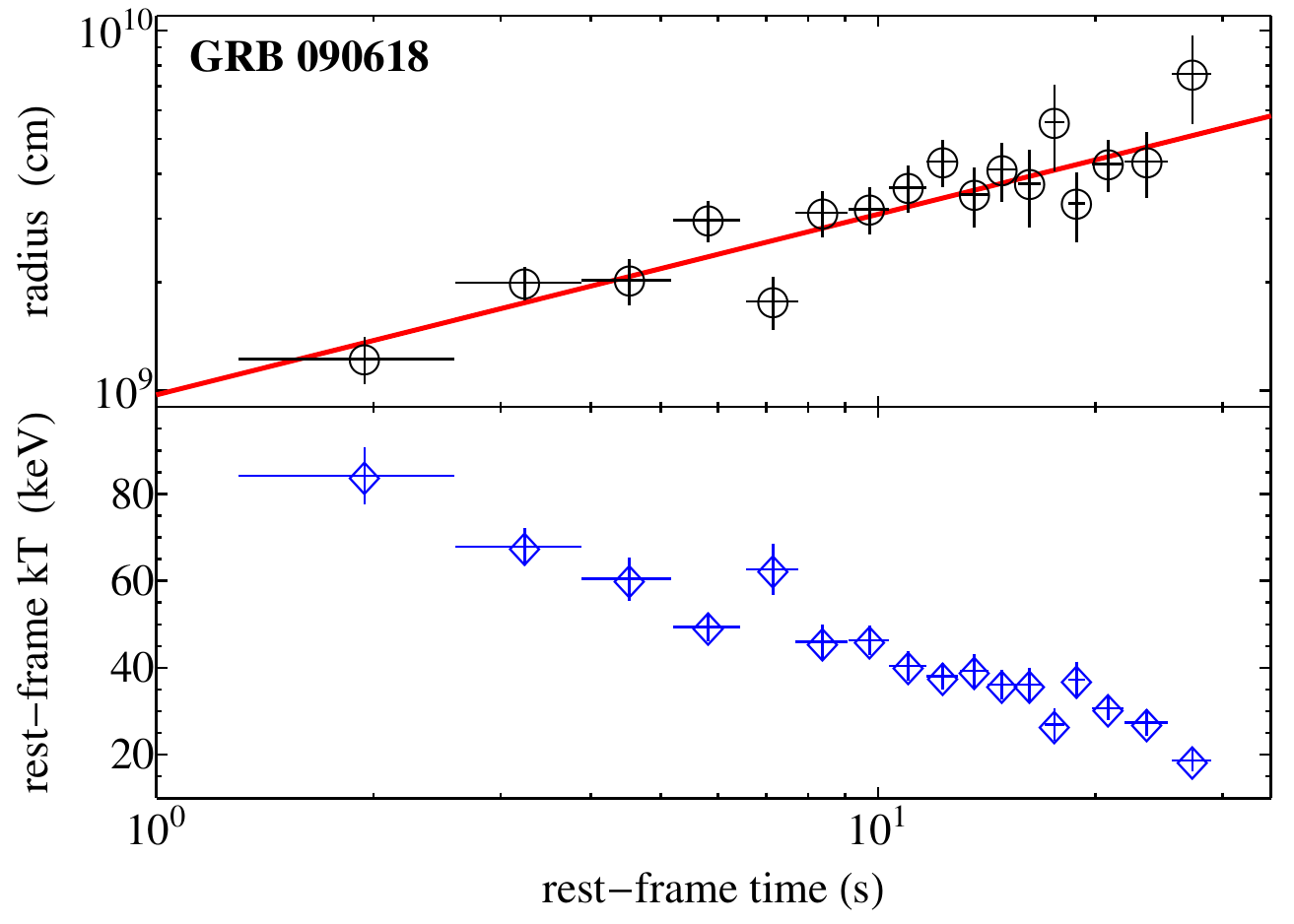}
(b)\includegraphics[width=0.45\hsize,clip]{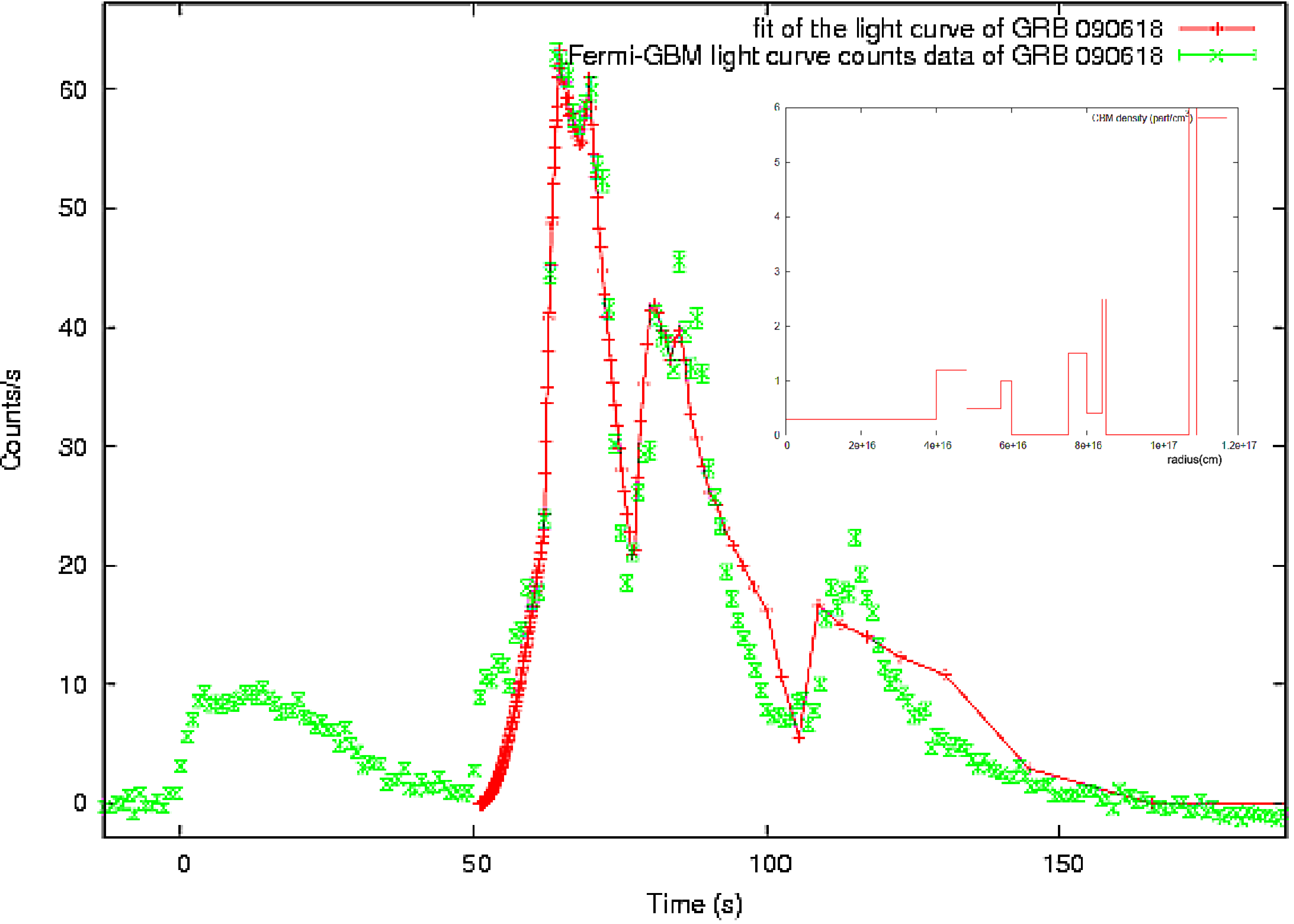}
\null\hfill\hfill\null\\
(c)\includegraphics[width=0.45\hsize,clip]{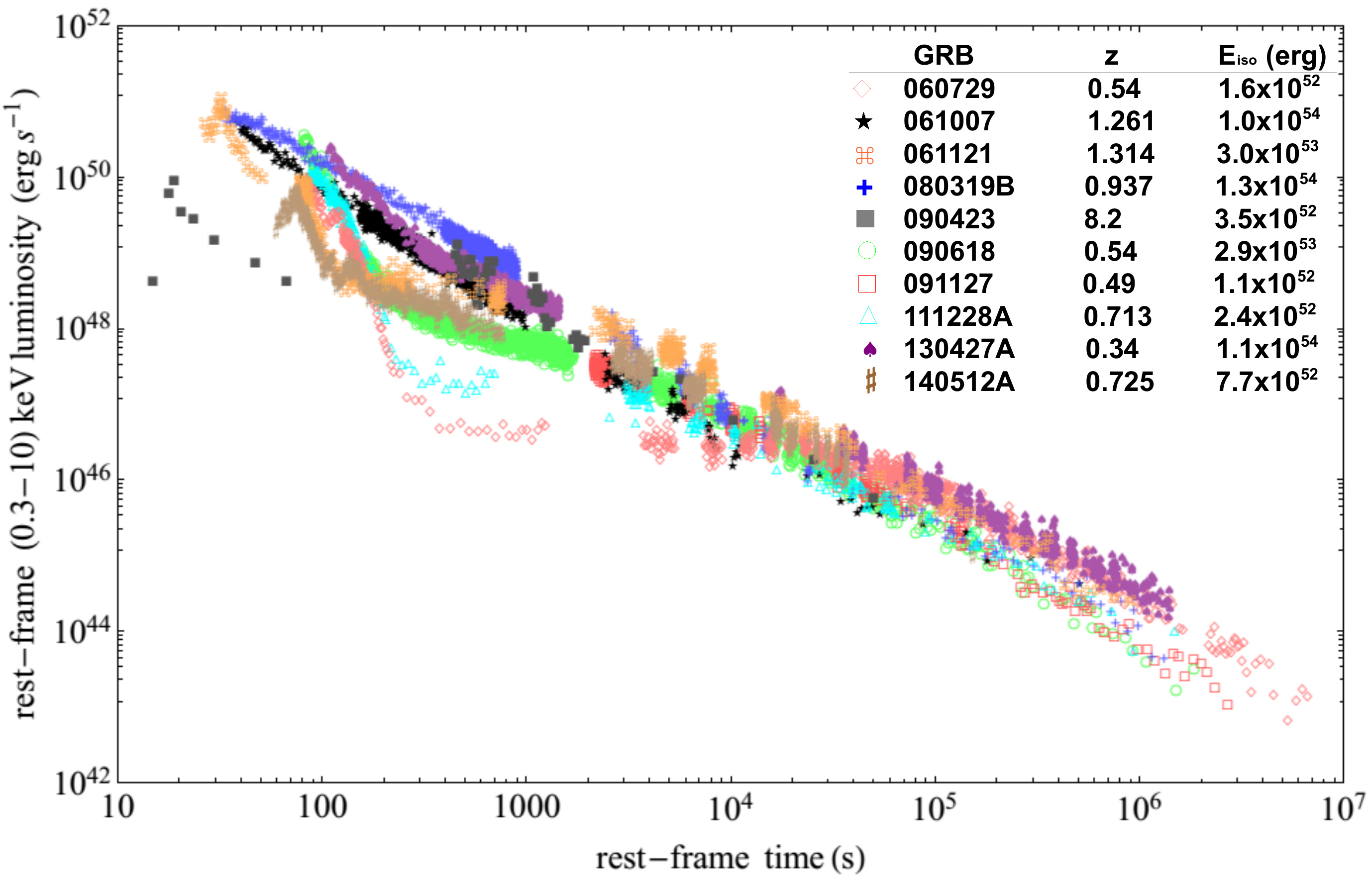}
(d)\includegraphics[width=0.45\hsize,clip]{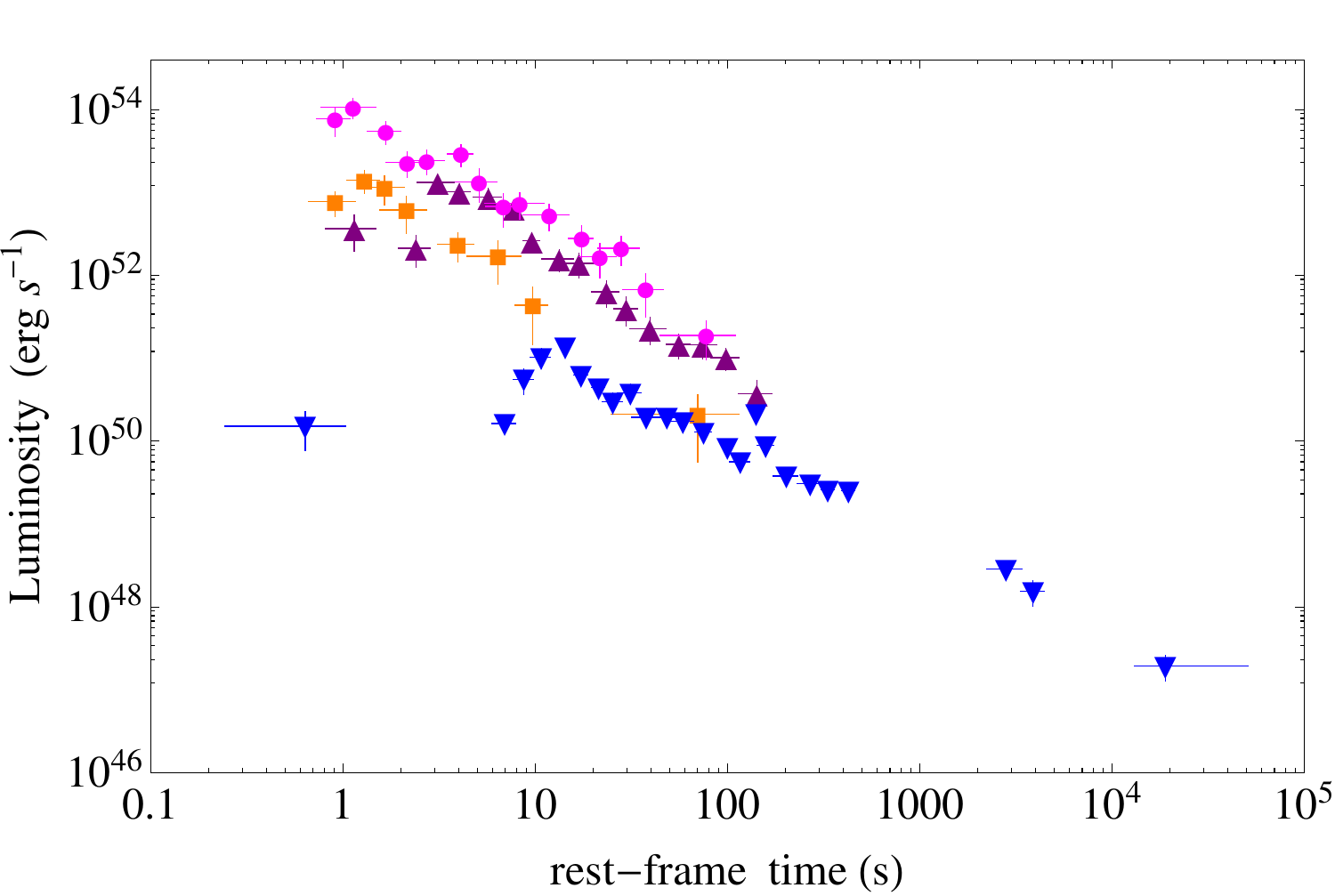}
\null\hfill\hfill\null
\caption{(a) Upper panel: the evolution of the radius of the thermal component detected in the Episode 1 of GRB 090618 (black circles) and its linear fit (solid red curve). Lower panel: the decay of the corresponding rest-frame temperature (blue diamonds). Reproduced from \citet{Izzo2012}. (b) The fireshell simulation (red line) of the light curve of Episode 2 of the prototype GRB 090618 (green data). The small inset reproduces the CBM profile required for the simulation. Reproduced from \citet{Izzo2012}. (c)The rest-frame $0.3$--$10$~keV luminosity light curves of selected BdHNe. All these sources exhibit the overlapping of the late power-law decay, outlined in \citet{Pisani2013}, and a nested structure, as outlined in \citet{Ruffini2014}. (d) The rest-frame $0.1$--$100$~GeV luminosity light curves of selected BdHNe (reproduced from \citealp{Ackermann2013}): GRB 080916C (magenta circles), GRB 090902B (purple triangles), GRB 110731A (orange squares), GRB 130427A (blue reversed triangles).}
\label{rad_ind_tot}
\end{figure*}

The observational features of long bursts with energy above $\approx10^{52}$~erg are listed below and summarized in Fig.~\ref{rad_ind_tot}. 
These bursts are interpreted within the theoretical framework of the IGC as a new class which we indicate as BdHNe.

The lower limit on the energetic of the BdHNe is $(9.2\pm1.3)\times10^{51}$~erg as measured in GRB 070611.

 The observed isotropic energies are in the range $(9.2\pm1.3)\times10^{51}\lesssim E_{\rm iso}\lesssim(4.07\pm0.86)\times10^{54}$~erg \citep[see Fig.~\ref{fig:ffff} and][]{2013IJMPD..2230028A,2015ApJ...798...10R} and are in principle dependent on the NS mass which we have assumed, as an example, $\approx 1.4$~M$_\odot$ (see sec.~\ref{sec:1052erg}).

The spectral peak energies are in the range $0.2\lesssim E_{\rm p,i}\lesssim 2$~MeV \citep[see Fig.~\ref{fig:ffff} and][]{2013IJMPD..2230028A,2015ApJ...798...10R} and increase monotonically with $E_{\rm iso}$.

The cosmological redshifts are in the range $0.169\leq z\leq9.3$, with an average value of $\approx2.42$ (see Table~\ref{tab:BdHNe}).

The prompt emission phase of BdHNe exhibits a more complex structure than that of XRFs. Indeed three different regimes are found:
\begin{itemize}
\item[a)] A thermal emission with a decreasing temperature following a broken power-law behavior, and an additional non-thermal spectral component (a power-law), dominate the early emission in selected BdHNe \citep[see, e.g.,][and Fig.~\ref{rad_ind_tot}~(a)]{Izzo2012}. The existence of this thermal component was first identified in the GRB BATSE data by \citet{Ryde2004,Ryde2005}. 
It has been then shown to occur in the case of BdHNe as GRB 090618 \citep[][and Fig.~\ref{rad_ind_tot}~(a)]{Izzo2012}, GRB 101023 \citep{Penacchioni2011}, GRB 110709B \citep{Penacchioni2013}, and GRB 970828 \citet{2015ARep...59..626R}.
The characteristic radii inferred from the cooling thermal component are of the order of $10^{9}$--$10^{10}$~cm and the average expansion speed is $\sim10^8$--$10^9$ cm s$^{-1}$ \citep{Izzo2012,Penacchioni2011,Penacchioni2013,2015ARep...59..626R}.
\item[b)] This early emission is followed by the characteristic GRB emission (see Fig.~\ref{rad_ind_tot}~(d)), encompassing a thermal precursor, the P-GRB \citep{RSWX2,RSWX}, followed by the prompt emission \citep{Ruffini2002,Ruffini2004,Ruffini2005}.
\item[c)] The prompt emission is followed by a steep decay, then by a plateau and a late power-law decay. These features have been first reported in \citet{Nousek2006} and \citet{Zhang2006}.
\end{itemize}

The late decay has typical slopes of $-1.7\lesssim\alpha_X\lesssim-1.3$ \citep{Pisani2013} and shows a characteristic power-law behavior both in the optical and in X-rays. When computed in the source cosmological rest-frame, the late power-law decay in X-rays exhibits new features: overlapping and  nesting (see Fig.~\ref{rad_ind_tot}~(c)). Overlapping has been proven in a sample of six BdHNe: GRBs 060729, 061007, 080319B, 090618, 091127, and 111228, \citep{IzzoRel,Pisani2013}. The nested property of the BdHNe has been discussed in \citet{Ruffini2014}, where it has been shown that the duration (the luminosity) of the plateau phase is inversely (directly) proportional to the energy of the GRB emission: the more energetic the source, the smaller (higher) the duration (the luminosity) of the plateau.

For all BdHNe at $z\lesssim1$, an optical SN with a luminosity similar to the one of SN 1998bw \citep{Galama1998}, occurs after $10$--$15$ days in the cosmological rest-frame.

A distinctive high energy emission observed up to $100$ GeV shows a luminosity light curve following a precise power-law behavior with index $\approx-1.2$ \citep[Fig.~\ref{rad_ind_tot}~(d) and][]{Nava2014}. The turn-on of this GeV emission occurs after the P-GRB emission and during the prompt emission phase.

\subsection{Theoretical interpretation of BdHNe within the IGC paradigm}\label{sec:BdHNe1b}

In the IGC paradigm a BdHN occurs when the CO$_{\rm core}$-NS binary is more tightly bound ($a\lesssim10^{11}$~cm, see e.g., \citealp{2015ApJ...812..100B}). The larger accretion rate of the SN ejecta, e.g., $\gtrsim10^{-2}$--$10^{-1}~M_\odot$~s$^{-1}$, leads the companion NS to easily reach its critical mass $M_{\rm crit}$ \citep{2012ApJ...758L...7R,2014ApJ...793L..36F,2015ApJ...812..100B}, leading to the formation of a BH. 
The electrodynamical conditions encountered in the final accretion phase explain the existence of a vacuum polarization process leading to the creation of  an $e^+e^-$ plasma and to the formation of a KNBH with a large variety of new astrophysical phenomena. For the sake of clarity and independence on the physical regime encountered, in the IGC paradigm we have divided the activities of the BdHNe in a numbered set of distinct Episodes.   

Episode 1 of BdHNe originates in the same hypercritical accretion process as the corresponding one of XRFs. The corresponding spectrum again exhibits an expanding thermal component and a power-law function \citep{Izzo2012,2015ARep...59..626R}. The typical radii inferred from the thermal component are of the order of $10^{9}$--$10^{10}$~cm and the average expansion speed is $\sim10^8$--$10^9$ cm s$^{-1}$ \citep[see Fig.~\ref{rad_ind_tot}~(a) and][]{Izzo2012,2015ARep...59..626R}.

Episode 2 corresponds to the authentic long GRB emission (see Fig.~\ref{rad_ind_tot}~(b)), stemming from the collapse of the companion NS to a BH. For its theoretical description we adopt the traditional fireshell model \citep[see][and Section~\ref{sec:fireshell}]{Ruffini2001c,Ruffini2001,Ruffini2001a}. The GRB emission occurs at Lorentz factor at the transparency of $\Gamma=10^2$--$10^3$ \citep{Izzo2012,2015ARep...59..626R} and the spatial extension of the interaction of the fireshell with the circumburst medium goes all the way up to $\sim10^{16}$--$10^{17}$~cm, reached at the end of Episode 2 \citep{Izzo2012}. The BdHNe have $E_{\rm iso}\gtrsim 10^{52}$~erg and their $E_{p,i}\gtrsim200$~keV is in the hard $\gamma$-ray domain.

Episode 3 in BdHNe originates from the SN ejecta \citep{2015ApJ...798...10R}. In this case an extra energy injection is delivered by the interaction of the GRB outflow with the SN ejecta resulting in an isotropic energy emission of $10^{51}$--$10^{52}$~erg. This interaction produces a flare at the beginning of Episode 3 (typically at a rest-frame time of $\sim10^2$~s) with the typical signature of an expanding thermal component in its spectrum. The radii inferred from this thermal component are $\sim10^{12}$--$10^{13}$~cm and their evolution reveals a mildly relativistic expansion at $\Gamma\approx2$ \citep{Ruffini2014,2015ApJ...798...10R}.
The rest-frame $0.3$--$10$ keV luminosity light is then followed by a plateau phase and a late power-law decay. 
The late decay has been shown to exhibit a common power-law behavior and a nested structure \citep[see, e.g.,][and Fig.~\ref{rad_ind_tot}~(c)]{Pisani2013,Ruffini2014}. 
The possibility of using the late X-ray emission as a distance indicator has been explored by inferring the redshifts of GRBs 101023 and 110709B \citep{Penacchioni2011,Penacchioni2013}, and has been applied to predict the occurrence of the SN associated to GRB 130427A after $\sim10$--$15$ days in the cosmological rest-frame before its discovery \citep{2013GCN..14526...1R}, later confirmed by the observations \citep{Postigo2013cq,2013GCN..14686...1L,2013GCN..14666...1W,GCN2013conf}.

Episode 4, as predicted in the IGC paradigm and in analogy to XRFs, corresponds to the optical SN emission observable in all BdHNe at $z\lesssim1$ after $\approx10$--$15$ days in the cosmological rest-frame. It is remarkable that these SNe have a standard luminosity, like the one of SN 1998bw \citep[see, e.g.,][]{2014A&A...567A..29M}.

A new Episode 5, here introduced, is identified with the long-lived GeV emission. This emission is conceptually distinct in its underlying physical process from that of Episode 3. When LAT data are available, the majority of BdHNe observed by the \textit{Fermi} satellite \citep{Atwood2009} exhibit such an emission, similar to the one observed in S-GRBs (see Section~\ref{sec:descr_S-GRFs}).
In \citet{2015ApJ...798...10R} the further accretion of matter onto the newly-formed BH has been indicated as the origin of this GeV emission.
An outstanding exception is GRB 151027A (Kovacevic et al., in preparation). 

Also for BdHNe the in-state is composed of an exploding CO$_{\rm core}$ and a companion NS. The out-state is again multiple system. First, there is a GRB composed of the P-GRB and its prompt emission. Then there is a newly-formed BH, produced by the hypercritical accretion of part of the SN ejecta onto the binary companion NS reaching $M_{\rm crit}$. Again, there is a $\nu NS$ originating from the SN explosion. Finally, there is the remaining part of the SN ejecta shocked by the GRB emission. The energy injection into the SN ejecta from both the hypercritical accretion phase and the GRB emission leads also in this case to the occurrence of a broad-lined SN Ic \citep[\textit{hypernova}, see, e.g.,][]{2003ApJ...598.1163M} with a kinetic energy larger than that of the traditional SNe Ic.

\subsection{Prototypes}\label{sec:BdHNe2}

In the following selected prototypes of BdHNe are given and illustrated in Fig.~\ref{rad_ind_tot}.

The first systematic time-resolved spectral analysis of an Episode 1 of a BdHN has been performed for GRB 090618 \citep{Izzo2012}.
In this source the typical radii inferred from the cooling thermal component are of the order of $10^{9}$--$10^{10}$~cm and the average expansion speed is $\sim10^8$--$10^9$ cm s$^{-1}$ (see Fig.~\ref{rad_ind_tot}(a)). Similar results have been obtained for GRB 101023 \citep{Penacchioni2011}, GRB 110907B \citep{Penacchioni2013}, and GRB 980828 \citep{2015ARep...59..626R}.

The selected prototypes of Episode 2 emission have isotropic energies ranging from $E_{\rm iso}=1.60\times10^{53}$~erg in GRB 970828 \citep{2015ARep...59..626R}, to $E_{\rm iso}=1.32\times10^{54}$~erg in GRB 080319B \citep{Patricelli}. The amount of baryonic matter loaded before the P-GRB emission, the baryon load $B\equiv M_B c^2/E^{\rm tot}_{e^+e^-}$, where $E^{\rm tot}_{e^+e^-}$ is the pair plasma energy and $M_B$ the engulfed baryon mass, is in the range from $B=1.98\times10^{-3}$ for GRB 090618 \citep{Izzo2012}, to $B=7.0\times10^{-3}$ in GRB 970828 \citep{2015ARep...59..626R}. Correspondingly, their transparency emission occurs at Lorentz factors at the transparency ranging from $\Gamma=143$ in GRB 970828 \citep{2015ARep...59..626R}, to $\Gamma=490$ in GRB 090618, \citep{Izzo2012}. 
The average density of the circumburst medium in these prototypes, inferred from description of the the interaction with the fireshell after its transparency  \citep{Ruffini2002,Ruffini2004,Ruffini2005}, vary from $0.6$ cm$^{-3}$ in GRB 090618 \citep{Izzo2012}, to $\approx10^3$~cm$^{-3}$ in GRB 970828 \citep{2015ARep...59..626R}. The size of the corresponding emitting region, $\sim10^{16}$--$10^{17}$~cm, is clearly incompatible with the radii inferred from Episode 1 and 3. This points to the different origins in the emission mechanisms of the above three Episodes. 

The radii inferred from the expanding thermal components observed in the spectra of the flares at the beginning of Episode 3 are typically $\sim10^{12}$--$10^{13}$~cm. This has been found in the cases of GRB 090618, \citep[see, e.g.,][]{Starling2010,Ruffini2014}, and GRB 130427A \citep{2015ApJ...798...10R}. In both these sources, the expansion of the thermal emitter of Episode 3 proceeds at $\Gamma\approx2$ \citep{Ruffini2014,2015ApJ...798...10R}. After the initial emission in the spike of Episode 3, the rest-frame $0.3$--$10$ keV luminosity light curve is then followed by a plateau phase and a late power-law decay. The overlapping of the late power-law decay and the nested structure are reproduced in Fig.~\ref{rad_ind_tot}~(c) for selected sources: GRB 060729, GRB 061007, GRB 080319B, GRB 090618, GRB 091127B, and GRB 111228A \citep[considered in][]{Pisani2013}, GRB 061121 and GRB 130427A \citep[considered in][]{Ruffini2014,2015ApJ...798...10R}, GRB 090423 \citep{2014A&A...569A..39R}, and GRB 140512A (introduced here). 

Episode 4 has been spectroscopically identified for the two closest BdHNe, e.g., GRB 091127--SN 2009 nz \citep{Cobb2010} and GRB 130427A-- SN 2013cq \citep{GCN2013conf}. In the cases of GRB 060729 \citep{Cano2010}, GRB 080319B \citep{Kann2008}, GRB 090618 \citep{Cano2010}, and GRB 111228A \citep{DAvanzo2012}, at $z\lesssim1$, the identification was possible through the detection of bumps in their Episode 3 optical light curves.

Turning now to the Episode 5 of BdHNe, the GeV emission has been studied in detail in the case of GRB 130427A \citep{2015ApJ...798...10R}, as well as in other selected BdHNe (see Fig.~\ref{rad_ind_tot}~(d) and \citealp{Ackermann2013}). The turn-on has been identified as the on-set of the emission from the newly-formed BH \citep{2015ApJ...798...10R}.

A complete list of BdHNe is shown in Table~\ref{tab:BdHNe}.

The ultra long GRBs \citep{2014ApJ...781...13L,2015ApJ...800...16B} are certainly BdHNe on the ground of their late x-ray rest frame luminosity (Pisani et al., submitted to ApJ).
\begin{longtable}{lcc|lcc}
\caption{List of the BdHNe considered in this work up to the end of 2014. For each source (first columns) the values of $z$ and $E_{\rm iso}$ (second and third columns, respectively) are listed.\label{tab:BdHNe}
}\\
\hline\hline
GRB & $z$ & $E_{\rm iso}/$($10^{52}$ erg) &  GRB & $z$ & $E_{\rm iso}/$($10^{52}$ erg)\\
\hline
\endfirsthead
\caption{continued.}\\
\hline\hline
GRB & $z$ & $E_{\rm iso}/$($10^{52}$ erg) &  GRB & $z$ & $E_{\rm iso}/$($10^{52}$ erg)\\
\hline
\endhead
\hline
\endfoot
970228	&	$0.695$	&	$1.65\pm0.16$	&	081008	&	$1.969$	&	$10.0\pm1.0$\\
970828	&	$0.958$	&	$30.4\pm3.6$	&	081028	&	$3.038$	&	$18.3\pm1.8$\\
971214	&	$3.42$	&	$22.1\pm2.7$	&	081029	&	$3.8479$	&	$12.1\pm1.4$\\
980329	&	$3.5$	&	$267\pm53$	&	081109	&	$0.9787$	&	$1.81\pm0.12$\\
980703	&	$0.966$	&	$7.42\pm0.74$	&	081118	&	$2.58$	&	$12.2\pm1.2$\\
990123	&	$1.6$	&	$241\pm39$	&	081121	&	$2.512$	&	$32.4\pm3.7$\\
990506	&	$1.3$	&	$98.1\pm9.9$	&	081203A	&	$2.05$	&	$32\pm12$\\
990510	&	$1.619$	&	$18.1\pm2.7$	&	081221	&	$2.26$	&	$31.9\pm3.2$\\
990705	&	$0.842$	&	$18.7\pm2.7$	&	081222	&	$2.77$	&	$27.4\pm2.7$\\
991208	&	$0.706$	&	$23.0\pm2.3$	&	090102	&	$1.547$	&	$22.6\pm2.7$\\
991216	&	$1.02$	&	$69.8\pm7.2$	&	090205	&	$4.6497$	&	$1.12\pm0.16$\\
000131	&	$4.5$	&	$184\pm32$	&	090313	&	$3.375$	&	$4.42\pm0.79$\\
000210	&	$0.846$	&	$15.4\pm1.7$	&	090323	&	$3.57$	&	$438\pm53$\\
000418	&	$1.12$	&	$9.5\pm1.8$	&	090328	&	$0.736$	&	$14.2\pm1.4$\\
000911	&	$1.06$	&	$70\pm14$	&	090418A	&	$1.608$	&	$17.2\pm2.7$\\
000926	&	$2.07$	&	$28.6\pm6.2$	&	090423	&	$8.1$	&	$8.8\pm2.1$\\
010222	&	$1.48$	&	$84.9\pm9.0$	&	090424	&	$0.544$	&	$4.07\pm0.41$\\
010921	&	$0.45$	&	$0.97\pm0.10$	&	090429B	&	$9.3$	&	$6.7\pm1.3$\\
011121	&	$0.36$	&	$8.0\pm2.2$	&	090516	&	$4.109$	&	$72\pm14$\\
011211	&	$2.14$	&	$5.74\pm0.64$	&	090519	&	$3.85$	&	$24.7\pm2.8$\\
020124	&	$3.2$	&	$28.5\pm2.8$	&	090529	&	$2.625$	&	$2.56\pm0.30$\\
020127	&	$1.9$	&	$3.73\pm0.37$	&	090530	&	$1.266$	&	$1.73\pm0.19$\\
020405	&	$0.69$	&	$10.6\pm1.1$	&	090618	&	$0.54$	&	$28.6\pm2.9$\\
020813	&	$1.25$	&	$68\pm17$	&	090715B	&	$3.0$	&	$23.8\pm3.7$\\
021004	&	$2.3$	&	$3.47\pm0.46$	&	090809	&	$2.737$	&	$1.88\pm0.26$\\
021211	&	$1.01$	&	$1.16\pm0.13$	&	090812	&	$2.452$	&	$47.5\pm8.2$\\
030226	&	$1.98$	&	$12.7\pm1.4$	&	090902B	&	$1.822$	&	$292\pm29.2$\\
030323	&	$3.37$	&	$2.94\pm0.92$	&	090926	&	$2.106$	&	$228\pm23$\\
030328	&	$1.52$	&	$38.9\pm3.9$	&	090926B	&	$1.24$	&	$4.14\pm0.45$\\
030329	&	$0.169$	&	$1.62\pm0.16$	&	091003	&	$0.897$	&	$10.7\pm1.8$\\
030429	&	$2.65$	&	$2.29\pm0.27$	&	091020	&	$1.71$	&	$8.4\pm1.1$\\
030528	&	$0.78$	&	$2.22\pm0.27$	&	091024	&	$1.092$	&	$18.4\pm2.0$\\
040912	&	$1.563$	&	$1.36\pm0.36$	&	091029	&	$2.752$	&	$7.97\pm0.82$\\
040924	&	$0.859$	&	$0.98\pm0.10$	&	091109A	&	$3.076$	&	$10.6\pm1.4$\\
041006	&	$0.716$	&	$3.11\pm0.89$	&	091127	&	$0.49$	&	$1.64\pm0.18$\\
050126	&	$1.29$ 	&	$2.47\pm0.25$	&	091208B	&	$1.063$	&	$2.06\pm0.21$\\
050315	&	$1.95$	&	$6.15\pm0.30$	&	100219A	&	$4.6667$	&	$3.93\pm0.61$\\
050318	&	$1.444$	&	$2.30\pm0.23$	&	100302A	&	$4.813$	&	$1.33\pm0.17$\\
050319	&	$3.243$	&	$4.63\pm.0.56$	&	100414A	&	$1.368$	&	$55.0\pm5.5$\\
050401	&	$2.898$	&	$37.6\pm7.3$	&	100513A	&	$4.8$	&	$6.75\pm0.53$\\
050502B	&	$5.2$	&	$2.66\pm0.22$	&	100621A	&	$0.542$	&	$2.82\pm0.35$\\
050505	&	$4.27$	&	$16.0\pm1.1$	&	100728A	&	$1.567$	&	$86.8\pm8.7$\\
050525A	&	$0.606$	&	$2.30\pm0.49$	&	100728B	&	$2.106$	&	$3.55\pm0.36$\\
050603	&	$2.821$	&	$64.1\pm6.4$	&	100814A	&	$1.44$	&	$15.3\pm1.8$\\
050730	&	$3.969$	&	$11.8\pm0.8$	&	100901A	&	$1.408$	&	$4.22\pm0.50$\\
050802	&	$1.71$	&	$5.66\pm0.47$ 	&	100906A	&	$1.727$	&	$29.9\pm2.9$\\
050814	&	$5.3$	&	$9.9\pm1.1$	&	101213A	&	$0.414$	&	$2.72\pm0.53$\\
050820	&	$2.615$	&	$103\pm10$	&	110128A	&	$2.339$	&	$1.58\pm0.21$\\
050904	&	$6.295$	&	$133\pm14$	&	110205A	&	$2.22$	&	$48.3\pm6.4$\\
050908	&	$3.347$	&	$1.54\pm0.16$	&	110213A	&	$1.46$	&	$5.78\pm0.81$\\
0509220	&	$2.199$	&	$5.6\pm1.8$	&	110213B	&	$1.083$	&	$8.3\pm1.3$\\
051022	&	$0.8$	&	$56.0\pm5.6$	&	110422A	&	$1.77$	&	$79.8\pm8.2$\\
051109A	&	$2.346$	&	$6.85\pm0.73$	&	110503A	&	$1.613$	&	$20.8\pm2.1$\\
051111	&	$1.55$	&	$15.4\pm1.9$	&	110715A	&	$0.82$	&	$4.36\pm0.45$\\
060115	&	$3.533$	&	$5.9\pm3.8$	&	110731A	&	$2.83$	&	$49.5\pm4.9$\\
060124	&	$2.296$	&	$43.8\pm6.4$	&	110801A	&	$1.858$	&	$10.9\pm2.7$\\
060202	&	$0.785$	&	$1.20\pm0.09$	&	110818A	&	$3.36$	&	$26.6\pm2.8$\\
060206	&	$4.056$	&	$4.1\pm1.9$	&	111008A	&	$4.9898$	&	$24.7\pm1.2$\\
060210	&	$3.91$	&	$32.2\pm3.2$	&	111107A	&	$2.893$	&	$3.76\pm0.55$\\
060306	&	$3.5$	&	$7.6\pm1.0$	&	111209A	&	$0.677$	&	$5.14\pm0.62$\\	
060418	&	$1.489$	&	$13.5\pm2.7$	&	111228A	&	$0.716$	&	$2.75\pm0.28$\\
060510B	&	$4.9$	&	$19.1\pm0.8$	&	120119A	&	$1.728$	&	$27.2\pm3.6$\\
060522	&	$5.11$	&	$6.47\pm0.63$	&	120326A	&	$1.798$	&	$3.27\pm0.33$\\
060526	&	$3.22$	&	$2.75\pm0.37$	&	120327A	&	$2.813$	&	$14.42\pm0.46$\\
060605	&	$3.773$	&	$4.23\pm0.61$	&	120404A	&	$2.876$	&	$4.18\pm0.34$\\
060607A	&	$3.075$	&	$11.9\pm2.8$	&	120624B	&	$2.197$	&	$319\pm32$\\
060707	&	$3.424$	&	$4.3\pm1.1$	&	120711A	&	$1.405$	&	$180\pm18$\\
060708	&	$1.92$	&	$1.06\pm0.08$	&	120712A	&	$4.175$	&	$21.2\pm2.1$\\
060714	&	$2.7108$	&	$7.67\pm0.44$	&	120716A	&	$2.486$	&	$30.2\pm3.0$\\
060814	&	$1.923$	&	$56.7\pm5.7$	&	120802A	&	$3.796$	&	$12.9\pm2.8$\\
060906	&	$3.6856$	&	$7.81\pm0.51$	&	120811C	&	$2.671$	&	$6.41\pm0.64$\\
060908	&	$1.884$	&	$7.2\pm1.9$	&	120815A	&	$2.358$	&	$1.65\pm0.27$\\
060926	&	$3.2086$	&	$2.29\pm0.37$	&	120909A	&	$3.93$	&	$87\pm10$\\
060927	&	$5.46$	&	$12.0\pm2.8$	&	121024A	&	$2.298$	&	$4.61\pm0.55$\\
061007	&	$1.262$	&	$90.0\pm9.0$	&	121027A	&	$1.773$	&	$3.29\pm0.17$\\
061110B	&	$3.4344$	&	$17.9\pm1.6$	&	121128A	&	$2.2$	&	$8.66\pm0.87$\\
061121	&	$1.314$	&	$23.5\pm2.7$	&	121201A	&	$3.385$	&	$2.52\pm0.34$\\
061126	&	$1.1588$	&	$31.4\pm3.6$	&	121229A	&	$2.707$	&	$3.7\pm1.1$\\
061222A	&	$2.088$	&	$30.0\pm6.4$	&	130408A	&	$3.758$	&	$35.0\pm6.4$\\
070110	&	$2.3521$	&	$4.98\pm0.30$	&	130418A	&	$1.218$	&	$9.9\pm1.6$\\
070125	&	$1.547$	&	$84.1\pm8.4$	&	130420A	&	$1.297$	&	$7.74\pm0.77$\\
070306	&	$1.4959$	&	$8.26\pm0.41$	&	130427A	&	$0.334$	&	$92\pm13$\\
070318	&	$0.84$	&	$3.64\pm0.17$	&	130427B	&	$2.78$	&	$5.04\pm0.48$\\
070411	&	$2.954$	&	$8.31\pm0.45$	&	130505A	&	$2.27$	&	$347\pm35$\\
070508	&	$0.82$	&	$7.74\pm0.29$	&	130514A	&	$3.6$	&	$52.4\pm9.2$\\
070521	&	$1.35$	&	$10.8\pm1.8$	&	130518A	&	$2.488$	&	$193\pm19$\\
070529	&	$2.4996$	&	$12.8\pm1.1$	&	130606A	&	$5.91$	&	$28.3\pm5.1$\\
070611	&	$2.0394$	&	$0.92\pm0.13$	&	130610A	&	$2.092$	&	$6.99\pm0.46$\\
070721B	&	$3.6298$	&	$24.2\pm1.4$	&	130701A	&	$1.155$	&	$2.60\pm0.09$\\
071003	&	$1.604$	&	$38.3\pm4.5$	&	130907A	&	$1.238$	&	$304\pm19$\\
071010B	&	$0.947$	&	$2.32\pm0.40$	&	130925A	&	$0.347$	&	$18.41\pm0.37$\\
071020	&	$2.145$	&	$10.0\pm4.6$	&	131105A	&	$1.686$	&	$34.7\pm1.2$\\
071031	&	$2.6918$	&	$4.99\pm0.97$	&	131117A	&	$4.042$	&	$1.02\pm0.16$\\
071117	&	$1.331$	&	$5.86\pm2.7$	&	140206A	&	$2.74$	&	$35.93\pm0.73$\\
080207	&	$2.0858$	&	$16.4\pm1.8$	&	140213A	&	$1.2076$	&	$9.93\pm0.15$\\
080210	&	$2.6419$	&	$4.77\pm0.29$	&	140226A	&	$1.98$	&	$5.8\pm1.1$\\
080310	&	$2.4274$	&	$8.58\pm0.90$	&	140304A	&	$5.283$	&	$13.7\pm1.1$\\
080319B	&	$0.937$	&	$118\pm12$	&	140311A	&	$4.954$	&	$11.6\pm1.5$\\
080319C	&	$1.95$	&	$14.9\pm3.0$	&	140419A	&	$3.956$	&	$186\pm77$\\
080325	&	$1.78$	&	$9.55\pm0.84$	&	140423A	&	$3.26$	&	$65.3\pm3.3$\\
080411	&	$1.03$	&	$16.2\pm1.6$	&	140428A	&	$4.7$	&	$1.88\pm0.31$\\
080413A	&	$2.433$	&	$8.6\pm2.1$	&	140430A	&	$1.6$	&	$1.54\pm0.23$\\
080413B	&	$1.1$	&	$1.61\pm0.27$	&	140506A	&	$0.889$	&	$1.12\pm0.06$\\
080514B	&	$1.8$	&	$18.1\pm3.6$	&	140508A	&	$1.027$	&	$23.24\pm0.26$\\
080603B	&	$2.69$	&	$6.0\pm3.1$	&	140509A	&	$2.4$	&	$3.77\pm0.44$\\
080604	&	$1.4171$	&	$1.05\pm0.12$	&	140512A	&	$0.725$	&	$7.76\pm0.18$\\
080605	&	$1.64$	&	$28\pm14$	&	140515A	&	$6.32$	&	$5.41\pm0.55$\\
080607	&	$3.036$	&	$200\pm20$	&	140518A	&	$4.707$	&	$5.89\pm0.59$\\
080710	&	$0.8454$	&	$1.68\pm0.22$	&	140614A	&	$4.233$	&	$7.3\pm2.1$\\
080721	&	$2.591$	&	$134\pm23$	&	140629A	&	$2.275$	&	$6.15\pm0.90$\\
080804	&	$2.205$	&	$12.0\pm1.2$	&	140703A	&	$3.14$	&	$1.72\pm0.09$\\
080805	&	$1.5042$	&	$5.05\pm0.22$	&	140801A	&	$1.32$	&	$5.69\pm0.05$\\
080810	&	$3.35$	&	$47.8\pm5.5$	&	140808A	&	$3.29$	&	$11.93\pm0.75$\\
080905B	&	$2.3739$	&	$4.55\pm0.37$	&	140907A	&	$1.21$	&	$2.29\pm0.08$\\
080913	&	$6.695$	&	$9.2\pm2.7$	&	141026A	&	$3.35$	&	$7.17\pm0.90$\\
080916A	&	$0.689$	&	$0.98\pm0.10$	&		        &		        &			      \\
080916C	&	$4.35$	&	$407\pm86$	&		        &		        &			     \\
080928	&	$1.692$	&	$3.99\pm0.91$	&		        &		        &			      \\
\end{longtable}

\section{The short gamma-ray flashes}\label{sec:descr_S-GRFs}

\subsection{General properties}\label{sec:SGRFs1}

\begin{figure*}
\centering
(a)\includegraphics[width=0.45\hsize,clip]{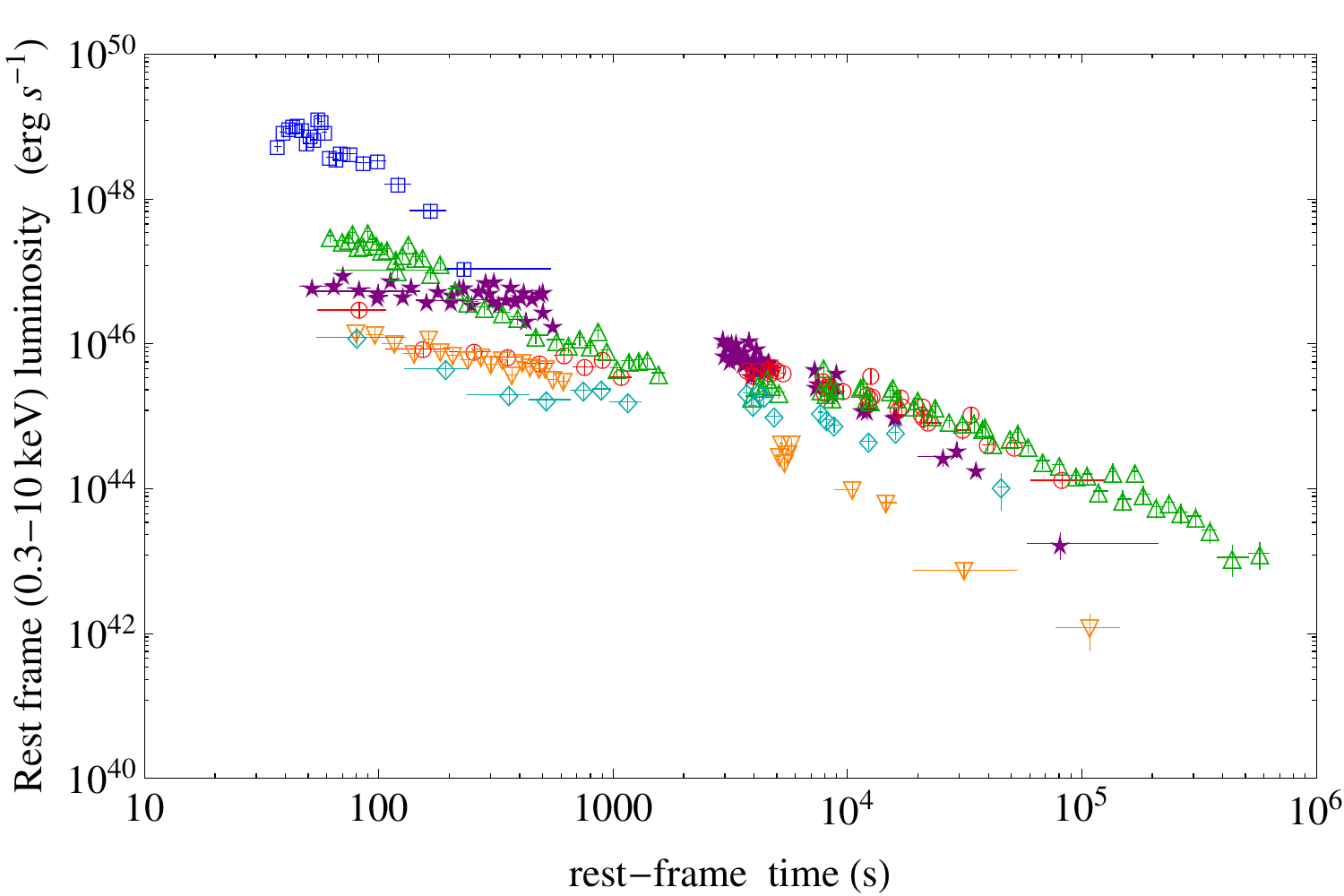}
(b)\includegraphics[width=0.45\hsize,clip]{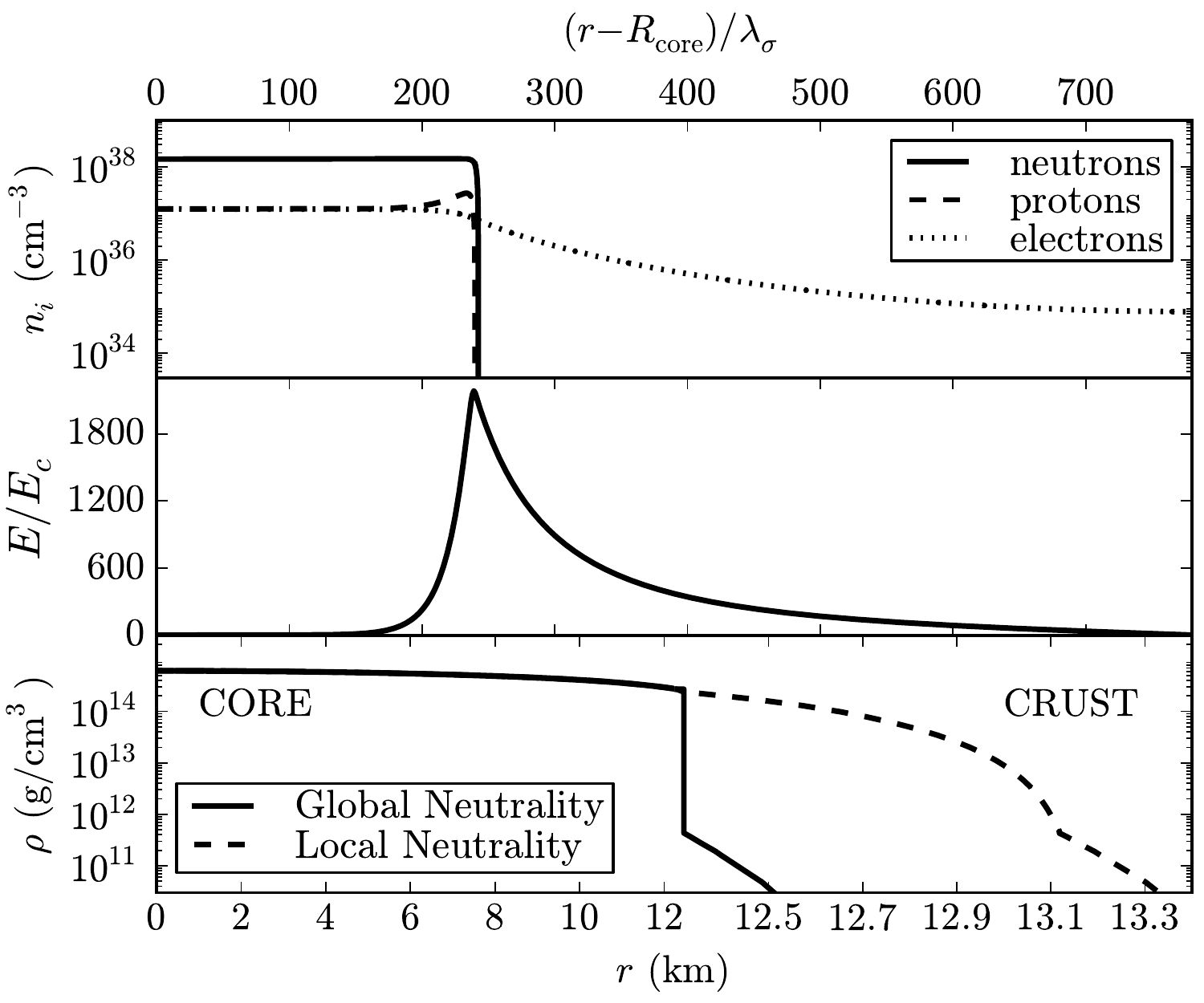}
\null\hfill\hfill\null\\
(c)\includegraphics[width=0.45\hsize,clip]{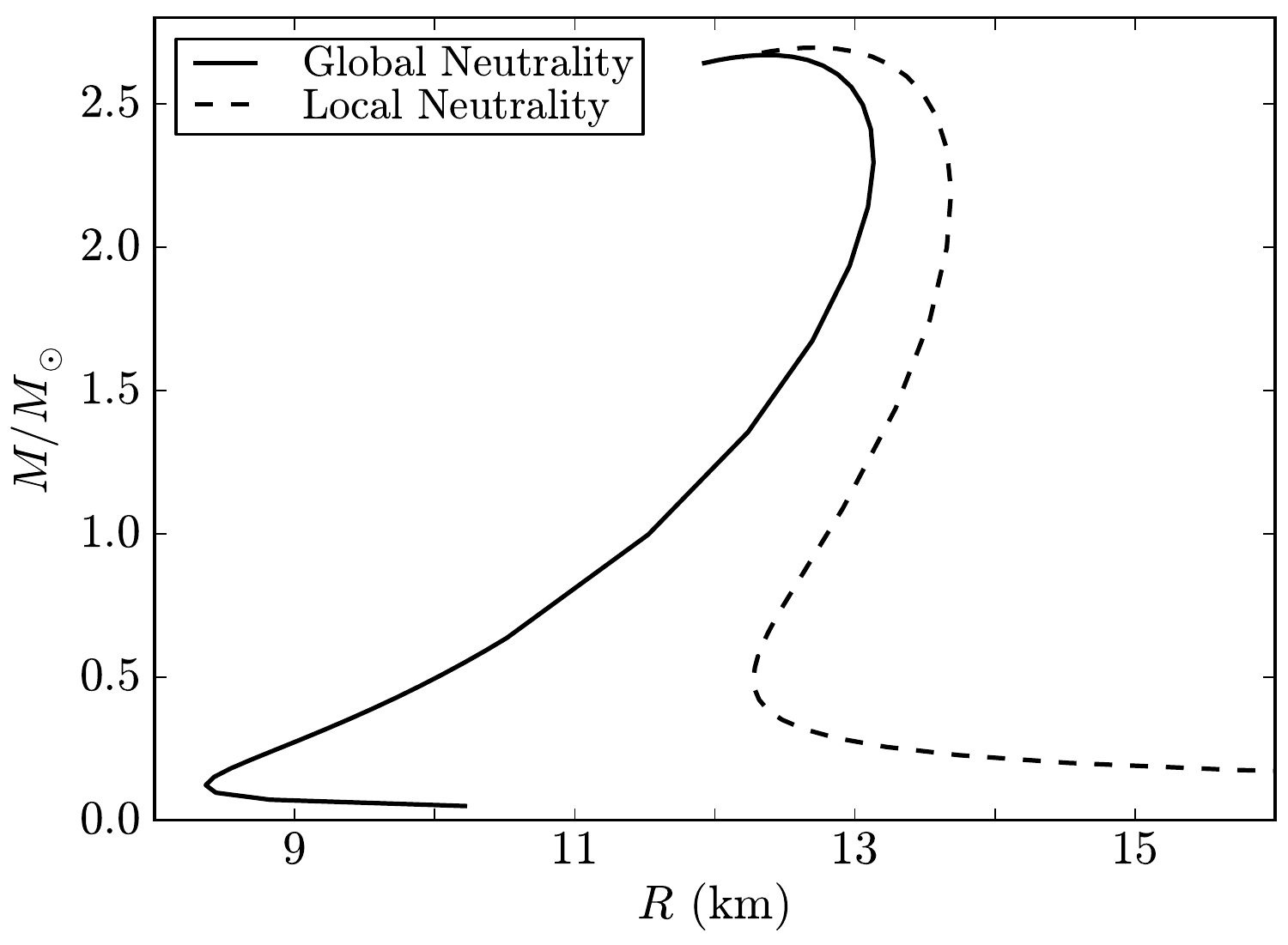}
(d)\includegraphics[width=0.45\hsize,clip]{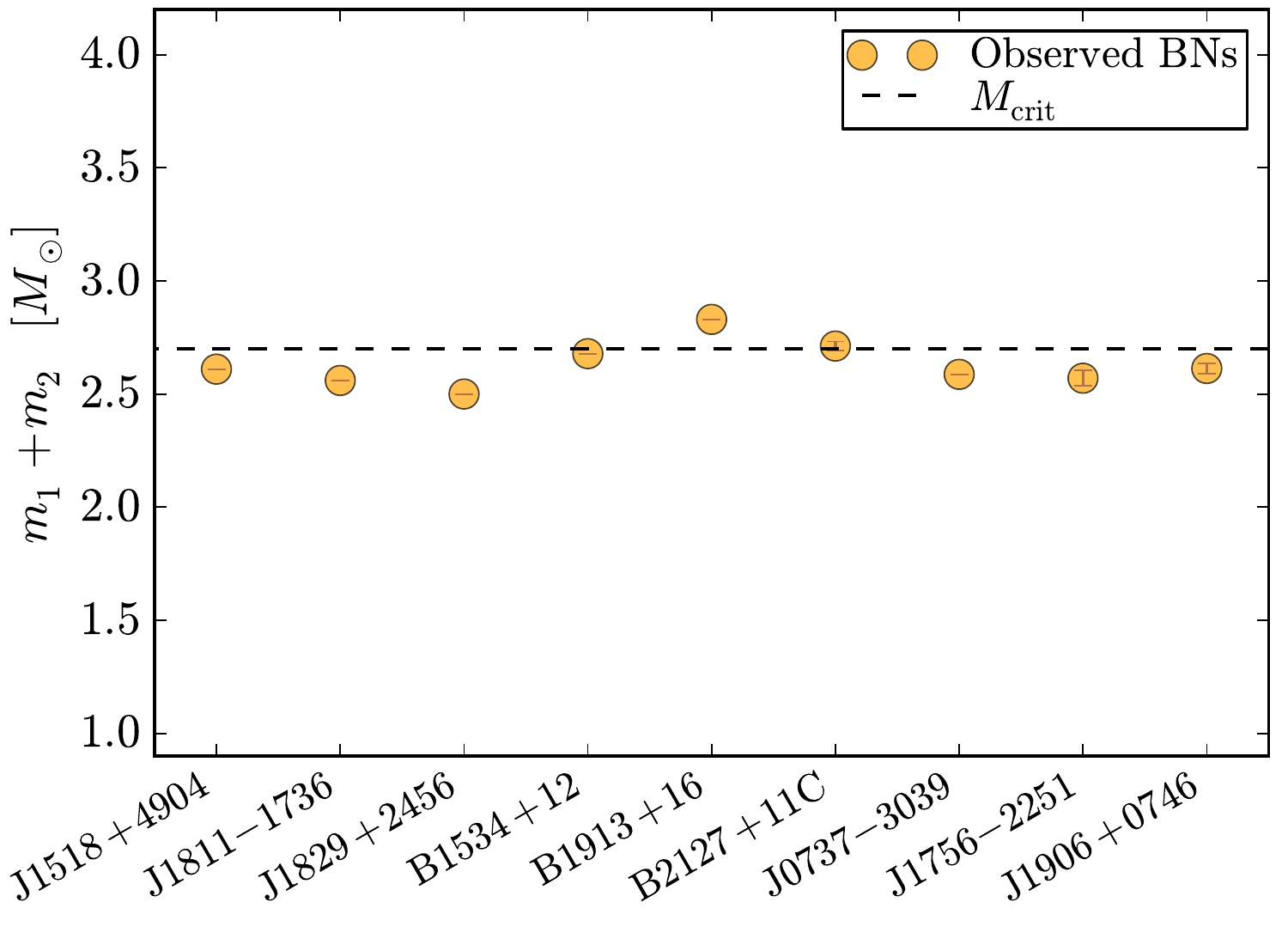}
\null\hfill\hfill\null
\caption{(a) Rest-frame $0.3$--$10$ keV luminosity light curves of some selected S-GRFs: GRB 051210 (blue squares), GRB 051221 (green triangles), GRB 061201 (orange reversed triangles),  GRB 070809 (light blue diamonds), GRB 130603B (purple stars), and GRB 140903A (red circles). See Table~\ref{tab:S-GRFs} for details on the sources. (b) Upper panel: particle density profiles in the NS core-crust boundary interface. Middle panel: electric field in the core-crust transition layer in units of $E_c$. Lower panel: density profile inside a NS star with central density $\rho\sim5\rho_{\rm nuc}$, where $\rho_{\rm nuc}$ is the nuclear density, from the solution of the TOV equations (locally neutral case) and the globally neutral solution presented in \cite{Belvedere}. Here the density at the edge of the crust is the neutron drip density $\rho_{\rm drip}=4.3\times10^{11}$~g~cm$^{-3}$ and $\lambda_\sigma=\hbar/(m_\sigma c)\sim0.4$~fm denotes the $\sigma$-meson Compton wavelength. Reproduced from \cite{Oliveira2014} with their kind permission.
(c) Mass-radius relation obtained with the local and the new global neutrality equilibrium configurations, by applying the NL3 nuclear model, with a critical mass of $2.67$ M$_\odot$ for non-rotating NS \citep{Belvedere}. Figure reproduced from \citep{Belvedere}. (d) Plot of the galactic binary NSs with known total masses ($m_1+m_2$, in solar masses). The horizontal dashed line marks the NS critical mass: systems beyond this value lead to BH formation. Reproduced from \citet{2015ApJ...808..190R}.}
\label{fig:S-GRF_Xray}
\end{figure*}

The observational features of short bursts with energy below $\approx10^{52}$~erg are listed below and summarized in Fig.~\ref{fig:S-GRF_Xray}. 
These bursts are interpreted within the theoretical framework of the NS-NS merger paradigm in the fireshell model as a new class which we indicate as S-GRFs.

The upper limit on the energetic of the S-GRFs is $(7.8\pm1.0)\times10^{51}$~erg as measured in GRB 100117A.

The isotropic energies are in the range $(8.5\pm2.2)\times10^{48}\lesssim E_{\rm iso}\lesssim(7.8\pm1.0)\times10^{51}$~erg \citep[see Fig.~\ref{fig:ffff} and][]{2012ApJ...750...88Z,Calderone2014,2015ApJ...808..190R}.

The spectral peak energies are in the range $0.2\lesssim E_{\rm p,i}\lesssim2$~MeV \citep[see Fig.~\ref{fig:ffff} and][]{2012ApJ...750...88Z,Calderone2014,2015ApJ...808..190R} and increase monotonically with $E_{\rm iso}$.

The cosmological redshifts are in the range $0.111\leq z\leq2.609$, with an average value of $\approx0.71$ (see Table~\ref{tab:S-GRFs}).

The prompt emission phase has a duration of a few seconds, and is expected to crucially be a function of the masses of the binary neutron stars.

The long lasting X-ray afterglow does not exhibit any specific common late power-law behavior (see Fig.~\ref{fig:S-GRF_Xray}~(a)).

For all S-GRFs no SN association is expected, nor observed.

No high energy GeV emission is expected nor observed in absence of BH formation.

\subsection{Theoretical interpretation of S-GRFs within the NS-NS merger paradigm in the fireshell model}\label{sec:descr_S-GRFs1b}

As noted in the Introduction, current paradigms indicate mergers of NS-NS or NS-BH binaries as progenitors.

The extension of the IGC paradigm considerations to NS--NS mergers has led to a new classification of short bursts into two sub-classes depending upon the mass of the merged core, namely whether or not a BH is formed out of the merger \citep[see Fig.~\ref{fig:S-GRF_Xray}~(d) and][]{2015ApJ...808..190R}. 
This, in turn, depends on the NS equation of state and on the adoption of a global neutrality model, as opposed to the case of absence of electromagnetic structures when local charge neutrality is imposed \citep[see, e.g.,][and references therein, and Fig.~\ref{fig:S-GRF_Xray}~(c)]{Belvedere}. Also relevant is the very different density distribution in the crust and in the core between these two treatments, which could play an important role in the NS--NS mergers (see Fig.~\ref{fig:S-GRF_Xray}~(d) and \citealt{Oliveira2014}).

S-GRFs originate from NS--NS mergers with initial total mass $m_1+m_2$ leading to merged core with mass smaller than $M_{\rm crit}$, therefore their outcomes are a MNS with additional orbiting material, or even a binary NS or WD companion \citep[see, e.g.,][and references therein]{1992ApJ...400..175B}, due to the energy and momentum conservation laws \citep{2015ApJ...808..190R}.  
As discussed in Sec.~\ref{sec:fireshell}, even though a BH is not formed out of the merger, also for these systems the general description of the fireshell model can be applied. A viable mechanism for S-GRFs can be the creation of a pair plasma via $\nu\bar{\nu}\to e^+e^-$ in a NS-NS merger \citep[see, e.g.,][]{Narayan1992,SalmonsonWilson2002,Rosswog2003}, where the maximum energy attainable in the process is $\approx10^{52}$ erg, which represents indeed the upper limit to the energetic of these systems.
Their energies are very similar to those emitted in XRFs. However, S-GRFs have $E_{p,i}$ as high as $\sim2$~MeV \citep[see, e.g.,][]{2015ApJ...808..190R,Calderone2014,2012ApJ...750...88Z}, therefore, in view of the hardness of their spectra, we adopted the name of S-GRFs to distinguish them from the corresponding XRFs.

S-GRFs coincide with the majority of the systems extensively discussed in \citet{2014ARA&A..52...43B}.
All S-GRFs have an extended X-ray afterglow \citep{2014ARA&A..52...43B,2015ApJ...808..190R}. 
Similarly to XRFs, the rest-frame $0.3$--$10$ keV luminosity light curve does not exhibit either a late common power-law behavior, or the nesting discovered in the BdHNe (see Fig.~\ref{fig:S-GRF_Xray}(a)). At the moment, there are still a large number of possible candidates for the description of the origin of the late X-ray afterglow emission: a) the interaction of the MNS with orbiting material or with a less massive binary NS or WD companion, b) the accelerated baryons interacting with the circumburst medium after the P-GRB emission, or c) the possible radioactive decay of heavy elements synthesized in the ejecta of a compact binary merger \citep{Li1998}.
In this light we recall the possibility of a \textit{macronova} emission, a near-infrared/optical transient (a bump) in the late afterglow (see the case of GRB 130603B in \citealp{2013ApJ...774L..23B} and \citealp{2013Natur.500..547T}).
 
As a general conclusion, in \citet{2015ApJ...808..190R} the necessary absence of a SN was indicated. It has been predicted there that, since no BH is produced in the merger, S-GRFs should never exhibit high energy GeV emission, which is expected to originate in the newly-born BH \citep{2015ApJ...808..190R}. No counterexample has been found as of today. In \citet{2016arXiv160702400E} it has been shown that the absence of detection of GeV emission, necessary within the fireshell model, is indeed supported by the observations. The entire section 6.5 of that paper is dedicated to the GeV emission of S-GRFs and S-GRBs. As evidenced there, it is concluded that S-GRFs, due to the upper limits of the LAT observations, have, if any, GeV fluxes necessarily $10^5$--$10^6$ times weaker than those of S-GRBs, although their $E_{iso}$ is only a factor $10^{2}$ smaller (see also Appendix~\ref{Appendix}).

\subsection{Prototypes}\label{sec:descr_S-GRFs2}

In Table~\ref{tab:S-GRFs} we indicate selected the prototypes of S-GRFs. For each of them, we list the values of $E_{\rm iso}$ and $z$ used in order to evaluate their rate.
\begin{table*}
\centering
\begin{tabular}{lcc|lcc}
\hline\hline
GRB & $z$ & $E_{\rm iso}/$($10^{50}$ erg) &  GRB & $z$ & $E_{\rm iso}/$($10^{50}$ erg)\\
\hline
050509B	&	$0.225$	&	$0.085\pm0.022$	&	090927	&	$1.37$	&	$27.6\pm3.5$\\
050709	&	$0.161$	&	$0.80\pm0.08$	&	100117A	&	$0.915$	&	$78\pm10$\\
051221A	&	$0.546$	&	$26.3\pm3.3$	&	100206A	&	$0.408$	&	$4.67\pm0.61$\\
060502B	&	$0.287$	&	$4.33\pm0.53$	&	100625A	&	$0.453$	&	$7.50\pm0.30$\\
061201	&	$0.111$	&	$1.51\pm0.73$ 	&	 100724A	&	$1.288$	&	$16.4\pm2.4$\\
061217	&	$0.827$ 	&	 $42.3\pm7.2$ 	&	101219A	&	$0.718$	&	$48.8\pm6.8$\\
070429B	&	$0.902$	&	$4.75\pm0.71$	&	111117A	&	$1.3$	&	$34\pm13$\\
070724A	&	$0.457$	&	$0.60\pm0.14$	&	120804A	&	$1.3$	&	$70\pm15$	\\
070729	&	$0.8$      	&	 $11.3\pm4.4$ 	&	130603B	&	$0.356$	&	$21.2\pm2.3$	\\
070809	&	$0.473$	&	$2.76\pm0.37$	&	131004A	&	$0.717$	&	$12.7\pm0.9$	\\
080123	&	$0.495$	&	$11.7\pm3.9$	&	140622A	&	$0.959$	&	$0.70\pm0.13$	\\
080905A	&	$0.122$	&	$6.58\pm0.96$	&	140903A	&	$0.351$	&	$1.41\pm0.11$	\\
090426	&	$2.609$	&	$44.5\pm6.6$         	&	141004A	&	$0.573$	&	$21.0\pm1.9$	\\
090515	&	$0.403$	&	$0.094\pm0.014$ 	&		&		&		\\
\hline
\end{tabular}
\caption{List of the S-GRFs considered in this work up to the end of 2014. For each source (first columns) the values of $z$ and $E_{\rm iso}$ (second and third columns, respectively) are listed.}
\label{tab:S-GRFs}
\end{table*}

\section{The short GRBs}\label{sec:descr_SGRBs}

\subsection{General properties}\label{sec:SGRBs1}

The observational features of short bursts with energy above $\approx10^{52}$~erg are listed below and summarized in Fig.~\ref{ShortXO}. 
These bursts are interpreted within the theoretical framework of the NS-NS merger paradigm in the fireshell model as a new class which we indicate as S-GRBs.

The lower limit on the energetic of the S-GRBs is $(2.44\pm0.22)\times10^{52}$~erg as measured in GRB 081024B.

The isotropic energies are in the range $(2.44\pm0.22)\times10^{52}\lesssim E_{\rm iso}\lesssim(2.83\pm0.15)\times10^{53}$~erg \citep[see Fig.~\ref{fig:ffff} and][]{2012ApJ...750...88Z,Muccino2012,Calderone2014,2015ApJ...808..190R}.

The spectral peak energies are in the range $2\lesssim E_{\rm p,i}\lesssim8$~MeV \citep[see Fig.~\ref{fig:ffff} and][]{2012ApJ...750...88Z,Muccino2012,Calderone2014,2015ApJ...808..190R} and increase monotonically with $E_{\rm iso}$.

The cosmological redshifts are in the range $0.903\leq z\leq5.52$, with an average value of $\approx2.48$ (see Table~\ref{tab:SGRBs}).

The P-GRB and the prompt emission components have a total duration of a few seconds, which is expected to crucially be a function of the masses of the binary neutron stars. (see Fig.~\ref{ShortXO}~(a) and (b)).

Only in the case of GRB 090510 an X-ray afterglow has been observed not conforming to any known afterglow (see Fig.~\ref{ShortXO}~(c)).

For all S-GRBs no SN association is expected nor observed.

In all S-GRBs an extremely high energy GeV emission has been observed (see Fig.~\ref{ShortXO}~(d)). It is interesting that even in one case, which was outside the nominal \textit{Fermi}-LAT field of view, evidence for high energy emission has been found (Ruffini \& Wang, in preparation, and \citealp{Ackermann2013}).

\subsection{Theoretical interpretation of S-GRBs within the NS-NS merger paradigm in the fireshell model}\label{sec:SGRBs1b}

S-GRBs originate from NS--NS mergers with initial total mass $m_1+m_2$ leading to a merged core with mass larger than $M_{\rm crit}$ so that a BH is formed \citep{2015ApJ...808..190R}.
In order to conserve energy and momentum, the outcome of such S-GRBs is a KNBH with additional orbiting material, or a binary companion \citep{1992ApJ...400..175B,2015ApJ...808..190R}.
If we compare and contrast the different Episodes encountered in the description of the BdHNe (see Section~\ref{sec:descr_BdHNe}) with those of S-GRBs, we find some remarkable analogies but also some differences in view of the simplicity of the underlying physical system of S-GRBs, which unlike the BdHNe, do not exhibit any of the extremely complex activities related to the SN (see Section~\ref{sec:descr_BdHNe}).

Episode 1 corresponds here to the activity of the NS--NS merger before the gravitational collapse into a BH. 
Because of the compactness of the systems this process at times is not observable or it possibly corresponds to faint precursors observed in some short bursts (see, e.g., \citealt{2010ApJ...723.1711T} and \citealt{2016arXiv160702400E}).

\begin{figure*}
\centering
(a)\includegraphics[width=0.45\hsize,clip]{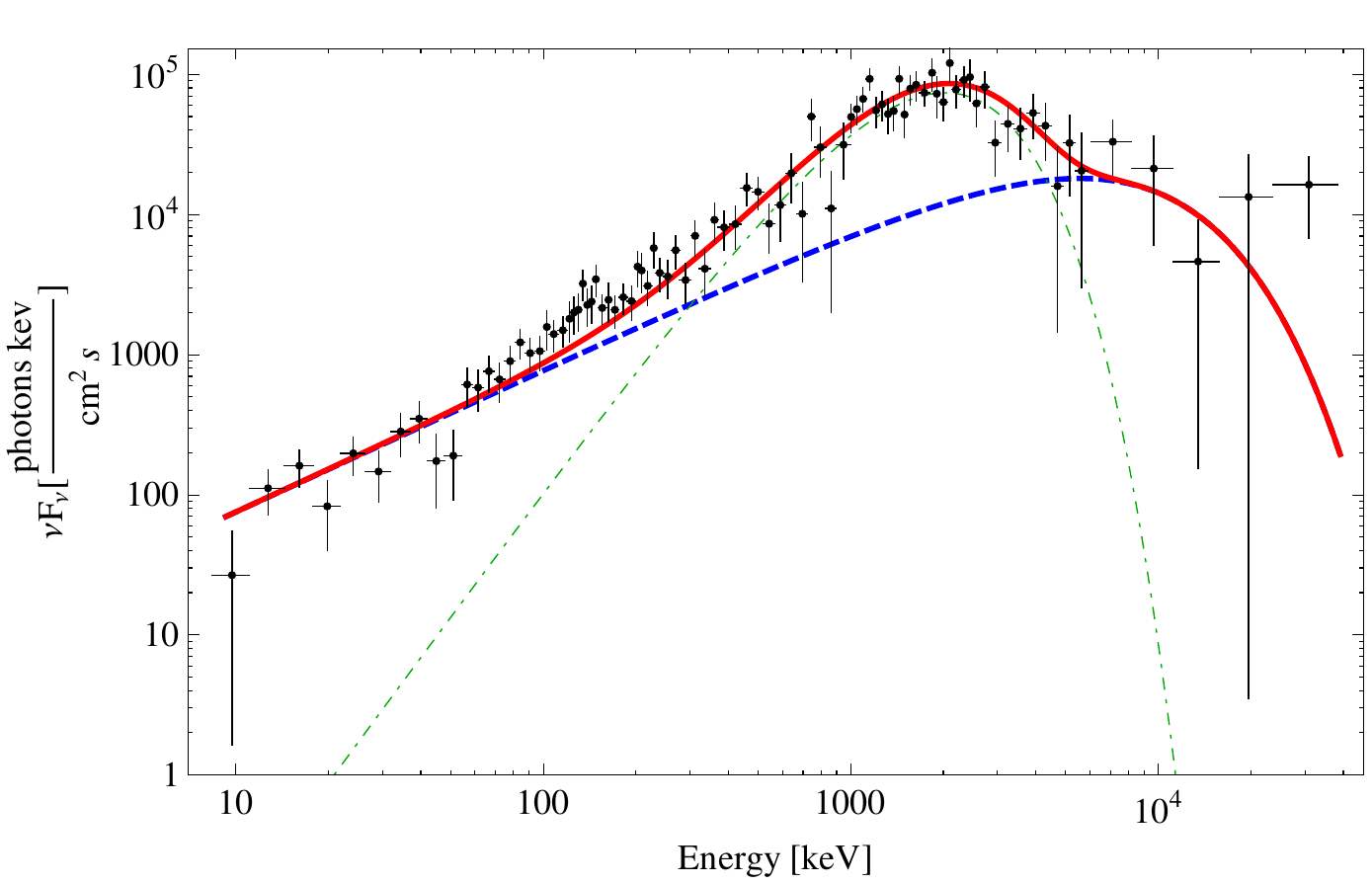}
(b)\includegraphics[width=0.45\hsize,clip]{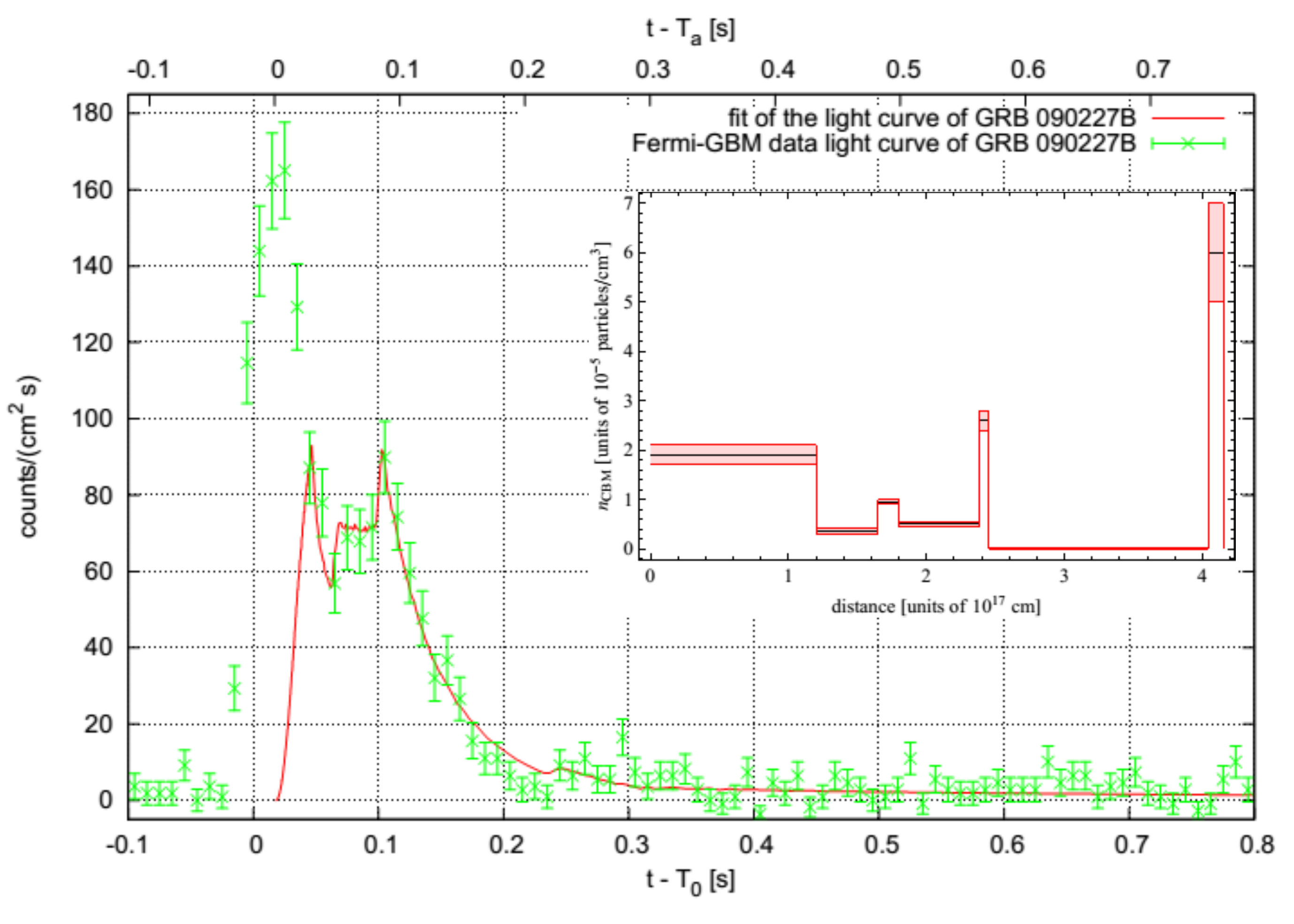}
\null\hfill\hfill\null\\
(c)\includegraphics[width=0.45\hsize,clip]{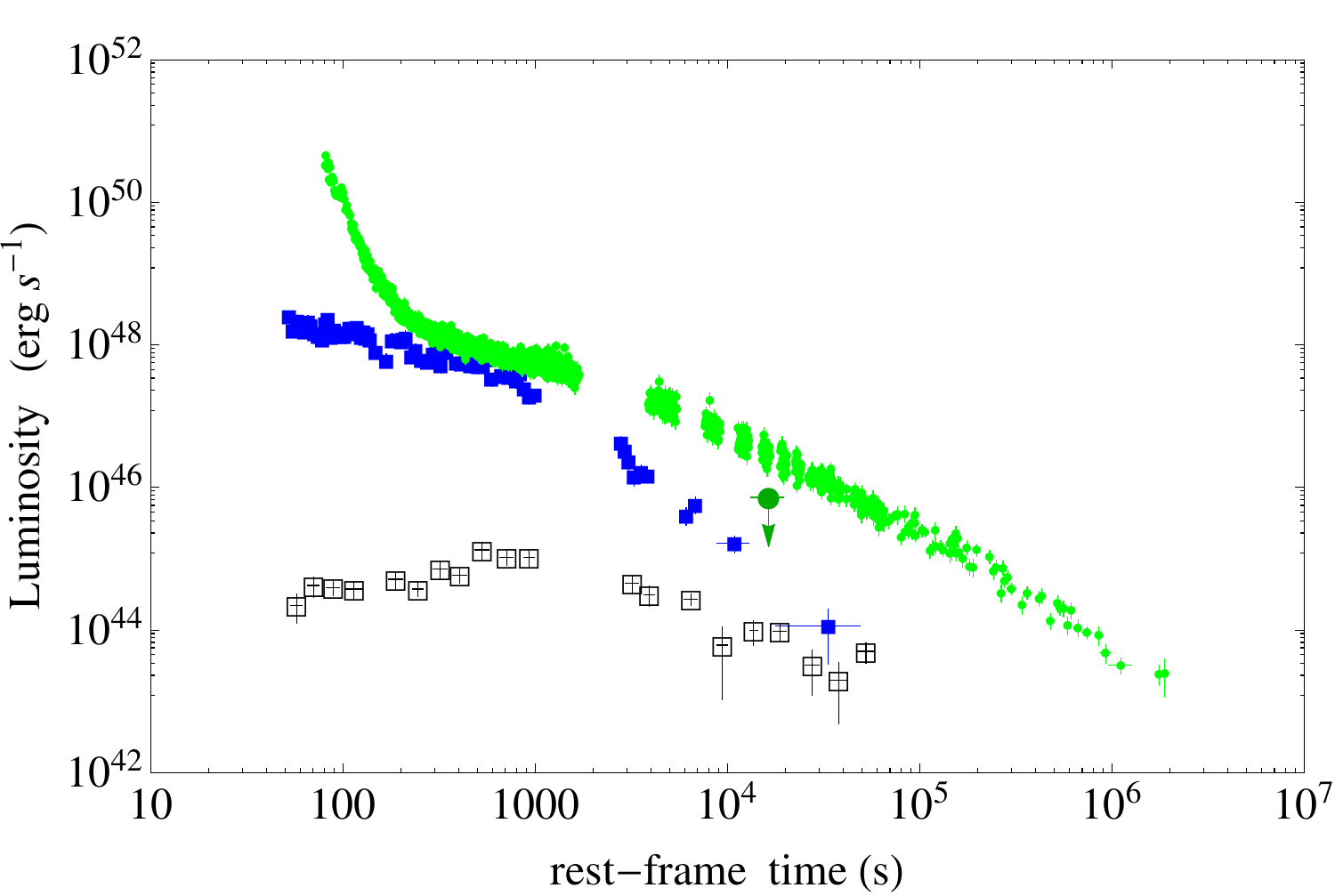}
(d)\includegraphics[width=0.45\hsize,clip]{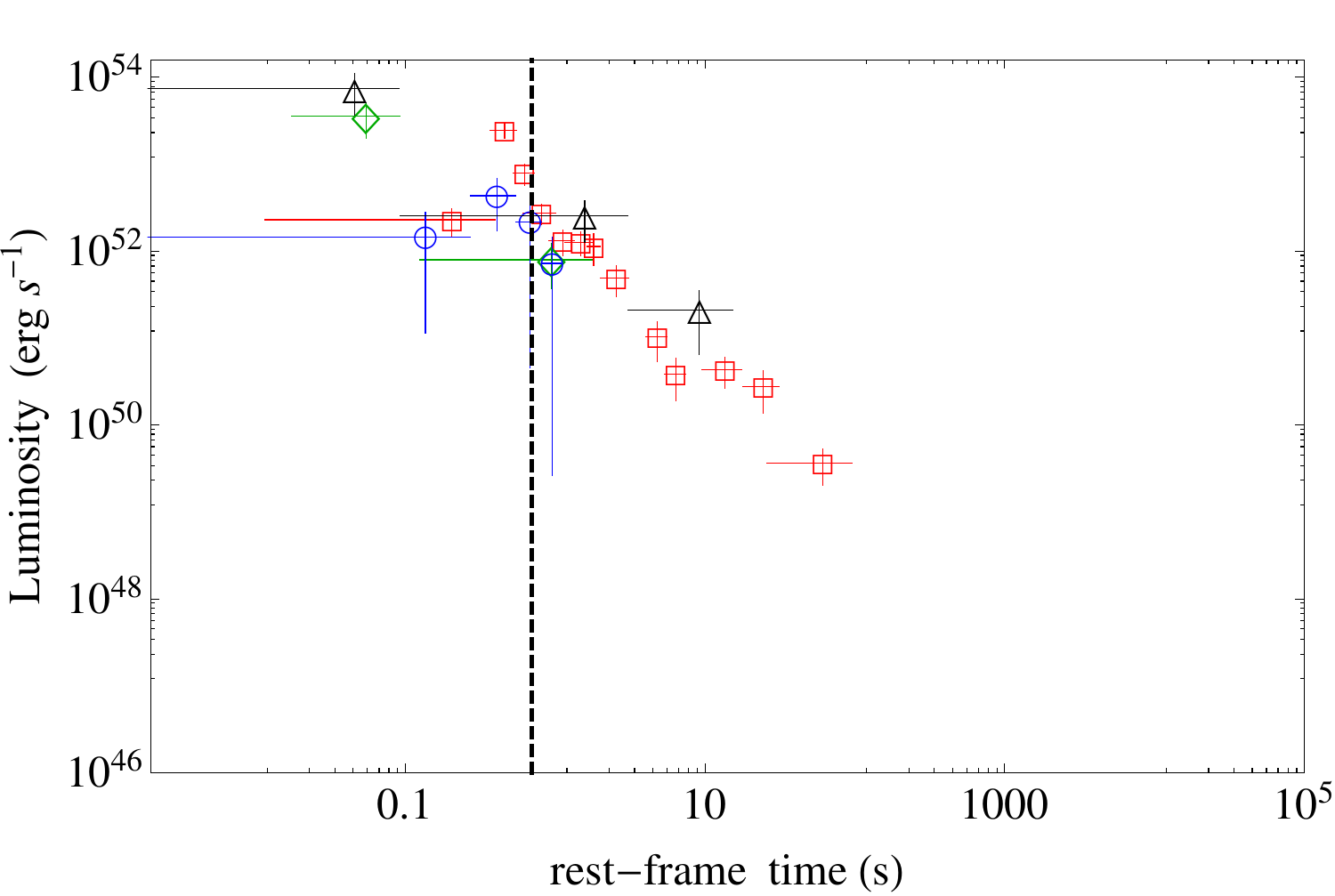}
\null\hfill\hfill\null
\caption{(a) The simulation of the observed P-GRB spectrum of the prototypical S-GRB 090227B: the dashed-dotted green line represents the BB emission, the dashed blue line corresponds to the early on-set of the prompt emission within the P-GRB computed from the fireshell simulation in the energy band $8$--$40000$ keV; the sum of the two components is shown as a solid red line. Reproduced from \citealp{Muccino2012}. (b) The NaI-n2 light curve of the prompt emission of the S-GRB 090227B (green data) and the simulation within the fireshell model (red curve). The small inset reproduces the CBM profile required for the simulation. Reproduced from \citealp{Muccino2012}. (c) The available X-ray and optical luminosities of S-GRBs: the X-ray rest-frame $0.3$--$10$ keV (blue filled squares) and the optical rest-frame $2$--$7$ eV (black empty squares, taken from \citealp{Depasquale2010}) luminosity light curves of GRB 090510, and the X-ray rest-frame $0.3$--$10$ keV upper limit of GRB 140619B (green filled circle, see, e.g., \citealp{2015ApJ...808..190R}). For comparison the rest-frame $0.3$--$10$ keV luminosity light curve of one of the prototypes of BdHNe, GRB 090618 (green circles) is shown. (d) The rest-frame $0.1$--$100$ GeV luminosity light curves of the S-GRBs 081024B (green diamonds), 090510 (red squares), 140402A (black triangles), 140619B (blue circles). The dashed vertical line marks the minimal turn-on time of the GeV emission of BdHNe. It is clear that in the case of S-GRBs the GeV emission turns on at shorter time scales and exhibits larger luminosities.}
\label{ShortXO}
\end{figure*}

Episode 2 corresponds to the GRB emission stemming from the NS-NS merger. It is described within the fireshell model as composed of two components (see Sec.~\ref{sec:fireshell}): the P-GRB emission, with a mainly thermal spectrum (see Fig.~\ref{ShortXO}~(a)), and  the prompt emission, with a characteristic non-thermal spectrum (see Fig.~\ref{ShortXO}~(b)).
Typically in all S-GRBs so far analyzed (see, e.g., GRB 090227B, \citealp{Muccino2012}, and GRB 140619B, \citealp{2015ApJ...808..190R}) the baryon load is standard, e.g., $B\approx10^{-5}$, and is consistent with the crustal masses of NS-NS mergers \citep{Belvedere2014,2015ApJ...808..190R}. 
The average densities of the circumburst medium where S-GRBs occur are $\langle n_{\rm CBM}\rangle\approx10^{-5}$ cm$^{-3}$, typical of the halos of GRB host galaxies \citep[see, e.g.,][]{Muccino2012,2015ApJ...808..190R}.
Most remarkable is that this model gives the theoretical explanation for the fulfillment of the $E_{\rm p,i}$--$E_{\rm iso}$ relation for the short bursts \citep[see Fig.~\ref{fig:ffff} and][]{2012ApJ...750...88Z,Calderone2014,2015ApJ...808..190R}.

Episode 3, which corresponds to the traditional X-ray afterglow, is missing here in view of the absence of the SN and of all the characteristic processes originating from the interaction between the GRB and the SN ejecta, as in the case of BdHNe (see Section~\ref{sec:descr_BdHNe}).
At times S-GRBs have nonprominent X-ray or optical emissions (see Fig.~\ref{ShortXO}~(c)).

Episode 4, identified with the optical emission of a SN, is here missing.

Episode 5 coincides with the long-lived GeV emission. 
All S-GRBs consistently exhibit this emission, which appears to be strictly correlated to the one observed in the BdHNe.
By analogy with BdHNe, we assume that the GeV emission originate from the activity of the newly-born KNBH produced in the merger \citep{2015ApJ...808..190R}.
Indeed the presence of a BH is the only commonality between BdHNe and S-GRBs.
By comparing and contrasting Figs.~\ref{rad_ind_tot}~(d) and \ref{ShortXO}~(d), we see that the turn-on of the GeV emission in S-GRBs occurs earlier and is energetically more prominent than the corresponding one of the BdHNe. To emphasize this point in Fig.~\ref{ShortXO}~(d) we have represented by a dashed line the minimal turn-on time of the GeV emission of BdHNe \citep[see][and Ruffini et al., in preparation]{2016arXiv160702400E}.
The very high angular momentum, expected to occur in NS-NS mergers, and the very high luminosities of the S-GRBs, originating in the corresponding BH formation, offer the great opportunity to analyze some of the features expected in a KNBH.

\subsection{Prototypes}\label{sec:SGRBs2}

In Table~\ref{tab:SGRBs} we list all the S-GRBs identified so far.
\begin{table*}
\centering
\begin{tabular}{lcc|lcc}
\hline\hline
GRB & $z$ & $E_{\rm iso}/$($10^{52}$ erg) &  GRB & $z$ & $E_{\rm iso}/$($10^{52}$ erg) \\
\hline
060801	&	$1.13$	&	$3.27\pm0.49$	&     090510  	&	$0.903$	&         $ 3.95\pm0.21$  \\ 
081024B	&	$3.05$	&	$2.44\pm0.22$	&     140402A	&	$5.52$	&     	$4.7\pm1.1$        \\
090227B       &	$1.61$	&	$28.3\pm1.5$         &     140619B	           &	$2.67$	&	$6.03\pm0.79$    \\	
\hline
\end{tabular}
\caption{List of the S-GRBs considered in this work up to the end of 2014. For each source (first columns) the values of $z$ and $E_{\rm iso}$ (second and third columns, respectively) are listed.}
\label{tab:SGRBs}
\end{table*}

The first identified S-GRB 090227B has been analyzed by \citet{Muccino2012}. The analysis of its P-GRB emission has found a baryon load $B=4.13\times10^{-5}$ and a Lorentz factor at the transparency condition $\Gamma=1.44\times10^4$.
The fit of the light curve of the prompt emission allowed the determination of the average number density of the circumburst medium, i.e., $\langle n_{CBM} \rangle=1.9\times10^{-5}$~cm$^{-3}$, which is typical of galactic halos where NS--NS mergers migrate, owing to natal kicks imparted to the binaries at birth \citep[see, e.g.,][]{2014ARA&A..52...43B}.
These values are strikingly similar to those inferred for other S-GRBs: GRB 081024B ($B=4.80\times10^{-5}$, $\Gamma=1.07\times10^4$, and $\langle n_{CBM} \rangle=5.0\times10^{-6}$~cm$^{-3}$, Aimuratov et al., in preparation), GRB 090510 ($B=5.54\times10^{-5}$, $\Gamma=1.04\times10^4$, and $\langle n_{CBM} \rangle=8.7\times10^{-6}$~cm$^{-3}$, \citealt{2016arXiv160702400E}), and GRB 140619B ($B=5.52\times10^{-5}$, $\Gamma=1.08\times10^4$, and $\langle n_{CBM} \rangle=4.7\times10^{-5}$~cm$^{-3}$, \citealp{2015ApJ...808..190R}).

With the exception of GRB 090227B, which was outside the nominal \textit{Fermi}-LAT field of view \citep{Ackermann2013}, the GeV luminosity light curves of the above four S-GRBs and that of the additional example recently identified, GRB 140402A (Ruffini et al., in preparation) follow a common behavior when computed in the source rest-frame (see Fig.~\ref{ShortXO}~(d)).

\section{Ultrashort GRBs}\label{sec:rate5}

As pointed out in the introduction, U-GRBs originate from the NS-BH binaries produced in the BdHNe and nearly 100$\%$ of these binaries remain bound \citep{2015arXiv150502809F}.
The lack of any observed source to date is mainly due to the extremely short duration of these systems \citep{2015arXiv150502809F}.

Interesting considerations, which may be of relevance for describing the U-GRB sub-class, can be found in \citet{1999ApJ...518..356P}.

\section{The gamma-ray flashes}\label{sec:DS}

\subsection{General properties}\label{sec:DSGRBs1}

\begin{figure*}
\centering
(a)\includegraphics[width=0.45\hsize,clip]{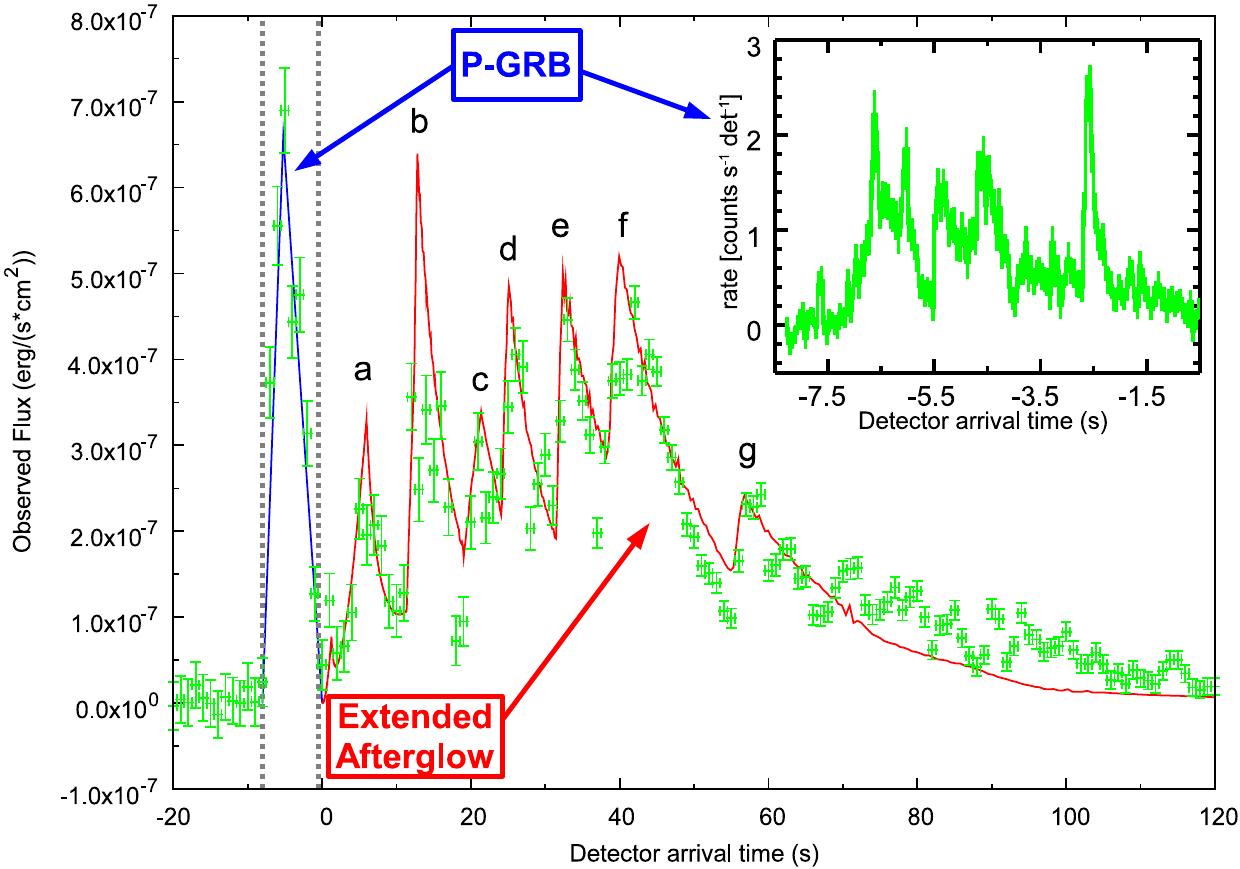}
(b)\includegraphics[width=0.45\hsize,clip]{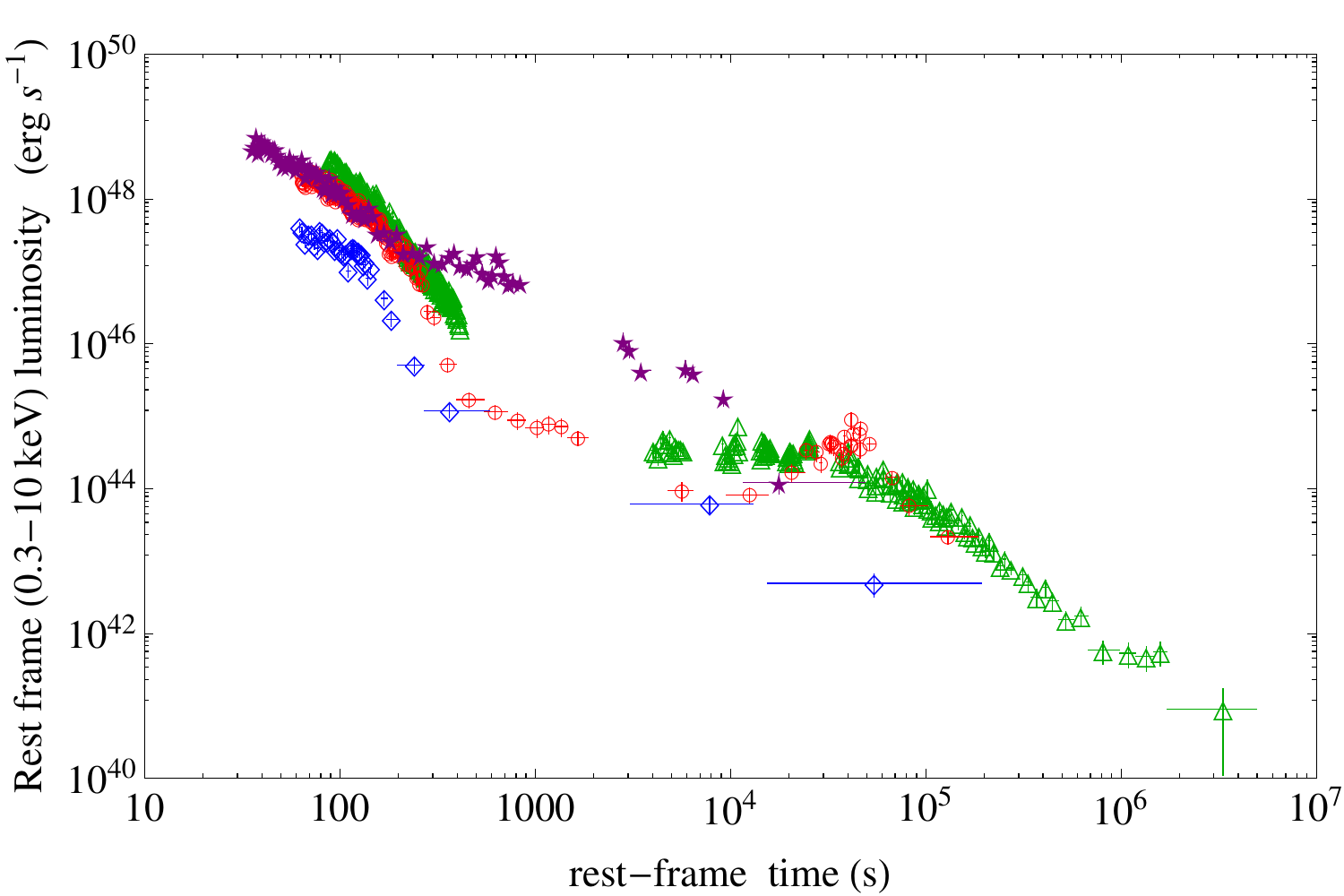}
\null\hfill\hfill\null
\caption{(a) The fireshell simulation of the \textit{Swift}-BAT prompt emission of GRB 060614  (taken from \citealp{Caito2009}). (b) The rest-frame $0.3$--$10$~keV luminosity light curves of selected GRFs: GRB 050724 (red circles), GRB 060614 (green triangles), GRB 070714B (purple stars), and GRB 071227 (blue diamonds).}
\label{fig:DSGRBs}
\end{figure*}

The observational features of short bursts followed by an extended emission with energy below $\approx10^{52}$~erg are listed below and summarized in Fig.~\ref{fig:DSGRBs}. 
These bursts are interpreted within the theoretical framework of a binary merger of a NS and a massive WD \citep{1992A&A...266..232D,1994A&A...287..403D} in the fireshell model as a new class which we indicate as GRFs.

The upper limit on the energetic of the GRFs is $(9.8\pm2.4)\times10^{51}$~erg as measured in GRB 070714B.

The isotropic energies are in the range $(8.9\pm1.6)\times10^{49}\lesssim E_{\rm iso} \lesssim(9.8\pm2.4)\times10^{51}$~erg.

The spectral peak energies are in the range $0.2\lesssim E_{\rm p,i}\lesssim2$~MeV.

The cosmological redshifts are in the range $0.089\leq z\leq2.31$, with an average value of $\approx0.54$ (see Table~\ref{tab:DSGRBs}).

The $\gamma$-ray emission is composed of: 1) an initial spike-like harder emission and 2) a prolonged softer emission observed up to a hundred seconds (see Fig.~\ref{fig:DSGRBs}(a)).

The long lasting X-ray afterglow does not exhibit any specific common late power-law behavior (see Fig.~\ref{fig:DSGRBs}(b)).

No SN association is expected, nor observed also in the case of nearby sources \citep{DellaValle2006sn}.

No high energy GeV emission is expected nor observed in absence of BH formation.

\subsection{Theoretical interpretation of GRFs within the NS-WD merger paradigm in the fireshell model}\label{sec:descr_DS-GRBs1b}

As we mentioned, the mergers of NS--WD binaries, notoriously very common astrophysical systems \citep{2015ApJ...812...63C}, can be the progenitors of another GRB sub-class: the GRFs. Possible evolutionary scenarios leading to NS--WD mergers have been envisaged e.g. in \citet{2014MNRAS.437.1485L,2000ApJ...530L..93T}. Another less likely but yet possible scenario is the merger of a NS--WD binary produced, as recalled in  Section~\ref{sec:descr_S-GRFs1b}, from an S-GRF. Namely, the merger of a not mass-symmetric NS--NS binary with total mass $m_1+m_2$ smaller than $M_{\rm crit}$, that produce a MNS with a low-mass WD companion \citep[see, e.g.,][and references therein]{1992ApJ...400..175B}, due to the energy and momentum conservation laws \citep{2015ApJ...808..190R}.

With the term GRFs we dubbed a class of long GRBs occurring in a CBM environment with low density, e.g., $\sim10^{-3}$~cm$^{-3}$, with a light curve in $\gamma$-rays composed of an initial spike-like hard emission, identified with the P-GRB, and prolonged softer emission, explained as the prompt emission \citep[see Fig.~\ref{fig:DSGRBs} (a) and][]{Caito2009,Caito2010}. 
No associated SN has been ever observed, although in the case of the low value of the cosmological redshift its detection would not have been precluded \citep{DellaValle2006}. 
The prototype of such systems is GRB 060614 \citep{Caito2009}.

Apart from the absence of any associated SN to a GRF, the identification of NS-WD binaries as progenitor systems of the GRFs comes from the following observational and theoretical evidences: a) the initial spike-like emission fulfills the $E_{\rm p,i}$--$E_{\rm iso}$ relation for S-GRFs and S-GRBs \citep{2012ApJ...750...88Z,Calderone2014,2015ApJ...808..190R}, both originating in NS--NS mergers \citep{2015ApJ...808..190R}; b) the value of the baryon load, $B\approx10^{-3}$ \citep{Caito2009,Caito2010} points to a system more baryon-contaminated than the simpler and more compact NS--NS merger \citep[$B\approx10^{-5}$, see, e.g.,][]{2015ApJ...808..190R}; c) the fit of the prompt emission within the fireshell model provides CBM with low density, e.g., $\sim10^{-3}$ cm$^{-3}$, typical of the halos of the GRB host galaxies \citep{Caito2009,Caito2010}; d) the presence of a macronova emission in the optical afterglow of the prototype GRF 060614 \citep{2015ApJ...811L..22J}.

In summary, we list below the different Episodes observed (or not) in GRFs.

Episode 1 does not exist due to the compactness of the NS--WD merger.

Episode 2 corresponds to the $\gamma$-ray emission stemming from the NS--WD merger.
The fireshell theory still applies to these systems in view of the considerations presented in Section~\ref{sec:fireshell}. Also in this case a viable mechanism consists in the the pair creation via $\nu\bar{\nu}\to e^+e^-$ during a NS--WD merger \citep[see, e.g.,][]{2011PhRvD..84j4032P}. This is in line with the upper limit to the energetic of these systems in $\gamma$-rays is $E_{\rm iso}\approx10^{52}$ erg.

Episode 3 in GRFs, like in the cases of XRFs and S-GRFs, does not exhibit either a late common power-law behavior, or the nesting discovered in the rest-frame $0.3$--$10$ keV luminosity light curves of BdHNe (see Fig.~\ref{fig:DSGRBs}(b)). Also for GRFs, possible candidates for the explanation of the late X-ray afterglow emission are: a) the accelerated baryons interacting with the CBM after the P-GRB emission, or b) the possible radioactive decay of heavy elements synthesized in the ejecta of a compact binary merger \citep{Li1998}.

Episode 4 is missing in view of the absence of the SN.

Episode 5, namely the GeV emission, does not occur for NS--WD mergers. 
This fact, together to the energetics of these systems, $E_{\rm iso}<10^{52}$~erg, implies that both of these necessary and sufficient conditions for the BH formation are not fullfilled. Therefore, in a NS--WD merger, in view of the limited mass of the WD component, the NS critical mass is never reached in the accretion process during the merger.

\subsection{Prototypes}\label{sec:DSGRBs2}

In Table~\ref{tab:DSGRBs} we list all the GRFs identified so far.
\begin{table}
\centering
\begin{tabular}{lcc|lcc}
\hline\hline
GRB & $z$ & $E_{\rm iso}/$($10^{50}$ erg) &  GRB & $z$ & $E_{\rm iso}/$($10^{50}$ erg) \\
\hline
050724	&	$0.257$	&	$6.19\pm0.74$	&	061021	&	$0.3462$	&	$50\pm11$	\\
050911	&	$0.165$	&	$0.89\pm0.16$	&	061210        	&	$0.409$	&	$0.24\pm0.06$	\\
060505	&	$0.089$	&	$2.35\pm0.42$	&	070506	&	$2.31$	&	$51.3\pm5.4$	\\
060614	&	$0.125$	&	$21.7\pm8.7$ 	&	070714B	&	$0.923$	&	 $98\pm24$	\\
061006	&	$0.438$	&	$17.9\pm5.6$ 	&	071227	&	$0.381$	&	$8.0\pm1.0$	\\
\hline
\end{tabular}
\caption{List of the GRFs considered in this work up to the end of 2014. For each source (first columns) the values of $z$ and $E_{\rm iso}$ (second and third columns, respectively) are listed.}
\label{tab:DSGRBs}
\end{table}

The prototype of GRFs is GRB 060614 and has been analyzed by \citet{Caito2009}. From the analysis of its P-GRB emission a baryon load $B=2.8\times10^{-3}$ and a Lorentz factor at the transparency condition $\Gamma=346$ have been found.
From the fit of the light curve of the prompt emission it has been infereed that this burst occurred in a CBM with density $n_{CBM}=2.3\times10^{-5}$--$4.8\times10^{-3}$~cm$^{-3}$, which is typical of galactic halos where NS--NS and NS--WD mergers occur \citep[see, e.g.,][]{2014ARA&A..52...43B}.
Analogous results were obtained for the GRF 071227: $B=2.0\times10^{-4}$ and $n_{CBM}=1.0\times10^{-4}$--$1.0\times10^{-2}$~cm$^{-3}$ \citep{Caito2010}.

Further analyses on other GRF examples will be presented elsewhere.

\section{The observed rates of short and long bursts}\label{sec:rates}

The observed GRB occurrence rate is defined by the convolution of both (likely redshift-dependent) luminosity function, which describes the fraction of bursts with isotropic equivalent luminosities in the interval $\log L$ and $\log L+d \log L$, and cosmic GRB occurrence rate, which gives the number of sources at different redshifts. 
The definition of both of these functions is still an open issue and depends on \textit{a priori} assumptions and some investigations have been carried out in the literature (see, e.g., \citealp{Soderberg2006Nature,Guetta2007,Liang07,Vir09,2010tsra.confE.204R,2010MNRAS.406.1944W,2011A&A...525A..53G,Kovacevic2014}, for long bursts, \citealp{Virgili2011,2015MNRAS.448.3026W}, for short bursts, and \citealp{2015ApJ...812...33S}, for both long and short bursts). To complicate the matter, also the instrumental sensitivity threshold, the field of view $\Omega_i$, and the operational time $T_i$ of the various detectors $i$ observing GRBs introduce additional uncertainties to the problem.

In the following we ignore the possible redshift-evolution of the luminosity function. Thus, if $\Delta N_i$ events are detected by various detectors in a finite logarithmic luminosity bin from $\log L$ to $\log L+\Delta\log L$, the total local event rate density of bursts between observed minimum and maximum luminosities, $L_{\rm min}$ and $L_{\rm max}$ respectively, is defined as \citep[see][]{2015ApJ...812...33S}
\begin{equation}
\label{eq:rate}
\rho_0\simeq \sum_{i}\sum_{\log L_{\rm min}}^{\log L_{\rm max}}\frac{4\pi}{\Omega_i T_i}\frac{1}{\ln10}\frac{1}{g(L)}\frac{\Delta N_i}{\Delta\log L}\frac{\Delta L}{L}\ ,
\end{equation}
where
\begin{equation}
\label{eq:rate2}
g(L)=\int_{0}^{z_{\rm max}(L)}\frac{f(z)}{1+z}\frac{dV(z)}{dz}dz\ ,
\end{equation}
and the comoving volume is given by
\begin{equation}
\frac{dV(z)}{dz}=\frac{c}{H_0}\frac{4\pi d_L^2}{(1+z)^2[\Omega_M(1+z)^3+\Omega_{\Lambda}]^{1/2}}\ ,
\end{equation}
where $d_L$ is the luminosity distance. The dimensionless function $f(z)$ describes the GRB cosmic redshift-dependent event rate density. In the following we assume no redshift dependency, therefore we set $f(z)=1$.
The maximum redshift $ z_{\rm max}(L)$ in Eq.~(\ref{eq:rate2}) defines the maximum volume inside which an event with luminosity $L$ can be detected. This redshift can be computed from the $1$~s-bolometric peak luminosity $L$, $k$-corrected from the observed detector energy band into the burst cosmological rest-frame energy band $1$--$10^4$~keV \citep{Schaefer2007}, and the corresponding $1$~s-threshold peak flux $f_{\rm th}$, which is the limiting peak flux to allow the burst detection \citep[see][for details]{Band2003}. Therefore, $z_{\rm max}$ can be defined via \citep[see, e.g.,][]{Zhang2,2014A&A...569A..39R}
\begin{equation}
f_{\rm th} = \frac{L}{4\pi d_{L}^{2} (z_{\rm max})k},
\end{equation}
where we duly account for the $k$-correction.

Within the assumptions that the GRB luminosity function does not evolve with redshift and that $f(z)=1$, we investigate the evolution with the redshift of the GRB rates by separating the bursts into several redshift bins. As suggested in \citet{2015ApJ...812...33S}, this can be done in each redshift interval $z_j\leq z\leq z_{j+1}$ by changing the integration limits of Eq.~(\ref{eq:rate2}) into $z_j$ and min$[z_{j+1}, z_{\rm max,j}(L)]$, where $z_{\rm max,j}(L)$ is the maximum redshift for the j$^{th}$ redshift bin. Finally, from Eq.~(\ref{eq:rate}) we derive an event rate $\rho_{\rm 0}^{\rm z}$ in each redshift bin around $z$.

In the following we adopt the following fields of view and operational times for various detectors: \emph{Beppo}-SAX, $\Omega_{\rm BS}=0.25$~sr, $T_{\rm BS}=7$~y; BATSE, $\Omega_{\rm B}=\pi$~sr, $T_{\rm B}=10$~y, HETE-2, $\Omega_{\rm H}=0.8$~sr, $T_{\rm H}=7$~y; \textit{Swift}-BAT, $\Omega_{\rm S}=1.33$~sr, $T_{\rm S}=10$~y; \textit{Fermi}-GBM, $\Omega_{\rm F}=9.6$~sr, $T_{\rm F}=7$~y. We assume no beaming correction in computing the rates of the GRB sub-classes.

\subsection{Rate of S-GRFs}\label{sec:SGRFs3}

The local rate of S-GRFs, obtained from the sample of sources listed in Table~\ref{tab:S-GRFs}, is $\rho_0=3.6^{+1.4}_{-1.0}$~Gpc$^{-3}$~y$^{-1}$ and it is in agreement with the estimates obtained from the whole short burst population detected by the \textit{Swift}-BAT detector (and, therefore, including also S-GRBs and GRFs) and reported in the literature \citep[$1$--$10$~Gpc$^{-3}$~y$^{-1}$, see, e.g.,][and references therein]{2015ApJ...809...53C}. In particular our local rates with $f(z)=1$ agrees with recent more precise estimates: a) $4.1^{+2.3}_{-1.9}$~Gpc$^{-3}$~y$^{-1}$ for $L_{\rm min}=5\times10^{49}$~erg/s and for $f(z)$ described by a power law merger delay model \citep{2015MNRAS.448.3026W}; b) $4.2^{+1.3}_{-1.0}$, $3.9^{+1.2}_{-0.9}$, and $7.1^{+2.2}_{-1.7}$~Gpc$^{-3}$~y$^{-1}$ for $L_{\rm min}=7\times10^{49}$~erg/s and $f(z)$ described as Gaussian, log-normal and power law merger delay models, respectively \citep{2015ApJ...812...33S}.

The evolution of the S-GRF rate in various redshift bins is shown in Fig.~\ref{fig:ratesfam}~(c).
This rate decreases as a power law from the local value in the interval $0.1\leq z\leq0.4$ to a value of $0.042^{+0.046}_{-0.025}$Gpc$^{-3}$ y$^{-1}$ in the interval $1.0\leq z\leq2.7$. 
Also in the case of S-GRFs the increasing sampled comoving Universe volume and the threshold of the detectors play a fundamental role in the observed decrease of their rate at larger distances.

\subsection{Rate of S-GRBs}\label{sec:SGRBs3}

Previously we have identified and described four S-GRBs in \citet{2015ApJ...808..190R}, e.g., GRB 081024B, GRB 090227B, GRB 090510, and GRB 140619B. 
Here we present two additional new members of this class: GRB 060801 (at $z\approx1.13$ and with $z_{max}\approx2.04$, in this work) and GRB 140402A (at $z\approx5.52$ and with $z_{max}\approx7.16$, Ruffini et al. in preparation). 
From these six S-GRBs detected by the \textit{Fermi} and the \textit{Swift} satellites, we obtain via Eqs.~(\ref{eq:rate})--(\ref{eq:rate2}) a local rate $\rho_0=\left(1.9^{+1.8}_{-1.1}\right)\times10^{-3}$~Gpc$^{-3}$~y$^{-1}$.

With only six sources, we could not build the evolution with the redshift of such systems.

\subsection{Rate of XRFs}\label{sec:XRFs3}

In \citet{Kovacevic2014}, we have estimated an updated observed rate for the XRFs at $z<0.1$ based on the method outlined in \citet{Soderberg2006Nature} and \citet{Guetta2007}. 
In this work, we consider the complete list of XRFs shown in Table~\ref{tab:XRFs} and the method outlined in \citet{2015ApJ...812...33S}.
From Eq.~(\ref{eq:rate})--(\ref{eq:rate2}) the local rate of XRFs is $\rho_0=100^{+45}_{-34} $Gpc$^{-3}$ y$^{-1}$, where the attached errors are determined from the $95\%$ confidence level of the Poisson statistic \citep{Gehrels1986}. Within the extent of our different classification criteria and different choices for $f(z)$, our estimate is in agreement with those reported for low-luminous long GRBs in \citet{Liang07} and \citet{Vir09}, and in particular with the value of $164_{-65}^{+98}$Gpc$^{-3}$ y$^{-1}$, obtained by \citet{2015ApJ...812...33S} with the same method.

In the IGC scenario the XRF out-states are NS--NS binary systems.
For a reasonable sets of binary initial conditions, populations synthesis simulations performed by \citet{1999ApJ...526..152F} provide a NS--NS formation rate $(0.2$--$1600)$~Gpc$^{-3}$ y$^{-1}$. 
NS--NS formation rate accounts for other possible channels in the population synthesis models, in addition to the one we considered from the XRFs.
It is interesting, nevertheless, that our predicted rate is consistent with that obtained by \citet{1999ApJ...526..152F} obtained.

For the same above reason, our rate of XFRs can be also compared with the NS--NS merger rate proposed by \citet{Eichler1989}. 
In this historical paper, the NS--NS merger rate is derived from the strong assumption that each merger ejects always the same amount of material r-process classified and the heavy r-process material. \citet{Eichler1989} obtain a rough estimate of $(140$--$14000)$~Gpc$^{-3}$ y$^{-1}$, which is marginally consistent to the upper value of the local XRF rate.

The evolution of the XRF rate in various redshift bins is shown in Fig.~\ref{fig:ratesfam}~(a). It decreases from a value of $95^{+123}_{-63}$Gpc$^{-3}$ y$^{-1}$ in the interval $0\leq z\leq0.1$ to a value of $0.8^{+1.1}_{-0.5}$Gpc$^{-3}$ y$^{-1}$ in the interval $0.7\leq z\leq1.1$. This effect is mainly due to the intrinsic low luminosities of the bulk of the XRF population ($10^{46}$--$10^{48}$~erg/s, see, e.g., \citealp{2011ApJ...739L..55B}) and to the threshold of the detectors: at increasing sampled Universe comoving volumes, these low luminous XRFs become undetectable, therefore at higher redshifts the total XRF rate decreases.

\subsection{Rate of BdHNe}\label{sec:BdHNe3}

We proceed now in estimating the rate of BdHNe from the total sample of $233$ sources (see Table~\ref{tab:BdHNe}). 
From Eq.~(\ref{eq:rate})--(\ref{eq:rate2}) the local rate of BdHNe is $\rho_0=0.77^{+0.09}_{-0.08}$~Gpc$^{-3}$~y$^{-1}$. 
Our estimate is in agreement with two recent estimates obtained from long bursts with $L\geq10^{50}$~erg/s and by assuming $f(z)\neq1$: a) the value of $1.3^{+0.6}_{-0.7}$~Gpc$^{-3}$~y$^{-1}$ obtained by \citet{2010MNRAS.406.1944W}, even though limited to the \textit{Swift} long bursts and including long some bursts with $E_{\rm iso}<10^{52}$~erg, obtained from a GRB inferred cosmic rate independent on the star formation rate; b) the value of $0.8^{+0.1}_{-0.1}$~Gpc$^{-3}$~y$^{-1}$ obtained by \citet{2015ApJ...812...33S} with the same method and including the star formation rate dependence.

In the IGC scenario the BdHNe out-states are NS--BH binary systems.
Following again the work by \citet{1999ApJ...526..152F}, populations synthesis simulations (which accounts also for alternative scenarios to that of the IGC model) provide a NS--BH formation rate of $(0.02$--$1000)$~Gpc$^{-3}$ y$^{-1}$.
Also in this case, even though a straightforward comparison is not possible, the BdHNe rate is consistent with the NS–BH formation rate obtained by \citet{1999ApJ...526..152F}.

The evolution of the BdHN rate in various redshift bins is shown in Fig.~\ref{fig:ratesfam}~(b). 
It slightly decreases from the local value in the interval $0.1\leq z\leq0.4$ to a value of $0.17^{+0.05}_{-0.04}$Gpc$^{-3}$ y$^{-1}$ in the interval $3.6\leq z\leq9.3$. 
As stated for the case of XRFs, this effect occurs because for increasing sampled Universe comoving volumes, only the most luminous BdHNe are detectable, even though in a less marked way than the case of the XRFs.

\subsection{Rate of U-GRBs}\label{sec:UGRBsrate}

As pointed out in Section~\ref{sec:rate5}, nearly 100$\%$ of the BdHNe leads to bound NS-BH binaries, which are the progenitor systems of U-GRBs \citep{2015arXiv150502809F}.
If we include the possibility of other channels of formation for these NS-BH binaries, we can safely assume the BdHN local rate as a lower limit for these U-GRBs, e.g., $\rho_0=0.77^{+0.09}_{-0.08}$~Gpc$^{-3}$~y$^{-1}$.
From this consideration, it appears that the U-GRBs have the second higher rate among the short bursts after the S-GRFs.

\subsection{Rate of GRFs}\label{sec:DSGRB3}

We proceed now in estimating the rate of GRFs from the total sample of $10$ sources (see Table~\ref{tab:DSGRBs}). 
From Eq.~(\ref{eq:rate})--(\ref{eq:rate2}) we obtain a local rate $\rho_0=1.02^{+0.71}_{-0.46}$~Gpc$^{-3}$~y$^{-1}$, and represent the first estimate for these kind for bursts originating from NS--WD mergers.

\begin{figure*}
\centering
(a)\includegraphics[width=0.45\hsize,clip]{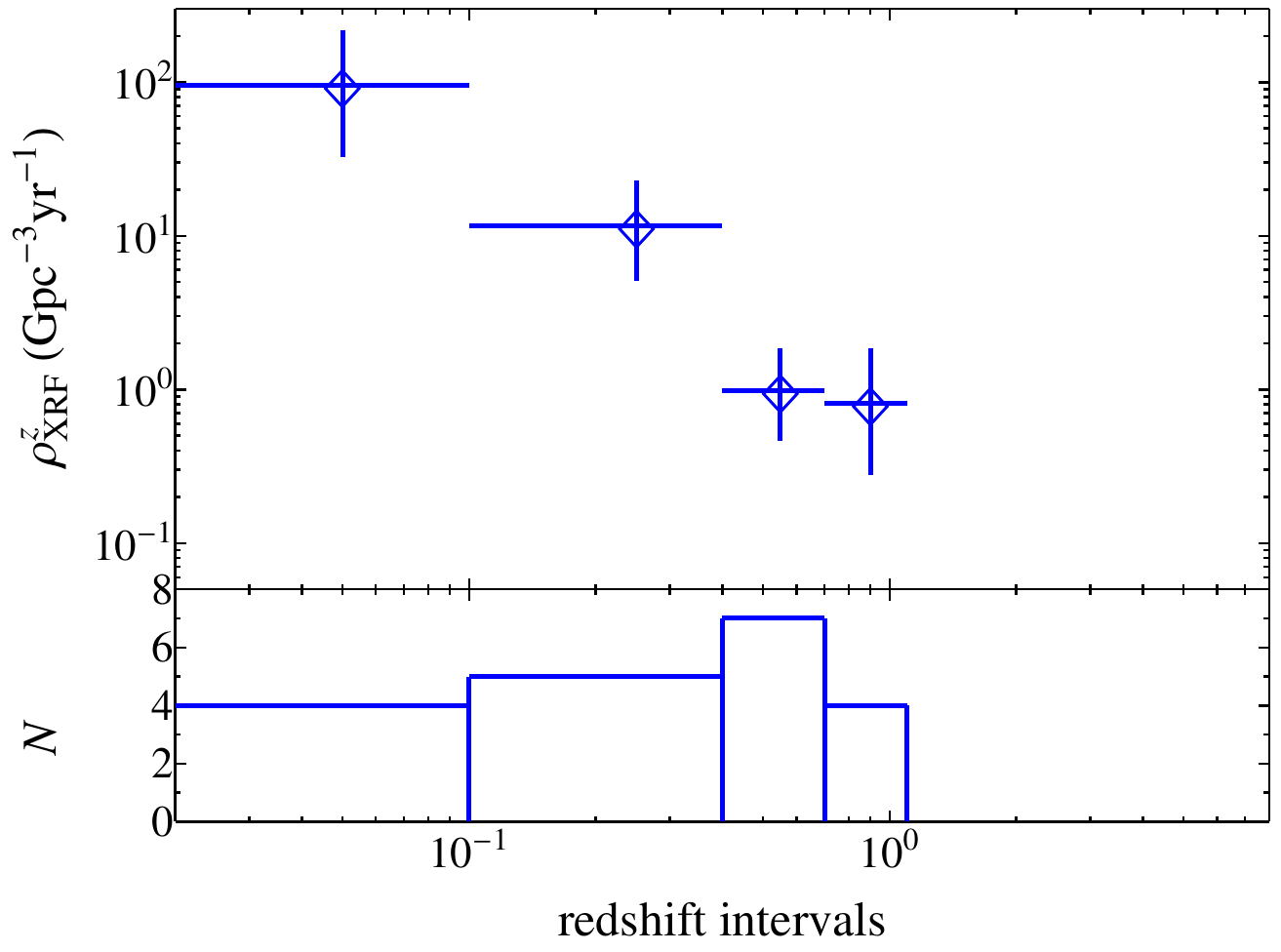}
(b)\includegraphics[width=0.45\hsize,clip]{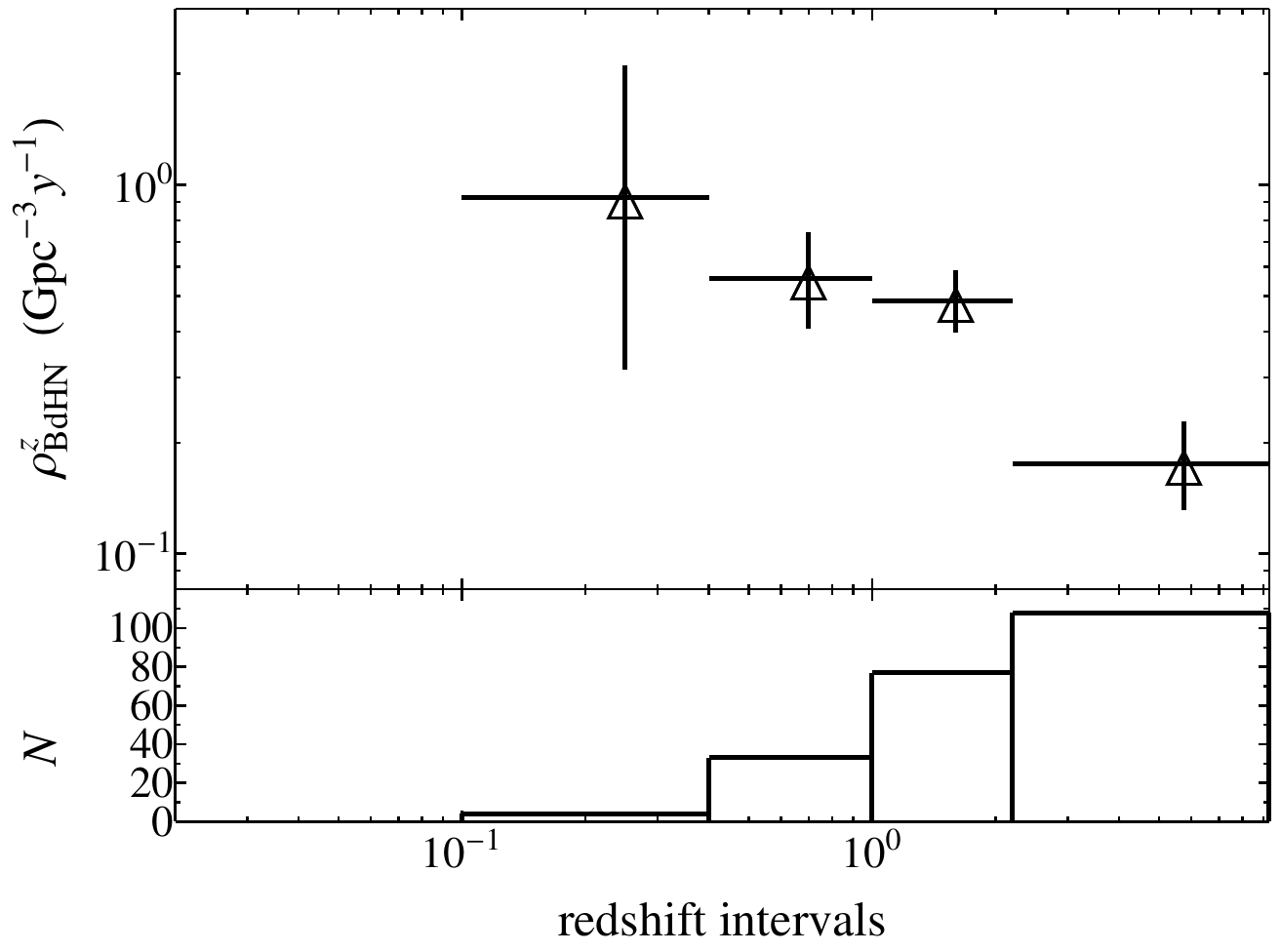}
(c)\includegraphics[width=0.45\hsize,clip]{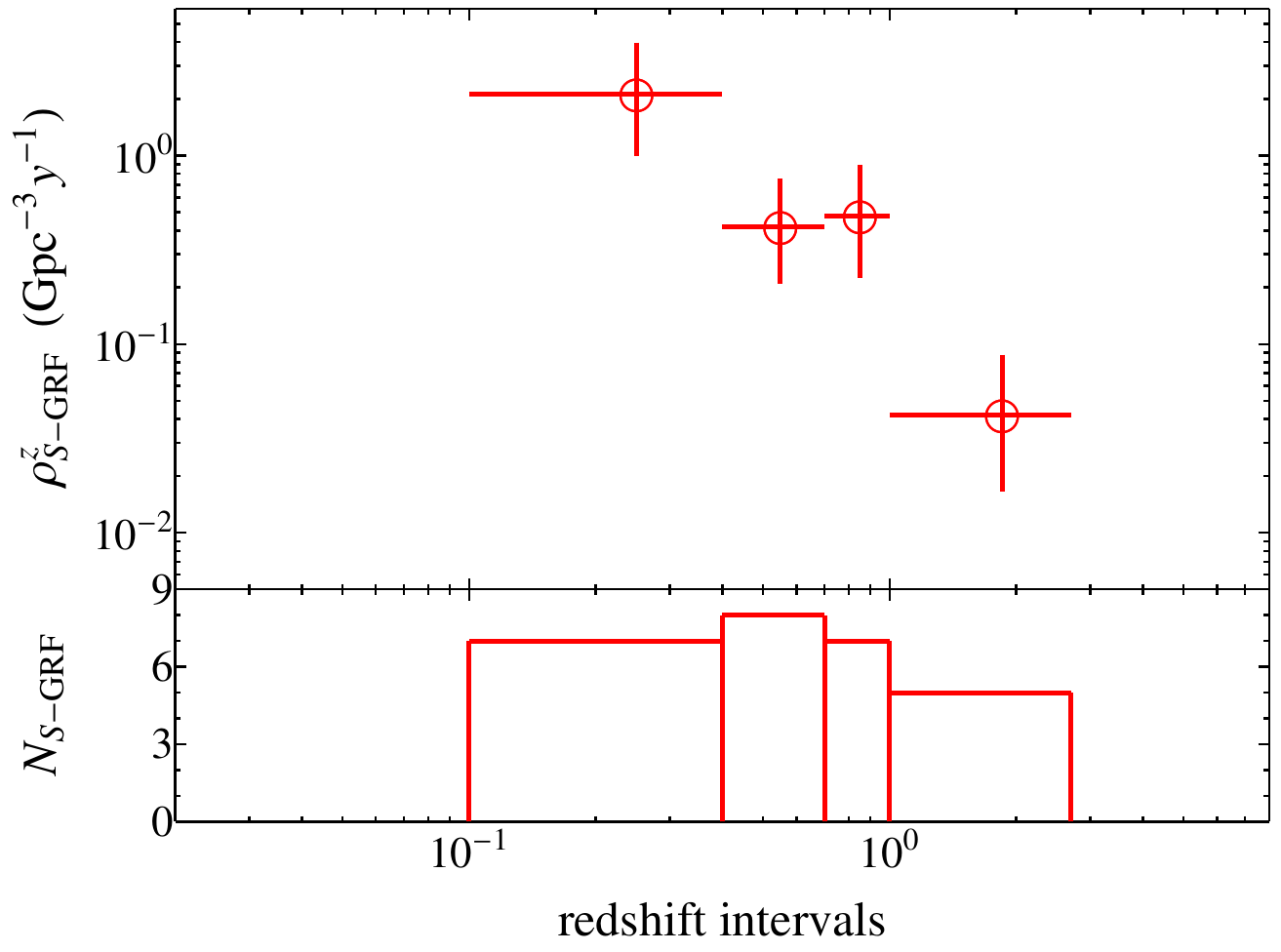}
(d)\includegraphics[width=0.45\hsize,clip]{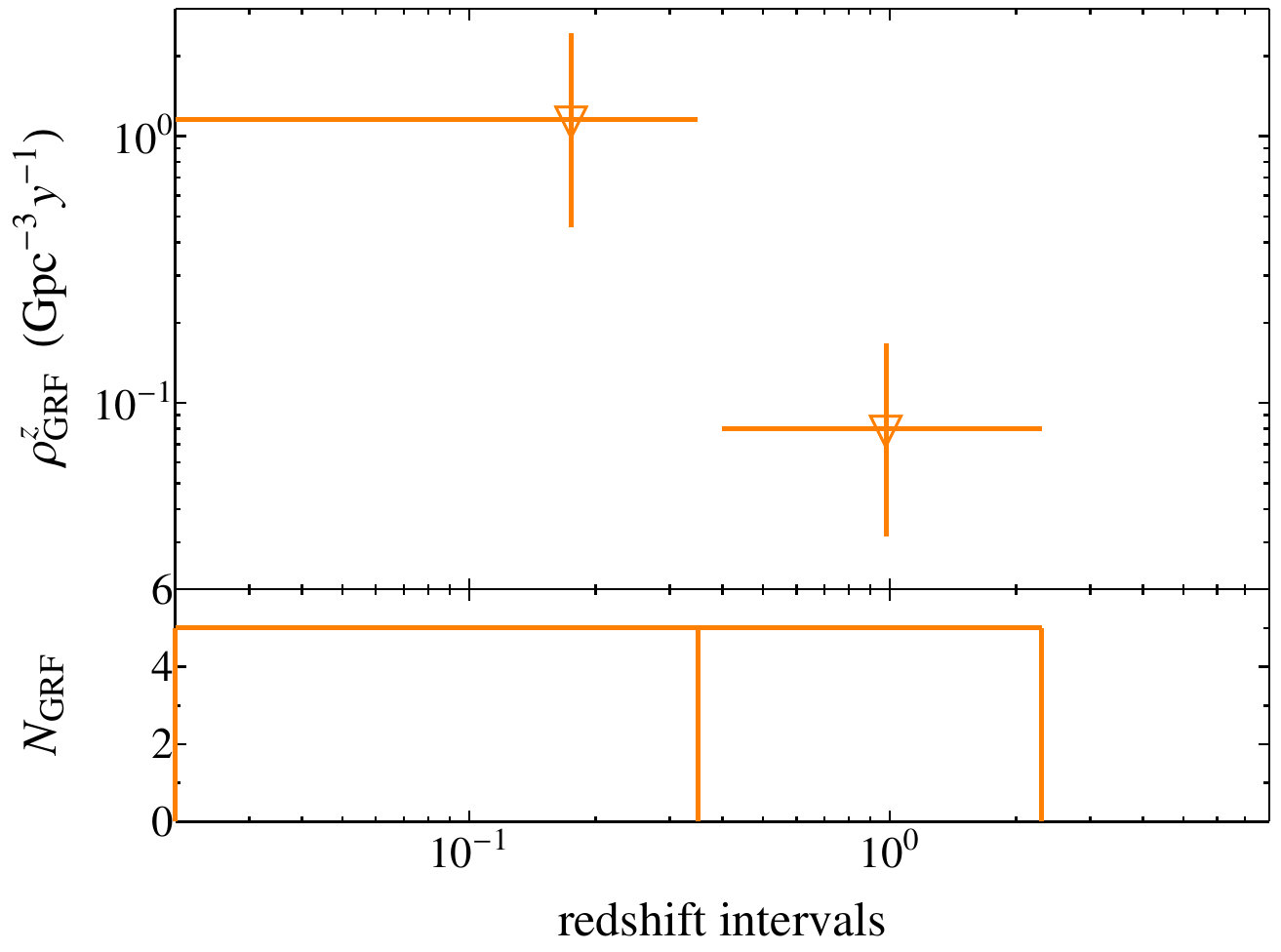}
\caption{The evolution of the rate with the redshift for the considered GRB sub-classes: (a) the XRF, (b) the BdHNe, (c) the S-GRFs, (d) the GRFs. In each plot the upper panel shows the evolution rate with the redshift, while the lower panel displays the number of observed sources in each redshift bin. Because of the limited amount of sources, for S-GRBs no redshift bin evolution is shown. In the case of U-GRBs, there are no current detections.}
\label{fig:ratesfam}
\end{figure*}

Due to the limited number of sources in our sample, we limited the study of the GRF rate evolution in two redshift bins, as shown in Fig.~\ref{fig:ratesfam}~(d). 
The rate starts from a value consistent with the above local rate, in the redshift interval $0\leq z\leq0.35$, to a value of $0.080^{+0.088}_{-0.048}$Gpc$^{-3}$ y$^{-1}$, in the interval $0.35\leq z\leq2.31$. 
Also for GRFs, the cutoff in the rate at higher redshift occurs because for increasing sampled Universe comoving volumes only the most luminous sources are detectable. However this effect, as in the case of S-GRFs, is more pronounced due to their intrinsically weaker luminosities, when compared to those of S-GRBs.

\section{Conclusions}\label{sec:conclusions}

The remarkable progress reached in understanding the GRB phenomenon has been made possible by the outstanding spectral and temporal information acquired from X-rays, $\gamma$-rays and high energy observatories, as well as from optical and radio data obtained by telescopes all over the planet. 
At the same time this result has been fostered by a novel deeper theoretical understanding in the physics and astrophysics of WD \citep[see, e.g.,][]{2013ApJ...762..117B}, NS \citep[see, e.g.,][]{Belvedere2014,2015PhRvD..92b3007C} and BH \citep[see, e.g.,][]{2010PhR...487....1R}. Consequently the understanding of the GRB phenomenon has evolved from an elementary paradigm based on a single jetted emission process as postulated in the fireball model \citep[see, e.g.,][and reference therein]{Sari1998b,Piran2005,Meszaros2006,Gehrels2009} to an authentic astrophysical laboratory involving many-body interactions between different astrophysical systems encountering previously unexplored regimes and observational evidence.

In the Introduction we review the increasing number of GRB observations which have led likewise to the theoretical progress in the understanding of the GRB phenomena. 
While the role of NS--NS (or NS-BH) binaries as ``in-states'' of short GRBs has been widely accepted and confirmed by strong observational and theoretical evidence \citep[see, e.g.,][]{Goodman1986,Paczynski1986,Eichler1989,Narayan1991,Narayan1992,MeszarosRees1997_b,Rosswog2003,Lee2004,2014ARA&A..52...43B,2015ApJ...808..190R}, the identification of the progenitor systems for long GRBs followed a more difficult path. 
Initially, theoretical models based on a single progenitor were proposed: a \textit{collapsar} \citep{Woosley1993}, or a \textit{magnetar} \citep[see,e .g.,][]{2001ApJ...552L..35Z}.
Then, the role of binary progenitor systems composed of two very massive stars for long GRBs was recognized by \citep{1999ApJ...526..152F}, where several different scenarios were there envisaged leading to a \textit{collapsar} \citep{Woosley1993}, as well as a few leading, alternatively, to a variety of binary compact systems.
These considerations were addressed by our group in a set of papers assuming that the birth of a SN and the occurrence of a GRB were qualitatively and quantitatively different astrophysical events in space and time. This led to the necessity of introducing the IGC paradigm \citep[see, e.g.,][]{Ruffini2001c,2006tmgm.meet..369R,Ruffini2007b,2008mgm..conf..368R,2012A&A...548L...5I,2012ApJ...758L...7R,2014ApJ...793L..36F,2015ApJ...798...10R}.
In the IGC paradigm the long GRB-SN coincidence originates from CO$_{\rm core}$--NS binary progenitors system.
This approach differs from alternative descriptions, e.g., the \textit{magnetars} and the \textit{collapsar} models, where the two events are coming from a single progenitor star.

In Section~\ref{sec:fireshell}, we review the fireshell model for GRBs \citep[see, e.g.][]{Ruffini2001c,Ruffini2001,Ruffini2001a} and its general description which can be applied to any source of an optically thick baryon-loaded $e^+e^-$ plasma, i.e., in the quantum-elctrodynamical process expected in the formation of a BH \citep[see, e.g.,][]{Preparata,RSWX2,RSWX,Cherubini,RRKerr}, as well as in the case of a pair plasma created via $\nu \bar{\nu}\leftrightarrow e^+e^-$ mechanism in a NS--NS merger \citep{Narayan1992,SalmonsonWilson2002,Rosswog2003}, or in the hyper-accretion disks around BHs \citep{Woosley1993,2011MNRAS.410.2302Z}.

In Section~\ref{sec:1052erg}, we discuss the role of the $10^{52}$~erg energy critical value introduced to discriminate between binary systems leading to the formation of a MNS (XRFs, S-GRFs and GRFs), with energy lower than the above critical value, and those leading to the formation of a BH (BdHNe, S-GRBs and U-GRBs), with energy larger than the above critical value. The value of $10^{52}$~erg is derived by considering the hypercritical accretion process onto a NS leading to an energy release in form of neutrinos and photons, given by the gain of gravitational potential energy of the matter accreted in the NS. This includes the change of binding energy of the NS while accreting both matter and angular momentum \citep{2016arXiv160602523B}. A typical NS mass of $\approx1.4$~M$_\odot$ has been assumed, as observed in galactic NS binaries \citep{2011A&A...527A..83Z,2014arXiv1407.3404A}. A NS critical mass in the range of $2.2$~M$_\odot$ up to $3.4~M_\odot$ depending on the equations of state and angular momentum \citep[see][for details]{2016arXiv160602523B,2015ApJ...812..100B,2015PhRvD..92b3007C} has been assumed.

\begin{figure*}
\centering
\includegraphics[width=0.8\hsize,clip]{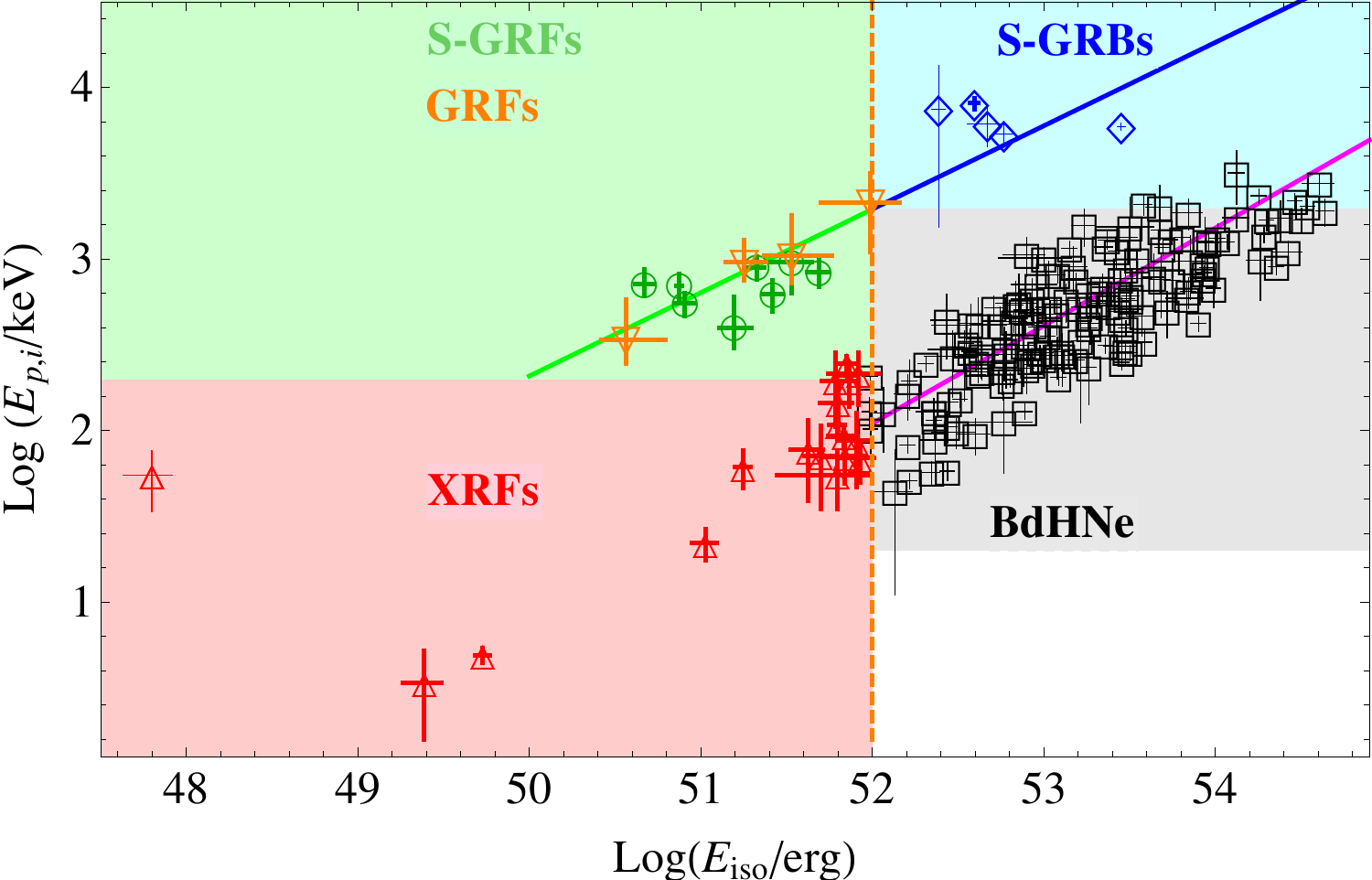}
\caption{The $E_{\rm p,i}$--$E_{\rm iso}$ plane for XRFs, BdHNe, S-GRBs, S-GRFs, and the initial spike-like emission of the GRFs. The XRFs (red triangles) cluster in the red shaded region ($E_{\rm p,i}\lesssim200$~keV and $E_{\rm iso}\lesssim 10^{52}$~erg), while the BdHNe (black squares) in the gray shaded one ($E_{\rm p,i}\gtrsim200$~keV and $E_{\rm iso}\gtrsim 10^{52}$~erg); the Amati relation \citep{2013IJMPD..2230028A} fulfilled by the BdHNe is plotted with a magenta solid line. The S-GRFs (green circles) and the initial spike-like emission of the GRFs (orange reverse triangles) cluster in the green shaded region ($E_{\rm p,i}\lesssim2$~MeV and $E_{\rm iso}\lesssim 10^{52}$~erg), while the S-GRBs (blue diamonds) in the blue shaded one ($E_{\rm p,i}\gtrsim2$~MeV and $E_{\rm iso}\gtrsim 10^{52}$~erg); the relation for short bursts \citep{2015ApJ...808..190R,Calderone2014,2012ApJ...750...88Z} is plotted with a green solid line for the S-GRFs and the GRFs, and in blue for the S-GRBs.}
\label{fig:ffff}
\end{figure*}

In Section~\ref{sec_3_prot_rat}, we describe the properties of XRFs (see Fig.~\ref{fig:ffff}). In these systems the distance between the CO$_{\rm core}$ and the NS companion is $a>10^{11}$~cm. The hypercritical accretion process is not sufficient to push the NS beyond its $M_{crit}$ and an MNS is formed \citep[see, e.g.,][]{2015ApJ...812..100B,2016arXiv160602523B}. In Table~\ref{tab:XRFs} we list the XRFs considered in this work, as well as the spectral, temporal and luminosity analysis of selected prototype, e.g., GRB 060218. The complete theoretical simulation of this prototype is presented in \citet{2016arXiv160602523B}.

In Section~\ref{sec:descr_BdHNe} we consider the BdHNe, for which the binary separation between the CO$_{\rm core}$ and the NS binary companion is $a<10^{11}$~cm and the hypercritical accretion process trigger the gravitational collapse of the NS into a BH \citep[see, e.g.,][]{2015ApJ...812..100B,2016arXiv160602523B}. We show here an updated list of BdHNe (see Table~\ref{tab:BdHNe}), as well as a diagram summarizing some of the key properties and prototypes (see Fig.~\ref{fig:ffff}), analyzed within the IGC paradigm and the fireshell model (see, e.g., GRB 090618 and GRB 130427A). 

In Section~\ref{sec:descr_S-GRFs}, we outline the properties of S-GRFs listed in Table~\ref{tab:S-GRFs} and shown in Fig.~\ref{fig:ffff}. These systems coincide with the short bursts considered in \citet{2014ARA&A..52...43B}. They originate in NS-NS mergers leading to the formation of a MNS and possibly a binary companion, in order to fulfill the conservation of energy and momentum \citep{2015ApJ...808..190R}. 

In Section~\ref{sec:descr_SGRBs} we present S-GRBs originating in NS-NS mergers leading to the formation of a BH (see Fig.~\ref{fig:ffff}).
We give, in Table~\ref{tab:SGRBs}, their updated list. We then describe their prototypes, analyzed within the fireshell model (see, e.g., GRB 090227B and GRB 140619B), and outline the key role of the P-GRB identification for their description, as well as the analysis of the GeV emission.

In Section~\ref{sec:rate5}, motivated by the results obtained by \citet{2015arXiv150502809F}, where it was shown that nearly 100$\%$ of the NS-BH binaries, namely the out-states of the BdHNe, remain bound, we add the description of this not yet observed, but theoretically predicted sub-class of U-GRBs, unaccounted for in current standard population synthesis analyses.

In Section~\ref{sec:DS}, we review the properties of the GRFs listed in Table~\ref{tab:DSGRBs} and shown in Fig.~\ref{fig:ffff}. 
We recall and describe the results obtained from the sources analyzed within the fireshell model (see, e.g., GRB 060614, \citealp{Caito2009} and GRB 071227, \citealp{Caito2010}).

The most important result of the present article is the estimate the rates of occurrence of the XRFs, BdHNe, S-GRFs, S-GRBs, U-GRBs, and GRFs sub-classes. 
In Section~\ref{sec:rates}, we introduce the procedure outlined in \citet{2015ApJ...812...33S} for estimating the local rates and their evolution with the redshift of the above sub-classes of long and short bursts, assuming no beaming (note: the recent observation of the absence of GeV emission associated to a BdHN may limit this assumption).
By ignoring possible redshift-evolution of the GRB sub-classes luminosity functions and assuming that the GRB cosmic event rate density is redshift-independent (e.g., $f(z)=1$), the above method duly takes into account observational constraints, i.e., the detector solid angle coverage of the sky $\Omega$ and sensitivities which in turn define a maximum volume of observation depending on the intrinsic luminosity of the sources (see Section~\ref{sec:rates} and  \citealp{Soderberg2006Nature,Guetta2007,Liang07,Vir09,Virgili2011,2010tsra.confE.204R,2010MNRAS.406.1944W,2015MNRAS.448.3026W,Kovacevic2014,2015ApJ...812...33S} for details).
We obtain:
\begin{itemize}
\item[-]{an S-GRF local rate of $\rho_0=3.6^{+1.4}_{-1.0}$~Gpc$^{-3}$~y$^{-1}$ (see Section~\ref{sec:SGRFs3});}
\item[-]{an S-GRB local rate of $\rho_0 =\left(1.9^{+1.8}_{-1.1}\right)\times10^{-3}$~Gpc$^{-3}$~y$^{-1}$ (see Section~\ref{sec:SGRBs3});}
\item[-]{an XRF local rate of $\rho_0=100^{+45}_{-34} $Gpc$^{-3}$ y$^{-1}$ (see Section~\ref{sec:XRFs3});}
\item[-]{a BdHN local rate of  $\rho_0=0.77^{+0.09}_{-0.08}$~Gpc$^{-3}$~y$^{-1}$ (see Section~\ref{sec:BdHNe3}; for the above reason this rate coincides with that of the U-GRBs, see Section~\ref{sec:UGRBsrate});}
\item[-]{a GRF local rate of $\rho_0=1.02^{+0.71}_{-0.46}$~Gpc$^{-3}$~y$^{-1}$ (see Section~\ref{sec:DSGRB3}).}
\end{itemize}
The local rates of S-GRFs, XRFs, and BdHNe, are in agreement, within the extent of the different classification criteria, with those reported in the literature.
The local rates of S-GRBs and GRFs are, instead, new ones following from the classification proposed in this work.
The evolution with the redshift of the rates of XRFs, BdHNe, S-GRFs, and GRFs is shown in Fig.~\ref{fig:ratesfam}.
It is certainly of interest to compare and contrast these results obtained from the direct observations of the sources in our new classification with the results computed from population synthesis models. Any possible disagreement will give the opportunity to identify possible missing links in the evolutionary phases within population synthesis analysis.

We are now in a position to apply the above rates of S-GRFs, S-GRBs and U-GRBs to assess the detectability and the expected number of gravitational wave detections by LIGO from NS-NS and NS-BH binaries (Ruffini et al., in preparation). We are also ready to apply the above BdHN rate to give an estimate of the contribution of GRBs to cosmic rays (Ruffini et al., in preparation).

\begin{figure*}[!hbtp]
\centering
\includegraphics[width=0.7\hsize,clip]{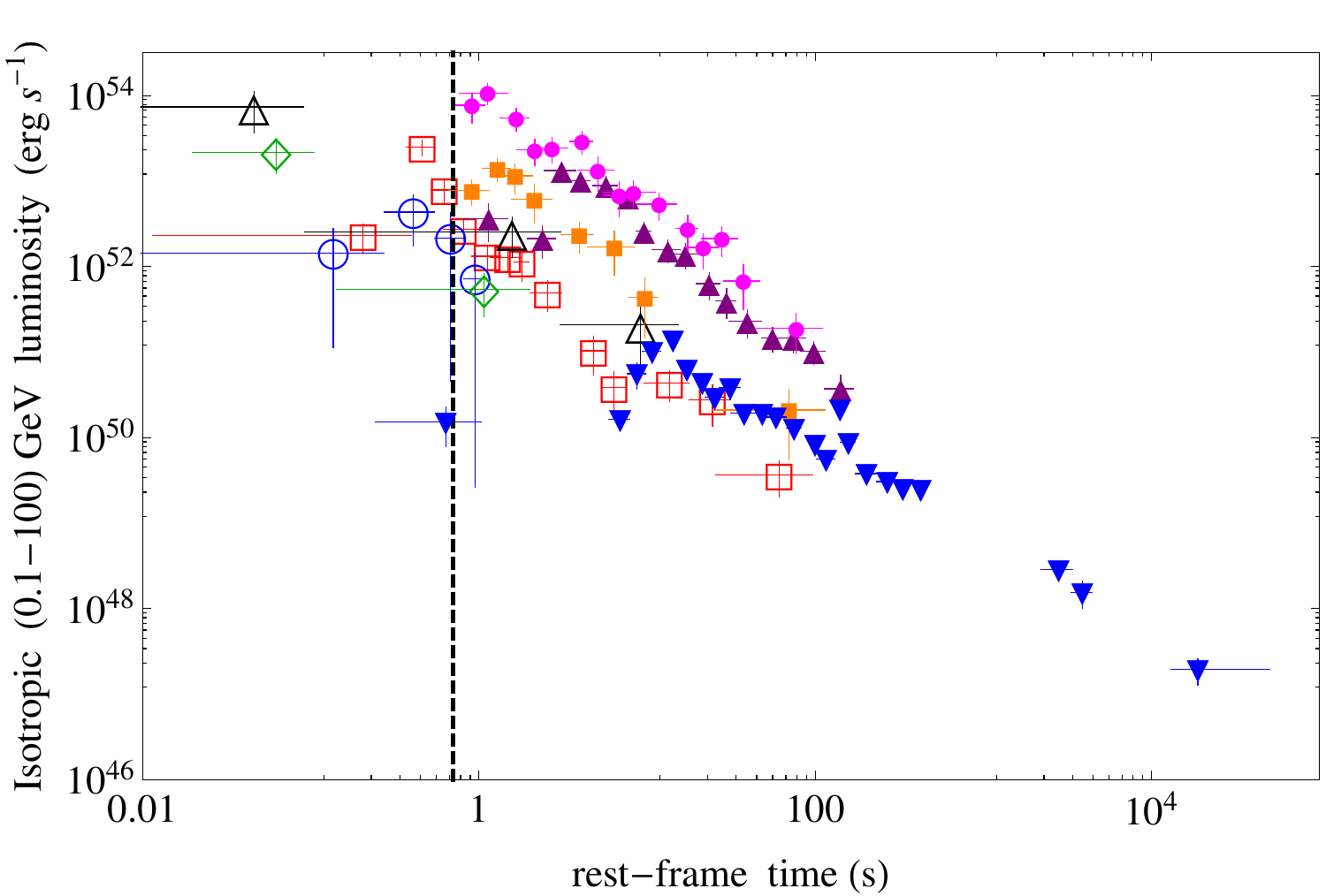}
\caption{The rest-frame $0.1$--$100$ GeV isotropic luminosity light curves of selected BdHNe (filled symbols) and all S-GRBs with available LAT data (empty symbols). BdHNe: GRB 080916C (magenta circles), GRB 090902B (purple triangles), GRB 110731A (orange squares), GRB 130427A (blue reversed triangles). S-GRBs: GRB 081024B (green diamonds), GRB 090510 (red squares), GRB 140402A (black triangles), GRB 140619B (blue circles). Also in this case the dashed vertical line marks the minimal turn-on time of the GeV emission of BdHNe.}
\label{gevtotal}
\end{figure*}

Before concluding, in support of the classification proposed in this article, we recall that the luminosity light curves of the GeV emission is uniquely observed in both BdHNe and S-GRBs. In both cases it follow a precise power-law behavior with time $\propto t^{-1.2}$ (see \citealt{Nava2014}, \citealt{2016arXiv160702400E} and Fig.~\ref{gevtotal}). 
An outstanding conclusion of this paper is that in both BdHNe and S-GRBs, where the presence of the BH is predicted, the turn-on of this GeV emission occurs after the P-GRB emission and at the beginning of the prompt emission phase (see Figs.~\ref{rad_ind_tot}~(d) and \ref{ShortXO}~(d)).
This commonality, in such different systems, as well as their energy requirements (see Tab.~\ref{tab:LAT} and Fig.~\ref{gevtotal}) are naturally explained if we assume, as indicated in \citet{2015ApJ...798...10R,2015ApJ...808..190R}, that this GeV emission originates by accretion processes in the newly-born BH.
We have pointed out in \citet{2016arXiv160702400E} how the total energy of the GeV emission can be expressed in term of the gravitational binding energy of matter accretion into Kerr BHs (see Ruffini \& Wheeler 1969, in problem 2 of $\S$ 104 in \citealt{LL2003}). This energetics requirement could not be fulfilled in the case of accretion onto a NS, in view of the much smaller value of the gravitational binding energy when compared to the case of a rotating BH \citep[see e.g.][]{2000AstL...26..772S}. On the general issue of the origin of the jetted GeV emission, and not just of its energetics, we refer to the last paragraph of the conclusions of the paper by \citet{2016arXiv160702400E}.

We have added a Table.~\ref{tab:LAT} with the values of the GeV emission for both the case of S-GRBs and BdHNe. These energy releases up to $\approx10^{54}$ erg can be explained by the occurrence of accretion onto a rotating BH with mass in the range of $3$--$10$~M$_\odot$.
It is also clear from Fig.~\ref{gevtotal} that S-GRBs and BdHNe have GeV emission sharing a common luminosity pattern and originating, in both cases, from a newly-born Kerr BH \citep{2015ApJ...798...10R,2015ApJ...808..190R}.
This picture includes also the first scenario of an IGC considered in \citet{Ruffini2001c} where an exploding $CO_{\rm core}$ is in a close binary system with an already formed BH companion. In view of the hypercritical accretion process of the SN ejecta onto an already formed BH, these systems have $E_{\rm iso}\gtrsim10^{54}$ erg and $E_{\rm p,i}\gtrsim2$ MeV. Their out-states are a binary composed of a more massive BH and a $\nu$NS. Such systems, which we refer to as BH-SNe, are expected to be the late evolutionary stages of X-ray binaries such as Cyg X-1 or Cyg X-3 \citep[see, e.g.,][]{1978pans.proc.....G}.
\begin{table}
\centering
\begin{tabular}{lcccc}
\hline\hline
source       &  z                          &  $E_{p,i}$          &    $E_{iso}$              &   $E_{LAT}$           \\
               &                              &  (MeV)               &    ($10^{52}$ erg)    &    ($10^{52}$ erg) \\
\hline
\multicolumn{5}{c}{S-GRBs}\\
\hline 
081024B  &  $2.6\pm1.6$          &  $8.7\pm4.9$      &  $2.44\pm0.22$      &  $2.70\pm0.93$      \\
090510    &  $0.903$                &  $7.89\pm0.76$  &  $3.95\pm0.21$     &  $5.78\pm0.60$    \\
140402A  &  $5.52\pm0.93$       &  $6.1\pm1.6$     &  $4.7\pm1.1$         &  $16.6\pm5.3$    \\
140619B  &  $2.67\pm0.37$       &  $5.34\pm0.79$  &  $6.03\pm0.79$      &   $2.34\pm0.91$      \\
\hline
\multicolumn{5}{c}{BdHNe}\\
\hline
080916C  &  $4.35$                  &  $2.76\pm0.37$   &  $407\pm86$          &  $440\pm47$    \\
090902B  &  $1.822$                 &  $2.19\pm0.22$  &  $292\pm29$          &  $110\pm5$    \\
110731A  &  $2.83$                  &  $1.16\pm0.12$   &  $49.5\pm4.9$         &  $42.5\pm7.4$    \\
130427A  &  $0.3399$               &  $1.25\pm0.15$   &  $92\pm13$         &  $19.9\pm2.9$    \\
\hline
\end{tabular}
\caption{List of the prompt and GeV emission properties of selected BdHNe and S-GRBs. We listed $z$, $E_{\rm p,i}$, $E_{iso}$ (in the rest-frame energy band $1$--$10000$~keV), and $E_{\rm LAT}$ (in the rest-frame energy band $0.1$--$100$~GeV).}
\label{tab:LAT}
\end{table}

In conclusion we have computed the occurence rate of short and long bursts following a new classification and obtaining figures in good agreement with the ones derived from population synthesis models. Essential to the classification have been the following new considerations:
\begin{itemize}
\item[1)] the binary nature of the progenitors and their separation;
\item[2)] the essential role of the hypercritical accretion process onto a NS member of a close binary system. The possible reaching of $M_{\rm crit}$ by the accretion process and the formation of a BH;
\item[3)] the activity of the newly-born BHs originating the energetic prominent GeV emission, which can be explained in terms of the gravitational energy release by accreting matter onto a Kerr BH.
\end{itemize}

This classification is now open to a verification by the addition of new GRBs sources and offer new possibilities of theoretical and observational activities including:
\begin{itemize}
\item[1)] the reaching of new observational constraints on the value of the NS critical mass $M_{\rm crit}$ and the minimum mass of a BH, which play a fundamental role in defining the separatrix among the different classes of our classification.
\item[2)] having elucidated the role of the activities of the newly-born BH in explaining the energetics of the GeV emission, in order to identify its microphysical process, the study of fundamental issues of general relativistic quantum electrodynamical processes appears to be open to further lines of inquiry \citep[see, e.g.,][and references therein]{2010PhR...487....1R};
\item[3)] it is conceivable that the sizable enlargement of the database of GRBs and of their spectral and luminosity time varability may open the possibilty of further enlarging the above classification.
\end{itemize}

\acknowledgments

We thank the Editor and the Referee for their comments which helped to improve the presentation and the contextualization of our results
This work made use of data supplied by the UK Swift Science Data Center at the University of Leicester.
J.A.R. acknowledges the support by the International Cooperation Program CAPES-ICRANet financed by CAPES-Brazilian Federal Agency for Support and Evaluation of Graduate Education within the Ministry of Education of Brazil. 
M.K. and Y. A. are supported by the Erasmus Mundus Joint Doctorate Program Grant N. 2013--1471 and 2014-0707, respectively, from EACEA of the European Commission.
M.M. acknowledges the partial support of the project N 3101/GF4 IPC-11, and the target program F.0679 of the Ministry of Education and Science of the Republic of Kazakhstan.

\appendix

\section{On the non-observed GeV emission in S-GRFs and XRFs}\label{Appendix}

\begin{figure*}
\centering
\includegraphics[width=0.49\hsize,clip]{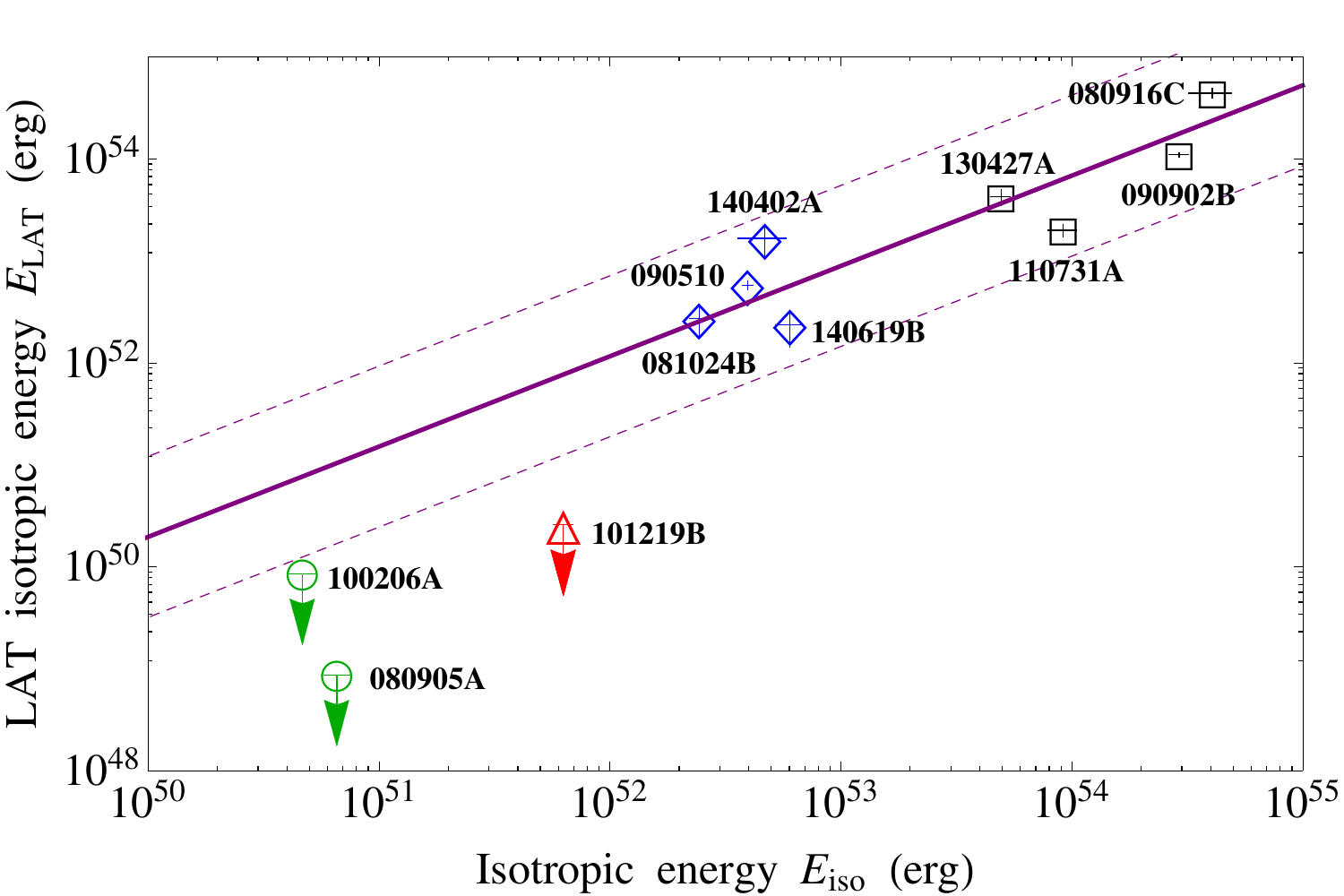}
\includegraphics[width=0.49\hsize,clip]{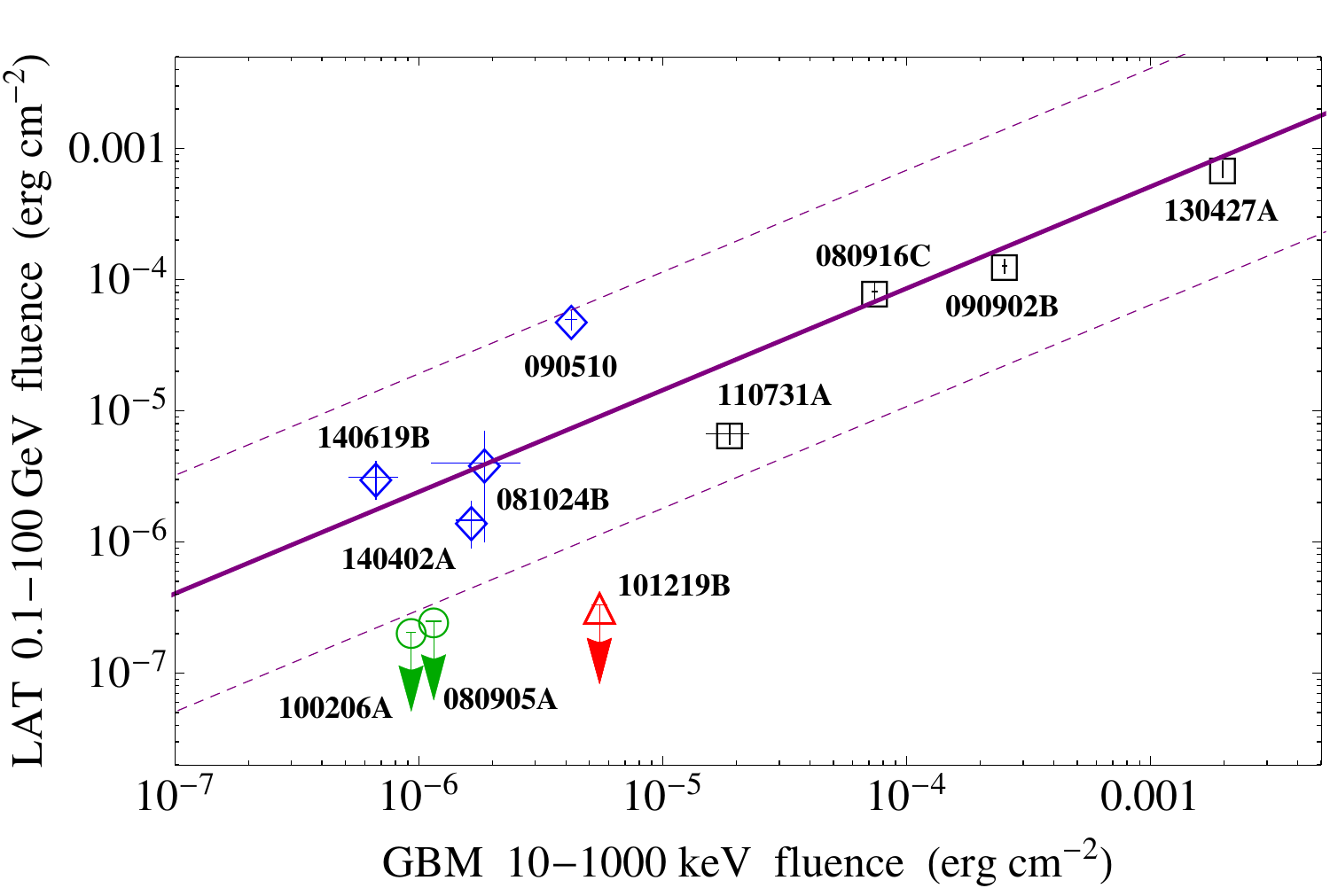}
\caption{For the sources in Tab.~\ref{tab:LAT} (BdHNe as black squares and S-GRBs as blue diamonds), as well as for 2 S-GRFs (green circles) and 1 XRF (red triangle) which did not exhibit GeV emission although they were in the optimal position ($<65^\circ$ from the LAT boresight) for its detection, we plot: (Left:) the relation between $E_{iso}$ of the prompt emission observed by the \textit{Fermi}-GBM instrument and the total isotropic energy in the $0.1$--$100$ GeV energy band observed by the \textit{Fermi}-LAT instrument (or the corresponding upper limit if not detected); and (Right:) the relation between the $10$--$1000$ keV fluence observed by the \textit{Fermi}-GBM instrument and the total $0.1$--$100$ GeV fluence observed by the \textit{Fermi}-LAT instrument (or the corresponding upper limit if not detected). The purple solid line is the relation between the plotted quantities of BdHNe and S-GRBs, and the dashed lines are the corresponding dispersion.}
\label{fig:fluence}
\end{figure*}

In Fig.~\ref{fig:fluence} we compare and contrast the sources in Tab.~\ref{tab:LAT}, all exhibiting a GeV emission, with 2 S-GRFs and 1 XRF which, as theoretically expected within the fireshell model, do not exhibit any GeV emission. All these sources were in the optimal position ($<65^\circ$ from the LAT boresight) for the detection of the GeV emission.

In the Left panel we plot the values of $E_{iso}$ and of the isotropic energy in the \textit{Fermi}-LAT energy band or the corresponding upper limits if not observed. These upper limits were obtained by using the unbinned likelihood analysis which was performed assuming an integration time of $100$~s after the flash trigger, a radius of the source region of $10^\circ$ and a zenith angle cut of $100^\circ$. This plot observationally supports the theoretical expectation, made in \citet{2016arXiv160702400E} and quoted in section~\ref{sec:descr_S-GRFs1b} above, that S-GRFs have, if any, GeV fluxes necessarily $10^5$--$10^6$ times weaker than those of S-GRBs, although their $E_{iso}$ is only a factor $10^{2}$ smaller.

Motivated by a request of the Referee, we also plotted in the Right panel, the values of the fluence observed by \textit{Fermi}-GBM and by \textit{Fermi}-LAT or the corresponding upper limits if not observed (computed as above).

From both plots it is clear that the upper limits to the GeV emission of S-GRFs and XRFs are much lower than what one may expect from the extrapolation to lower energies of the one observed in BdHNe and S-GRBs. This is a further clear observational support to the absence, theoretically implied by the fireshell model, of any GeV emission associated to S-GRFs and XRFs (see sections~\ref{sec:XRFs1b} and \ref{sec:descr_S-GRFs1b} above).

\end{document}